\documentclass[superscriptaddress,preprint,aps,onecolumn]{revtex4-1}
\usepackage{graphicx}
\usepackage{bm}
\usepackage{subfigure}
\usepackage{url}
\usepackage{float}

\begin{document}

\title{The Conformational Space of a Flexible Amino Acid at Metallic Surfaces}

\author{Dmitrii Maksimov}
\email{maksimov@fhi-berlin.mpg.de}
\affiliation{Fritz-Haber-Institut der Max-Planck-Gesellschaft, Faradayweg 4-6, 14195 Berlin, Germany}
\affiliation{Max Planck Institute for Structure and Dynamics of Matter, Luruper Chaussee 149, 22761 Hamburg, Germany}

\author{Carsten Baldauf}
\email{baldauf@fhi-berlin.mpg.de}
\affiliation{Fritz-Haber-Institut der Max-Planck-Gesellschaft, Faradayweg 4-6, 14195 Berlin, Germany}

\author{Mariana Rossi}
\email{mariana.rossi@mpsd.mpg.de}
\affiliation{Fritz-Haber-Institut der Max-Planck-Gesellschaft, Faradayweg 4-6, 14195 Berlin, Germany}
\affiliation{Max Planck Institute for Structure and Dynamics of Matter, Luruper Chaussee 149, 22761 Hamburg, Germany}

\begin{abstract}
In interfaces between inorganic and biological materials relevant for technological applications, the general challenge of structure determination is exacerbated by the high flexibility of bioorganic components, chemical bonding, and charge rearrangement at the interface. 
In this paper, we investigate a chemically complex 
 building block, namely, 
the arginine (Arg) amino-acid interfaced with Cu, Ag and Au (111) surfaces. 
We investigate how the environment changes the accessible conformational space of this amino acid, by building and analyzing a database of thousands of structures optimized with the PBE functional including screened pairwise van der Waals interactions. 
When in contact with metallic surfaces, the accessible space for Arg is dramatically reduced, while the one for Arg-H$^+$ is instead increased if compared to the gas-phase. This is explained by the formation of strong bonds between Arg and the surfaces and by their absence and charge screening on Arg-H$^+$ upon adsorption.
We also observe protonation-dependent stereoselective binding of the amino acid to the metal surfaces:
Arg adsorbs with its chiral C$_\alpha$H center pointing H away from the surfaces while Arg-H$^+$ adsorbs with H pointing toward the surface.
\end{abstract}

\clearpage
\maketitle






\section{Introduction \label{sec:Introduction}}


The organic-inorganic interfaces formed between peptides and surfaces 
are of interest to diverse fields like medicine, optoelectronics and energy storage \cite{WangGiovanni2013, GuoCahen2016, Khatayevich, Mannoor, Guy, Zhao, Sarikaya:2003-biomim, Costa:2015:IMPORTANT, Heinz:2016:Review, Walsh:2017ci}.
Amino acids and their oligomers -- i.e. peptides -- are particularly interesting materials because they are naturally biocompatible and offer a rich functional space already at the amino acid level, that can be extended by the combinatorial increase of molecular motifs available through the peptide bond formation. In these setups, the inorganic component offers a platform to immobilize and template the bioorganic counterpart, as well as to record electronic signals from interactions or reactions.
However, emergent behavior makes it impossible to base the description of such interfaces solely on the study of its independent components. 

On the experimental front, progress on the fundamental understanding of these interfaces has been achieved with recent soft landing techniques \cite{Rauschenbach2017} that allow amino acids and peptides of a few tens of building blocks to be deposited on solid surfaces and subsequently imaged in ultra-high vacuum conditions employing scanning tunneling microscopy (STM) \cite{Rauschenbach2016}. 
These experiments have shown that small sequence modifications, changes in the protonation of a peptide, and charge transfer at the interface can yield drastic changes in the two-dimensional structure and self-assembly~\cite{Rauschenbach2017, Abb2016, Humblot2016}.  

Explaining the physical mechanisms that govern adsorption and pattern formation calls for systematic theoretical investigations with atomistic resolution of such interfaces. 
However, such studies are challenging as they require (i) accurate energetics for a system build-up of elements across the periodic table and where considerable charge rearrangement and chemical reactions can occur (ii) sampling and representing a large conformational space, and (iii) dealing with structure motifs that can only be represented by unit-cells containing hundreds of atoms. Pioneering work that used density-functional theory (DFT) to study amino-acid at inorganic substrates has focused on small or rigid amino acids, and a limited portion of their conformational space \cite{DiFelice2003, DiFelice2004, Ghiringhelli2006, Arrouvel2007, Iori2008}. 
One of the first studies that was dedicated to larger amino acids was performed by Hong and coworkers \cite{Hong2009} and illustrates the challenge of properly sampling the large structure space of flexible biomolecules. These studies have clarified that an accurate energy function is only one of the ingredients needed to properly predict the structure of peptides at surfaces, with the sampling of structure space being just as important.

We propose to analyze how the structure space of a complex building block with multiple functional groups is affected by changes in its (non-biological) environment. To achieve this goal, we analyze a database of the arginine (Arg) amino-acid and its protonated counterpart (Arg-H$^+$) in the gas-phase and interfaced with Cu(111), Ag(111) and Au(111). Arginine is a good test bed because it is a relatively small molecule, but yet contains a very flexible side-chain and allows for different stable protonation states. We show that it is possible to perform an exhaustive conformational space exploration while still treating the potential-energy surface (PES) with DFT, thus capturing the charge rearrangements at the interface.
The outcome of such a computational structure-search is a large number of stationary-states on the respective PES, where not only the global minimum is of interest. Also the relative positioning of other local minima with respect to the structural degrees of freedom, as well as with respect to the energy scale of the PES, are relevant because they can reveal different basins and structures of interest under different conditions~\cite{ropo2016trends, Rossi_2013, Rossi_2014, Schubert_2015, Baldauf_2015}.
In order to handle such high-dimensional data, we make use of recent advances in machine-learning methods that can help to visualize conformational preferences of adsorbed molecules~\cite{Ceriotti2011, Tribello2012, Ceriotti2013, De2016, Bartok2017, De2017}.

In the following, we discuss the impact of the protonation state and the presence or absence of the surface on the accessible conformational space of arginine. We start by describing the procedure we followed to build the database of Arg and Arg-H$^+$ on Cu(111), Ag(111), and Au(111), based on thousands of first-principles structure optimizations which we make available to the community~\cite{DOI_of_the_data_set_at_NOMAD}.
We then analyze this database and show how different patterns of bonding and charge transfer induce fundamental changes in the accessible conformational space. We also provide an analysis of property trends across the different metallic surfaces, including protonation-dependent stereoselective binding to the surfaces and deprotonation propensities.

\section{Computational methods \label{sec:CompMethods}}

\begin{figure}[htb]
\centering
\includegraphics[width=0.6\linewidth]{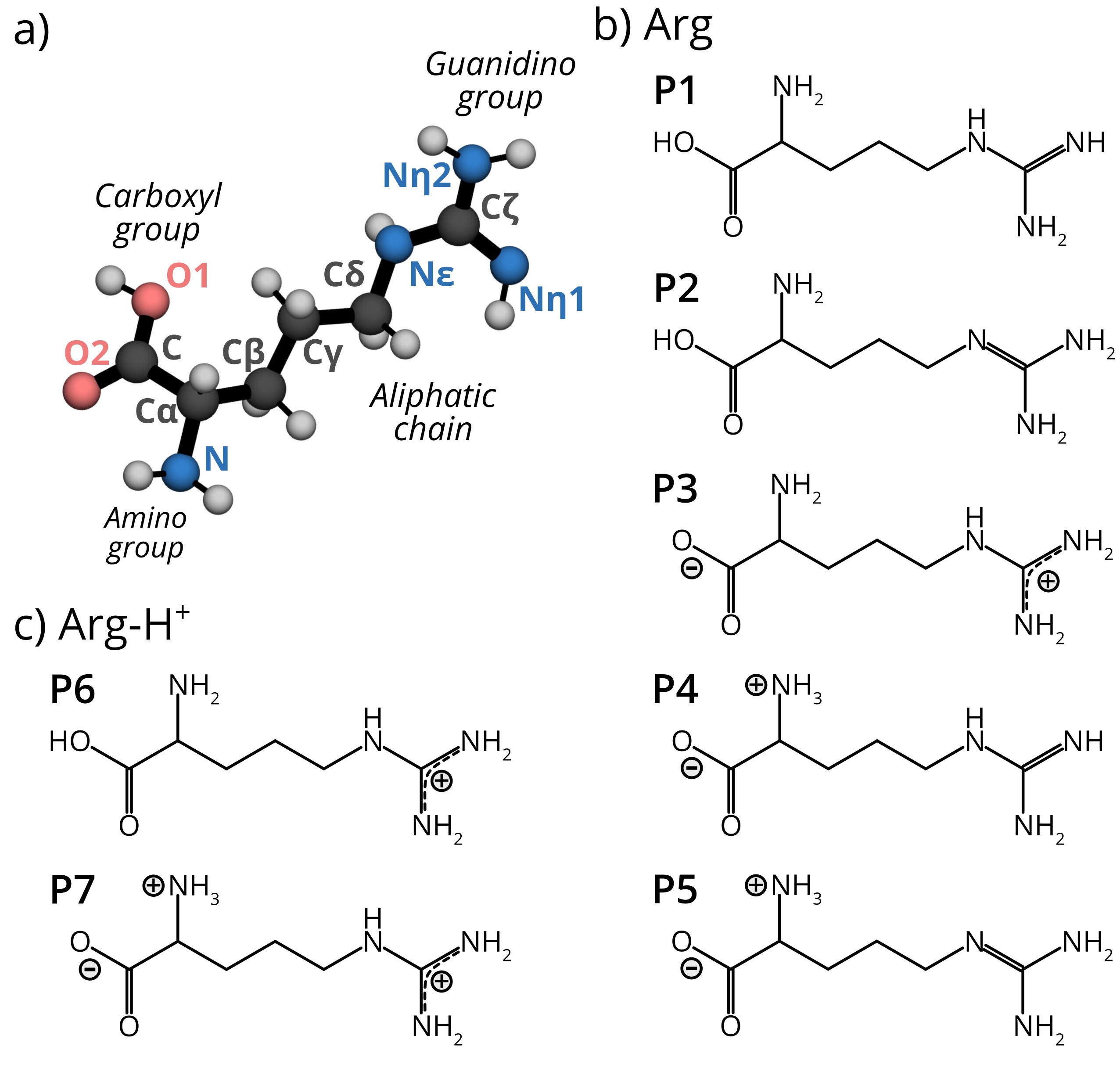}
\caption{a) Pictorial representation of the arginine amino acid, including labels of chemical groups and atoms. b) Protomers of Arg that are addressed in this work. c) Protomers of Arg-H$^+$ that are addressed in this work. }
\label{fig:protonationsketch}
\end{figure}

Arginine is an amino acid with a flexible side-chain of three (aliphatic) CH$_2$ groups and a guanidino group. 
A depiction of the Arg molecule including the labeling of the different chemical groups and specific atoms we will refer to in the manuscript is shown in Fig. \ref{fig:protonationsketch}(a). 
In the context of this publication we use the term \textit{protonation state} to distinguish between Arg and its singly-protonated form Arg-H$^+$. We use the word \textit{protomers} to distinguish different arrangements of protons within molecules of the same sum formula, for example the protomers \textbf{P1} to \textbf{P5} of Arg or the protomers \textbf{P6} and \textbf{P7} of Arg-H$^+$, shown in Fig. \ref{fig:protonationsketch}(b) and (c). In this section, we describe the computational setup, convergence tests, and sampling strategies we employed in order to build the database of Arg and Arg-H$^+$ adsorbed on different metal surfaces.

All the electronic structure calculations were carried out using the numeric atom-centered basis set all-electron code FHI-aims \cite{Blum2009, Havu2009}. We used the standard \textit{light} settings of FHI-aims for all species, except when stated otherwise. For modeling the adsorbed molecules, a $5\times6$ surface unit cell with $4\times4\times1$ $k$-point sampling was employed. The slab contains 4 layers and we added a 50 \AA~ vacuum in the $z$ direction in order to separate periodic images of the system. 
A surface unit cell of this size does not completely isolate neighboring molecules on the surface plane.
In order to estimate the magnitude of this spurious interaction, we calculated binding energies for three  Arg and three Arg-H$^+$ structures adsorbed on Cu(111) using different surface unit cell sizes. As shown in the SI (Tables S3 and S4), the relative binding energies change by no more than 50 meV when reaching a $10\times12$ cell. 

An accurate description of the dispersion interactions that play a role in weakly-bonded adsorbates on metallic surfaces can be achieved with newly-proposed methods that take into account electronic screening and the the many-body nature of the dispersion term \cite{Hermann2019}; however, such methods are not yet computationally feasible for large-scale studies.
Considering dispersion interactions is, however, crucial to model such interfaces \cite{Ruiz2016, Ruiz-thesis, Liu2013, Liu2012, Al-Saidi2012, VanRuitenbeek2012, Wagner2012, Carrasco2014}. We thus employ the PBE exchange-correlation functional augmented by the TS-vdW$^{\text{surf}}$ \cite{Ruiz2016} dispersion correction between pairs of atoms that involve the adsorbed molecule (molecule-molecule and molecule-surface dispersion). 
Adding pairwise vdW corrections for pairs of metal atoms did not result in a systematic improvement of the lattice constants due to the complex many-body nature of the electronic screening within the metallic bulk.  
Because the PBE lattice constants for Cu, Ag, and Au are in good agreement with experimental data (see Table S1 of the SI) and the electronic structure itself is not changed by the inclusion of these types of vdW interactions, we chose to use the simplest setup. 
We optimized all structures 
until all forces in the system were below 0.01\,eV/\AA\,. 
We also fixed the two bottom layers of the slabs in all optimizations. 
A dipole correction was applied in $z$ direction to compensate for the dipole formed by the asymmetric surface configurations. 

\subsection{Database Generation \label{sec:ModelGeenration}}

\begin{figure}[htb]
\centering
\includegraphics[width=\textwidth]{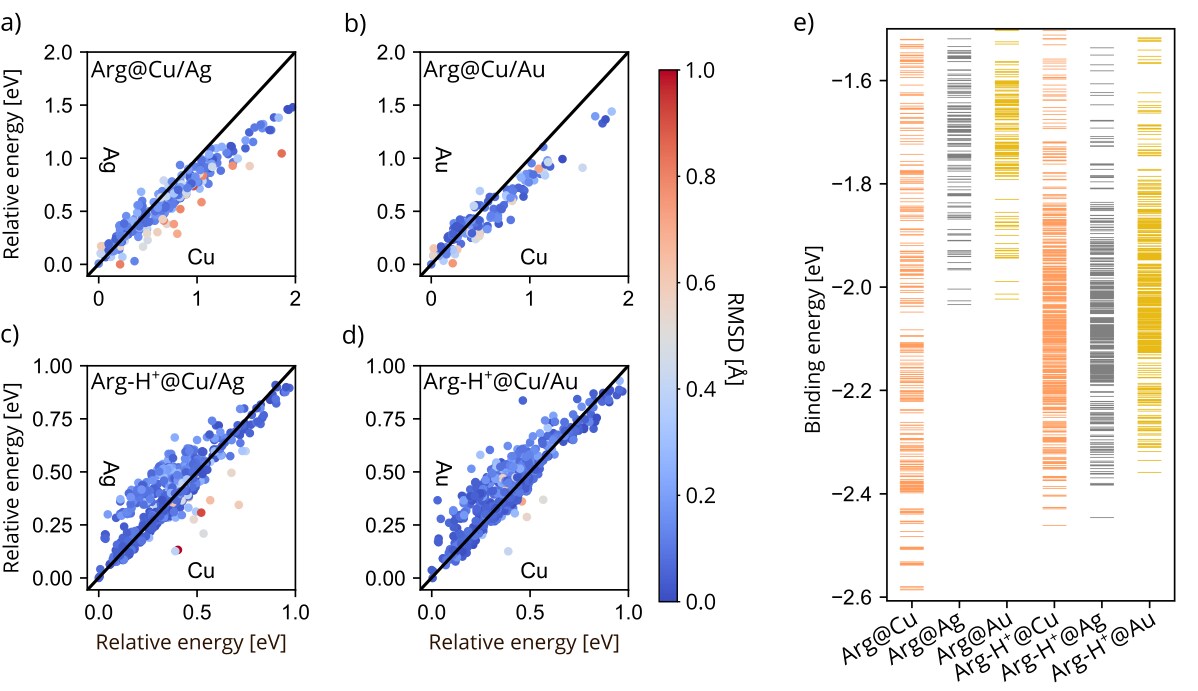}
\caption{(a-d) Correlation plots of relative energies of Arg or Arg-H$^+$ conformers on Cu, Ag, and Au (111) surfaces. Each dot corresponds to the same conformer optimized on the two surfaces addressed in each panel, color coded with respect to the RMSD (heavy atoms only) between the superimposed optimized structures without taking surface atoms into consideration. (e) Binding energies of Arg and Arg-H$^+$ on Cu(111), Ag(111) and Au(111) surfaces.}
\label{fig:relrel}
\end{figure}

The sampling of the structure space of the amino acid in two protonation states on metallic surfaces was performed by starting from a previously published data-set comprising stationary points of isolated amino acids and dipeptides \cite{Ropo2016,ropo2016trends}. 
For Arg, 1206 structures were present in the database.
In order to reduce the number of possibilities, but keeping a representative share of the structures, we considered the 300 lowest energy conformers, the 27 highest energy conformers, and 125 conformers uniformly spanning the energy range in between. 
For the Arg-H$^+$ amino acid, all 215 structures present in the gas-phase data set were used in this study.

We distinguish \textit{upstanding} positions of the molecules where the largest principal axis of the molecule is approximately perpendicular to the surface plane, from \textit{flat lying} positions with an arrangement parallel to the surface.
For Arg, 3 lying configurations per structure were generated by randomly placing the molecule flat on the Cu(111) surface and then rotating it by $120^{\circ}$ around the principal axis. 
Two upstanding configurations were generated for the 25 lowest-energy gas-phase structures by first placing the molecule in a random upright orientation and then flipping it. 
For Arg-H$^+$ a similar procedure was adopted: flat lying positions were created by $90^{\circ}$ rotations around the principal axis and upstanding configurations were created for 27 structures that were uniformly selected from the whole energy range. 
In summary, we considered a total of 1156 conformers of Arg@Cu(111) and 914 conformers of Arg-H$^+$@Cu(111).

All optimized structures that fell within a range of 0.5 eV from the global minimum on Cu(111) were transferred to Ag(111) and Au(111) and further optimized. In addition, we randomly picked 105 Arg-H$^+$ structures representing the higher energy range on Cu(111) to be further optimized on Ag(111) and Au(111). Moreover, for Arg, 180 randomly picked structures representing the higher energy range were considered on Ag(111) and 61 on Au(111). The total amount of calculated structures for each case is summarized in Table \ref{tbl:totalnmbrofstructs}. 

\begin{table}
\center
 \caption{Number of calculated Arg and Arg-H structures in isolation and adsorbed on Cu(111), Ag(111) and Au(111).}
 \label{tbl:totalnmbrofstructs}
\begin{tabular}{|r|c|c|c|c|}
\hline
     & Gas phase & Cu(111) & Ag(111) & Au(111) \\
\hline
Arg  & 1206      & 1156    & 327     & 209  \\
\hline
Arg-H$^+$ & 215       & 914      & 718       & 721         \\
\hline
\end{tabular}
\end{table}

We checked that this strategy ensured a sufficient sampling of the low-energy range of both Arg and Arg-H$^+$ on Ag(111) and Au(111) by analyzing the alterations in relative energy hierarchies on the different surfaces. In Fig. \ref{fig:relrel}(a-d) each dot corresponds to a conformer that was optimized first on the Cu(111) surface and then post-relaxed on Ag(111) or Au(111). Within the lowest 0.5 eV range, we do not observe any significant rearrangement of the energy hierarchy with respect to the Cu(111) surface. The energy hierarchies of both Arg and Arg-H$^+$ on the Ag(111) and Au(111) surfaces are almost identical. The most pronounced outliers in all plots correlate with a higher root mean square displacement (RMSD) of the molecular atoms (i.e. disregarding the surface-adsorption site), thus pointing to a structural rearrangement of the molecule.

\subsection{Structure space representation and analysis \label{sec:soap-map}}

We analyse the structure space of all systems considered, as well as its alterations by employing a dimensionality reduction procedure that makes it more intuitive to understand the high-dimensional space. Following Ref. \cite{De2017}, we represent the local atom-centered environments of the structures through the smooth overlap of atomic positions (SOAP)\cite{Bartok2017} descriptors. We then obtain the similarity matrix between different conformers with the regularized entropy match kernel (REMatch) \cite{De2016}. We used SOAP descriptors with a cutoff of 5.0 \AA, a Gaussian broadening of $\sigma=0.5$ \AA~ and an intermediate regularization parameter $\gamma$=0.01. SOAP kernels were calculated only for heavy atoms in the molecule (disregarding metal and hydrogen atoms) and were obtained using the GLOSIM package \cite{De2016, GLOSIM}. 

For the dimensionality-reduced representation, we here chose to use the multi-dimensional scaling (MDS) algorithm as implemented in the \texttt{scikit-learn} package\cite{scikit-learn}. This algorithm is similar to the Sketchmap algorithm previously employed in Ref. \cite{De2017}, but we found it more suitable for the data at hand, which is composed of decorrelated local stationary-points, instead of structures generated from molecular dynamics trajectories. In short, the low-dimensional map was obtained considering all calculated structures of Arg in the gas-phase and through minimizing the stress function
\begin{equation}
\sigma = \sum_{i\neq j} (D_{ij}-d_{ij})^2,  
\end{equation}
where $D_{ij}$ is the similarity between structures $i$ and $j$ in the high-dimensional space and $d_{ij}$ the Cartesian distance in the low-dimensional (2D) space. 
We then projected structures in different environments onto the pre-computed map of gas-phase Arg by fixing the parameters of the map and finding the low-dimensional coordinates $x$ that minimize  
\begin{equation}
\sigma_P(x)= \sum_{i=1}^{N_{\text{Arg}}} \left(|X-X_i|-|x-x_i|\right)^2 
\label{eq:proj}
\end{equation}
where the sum runs over all structures in the optimized gas-phase Arg map and $X$ represents the high-dimensional representation of the new structure to be projected. In order to classify structural patterns, we employ the following notations:
\begin{itemize}
    \item We represent the protomers by the labels shown in Fig. \ref{fig:protonationsketch}(b) and (c). 
    \item We identify the presence of strong intramolecular hydrogen bonds (H-bonds) whenever the distances between the hydrogen connecting donor and acceptor are below 2.5 \AA. We label the H-bond pattern between two atoms in the molecules according to the nomenclature shown in Fig. S3 in the SI.
    \item We further classify the structures according to the longest distance between two heavy atoms in the molecule.
\end{itemize}

\subsection{Binding energies and charge rearrangement \label{sec:BindingEnergies}}

The binding energies for each structure were calculated as follows. For neutral Arg we computed 
$
E_{\textsf{b}} =  E_{\textsf{mol@surf}} - E_{\textsf{surf}} - E_{\textsf{mol}}, 
$
where $E_{\textsf{mol@surf}}$ corresponds to the total energy of the interface, $E_{\textsf{surf}}$ is the total energy of the pristine metallic slab and $E_{\textsf{mol}}$ the total energy of the lowest energy gas-phase conformer.

For charged Arg-H$^+$, we considered the binding energy of a two-step reaction. \textit{First}, the interface is formed between the charged molecule and the clean surface:
$
E_{\textsf{b1}} = E_{\textsf{mol}^+\textsf{@surf}} - E_{\textsf{surf}} - E_{\textsf{mol}^+},
$
where $E_{\textsf{mol}^+}$ the total energy of the most stable gas-phase conformer of the isolated charged molecule. \textit{Second}, an electron from the metal neutralizes the unit cell where the adsorbed molecule is located, yielding
$
E_{\textsf{b2}} = E_{\textsf{mol@surf}} - E_{\textsf{mol}^+\textsf{@surf}} - E_{f},
$
where $E_f$ corresponds to the Fermi energy of the metallic surface. The final binding energy is thus considered to be
\begin{equation}
E_{\textsf{b}}^+ = E_{\textsf{b1}} + E_{\textsf{b2}} = E_{\textsf{mol@surf}} - E_{\textsf{surf}} - E_{f}- E_{\textsf{mol}^+}.
\end{equation}
For reference, we report the values we used for $E_f$ at each surface in Table S5. These binding energies are shown in Fig. \ref{fig:relrel}d, and will be discussed in detail in Section \ref{sec:Results}.

In addition,  for conformers within 0.1 eV of the respective global minimum of each system, we have calculated free energies at finite temperatures within the harmonic approximation \cite{McQuarrie, Fultz2010}, in order to address question about thermal stability of such structures. We have calculated
$
F_{\textsf{harm}}(T) = E_{\textsf{PES}} + F_{\textsf{vib}}(T),
$
where $E_{\textsf{PES}}$ is the total energy obtained from DFT, and we have used textbook expressions for the harmonic vibrational Helmholtz free energy $F_{\textsf{vib}}(T)$. We have calculated the Hessian matrix only taking into account displacements of the adsorbate. The surface was only considered as an external field, which is a good approximation for physisorbed systems. 

To address charge rearrangements after adsorption on the surface, we calculated electron density differences for selected  structures on each surface by computing
$
\Delta\rho = \rho_{\textsf{mol@surf}} - \rho_{\textsf{surf}} - \rho_{\textsf{mol}},  
$
and in the case of Arg and 
$
\Delta\rho^{(+)} = \rho_{\textsf{mol@surf}} - \rho_{\textsf{surf}} - \rho_{\textsf{mol}^{(+)}},
$
in the case of Arg-H$^+$. In these expressions, $\rho_{\textsf{mol@surf}}$ is the total electron density of the interface, $\rho_{\textsf{surf}}$ is the electron density of the slab without molecule, and $\rho_{\textsf{mol}}$ and $\rho_{\textsf{mol}^+}$ are electron densities of neutral Arg and charged Arg-H$^+$ molecules with the same geometries as in interface. The $+$ sign denotes that the final density difference integrates to +1 electron in the case of Arg-H$^+$.

\section{Results and discussion \label{sec:Results}}

\subsection{The unconstrained structure space: Arg in isolation \label{sec:Isolation}}

\begin{figure}[htbp]
\center
\includegraphics[width=0.87\linewidth]{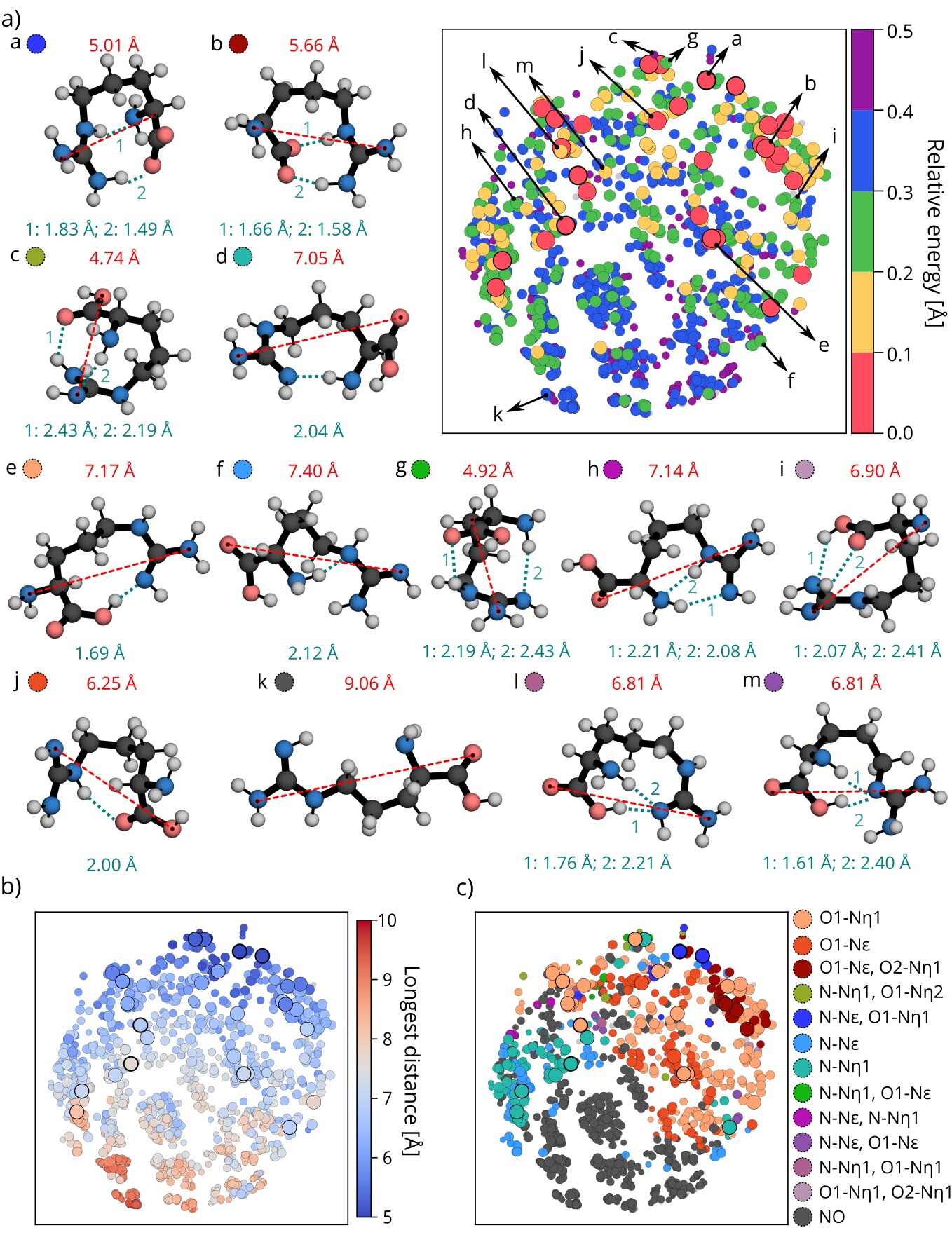}
\caption{Low-dimensional map of Arg stationary points on the PES. Only points linked to structures with a relative energy of 0.5 eV or lower are colored. Representative structures of all conformer families are visualized as well as their H-bond distances (in turquoise) and longest distance between two heavy atoms (in red) of the molecule. The maps are colored with respect to a) relative energy, b) longest distance, and c) H-bond pattern. The size of the dots also reflect their relative energy, with larger dots corresponding to lower energy structures.}
\label{fig:Arg_Sketchmap}
\end{figure}

We start by analysing the unconstrained conformational space of Arg in isolation, which is formed by more than 1200 local stationary states \cite{Ropo2016,ropo2016trends}. In order to rationalize different structural arrangements in this space, we utilize the dimensionality-reduction algorithm discussed in Section \ref{sec:soap-map} and build a two-dimensional map. On this map, shown in Figure \ref{fig:Arg_Sketchmap}(a), each dot represents one structure. A close proximity of the dots implies similarities of the heavy-atom arrangement between the conformations. This is the low-dimensional map that is taken as a reference for comparison throughout this manuscript. 

We proceed to color-code the dots on the map according to different properties. In Fig. \ref{fig:Arg_Sketchmap}(a) we show the map colored by the relative energy of each structure with respect to the global minimum $\Delta E_{\text{rel}}$. We only color structures with $\Delta E_{\text{rel}} < 0.5$ eV. The region with $\Delta E_{\text{rel}} < 0.1$ eV is colored red and is represented by 32 different structures that occupy different parts of the map. The dominant protomer (98.6\%) among these conformers is the one labeled \textbf{P1} in Fig. \ref{fig:protonationsketch}, i.e. non-zwitterionic. 
However, the lowest energy structure, labeled \textit{a} in panel (a) of Fig. \ref{fig:Arg_Sketchmap}, is protomer \textbf{P3}, with a shared proton between the carboxylic and the guanidino group. 
This structure is compact, with the longest distance within the molecule of only 5.01 \AA~ and presenting two strong intramolecular H-bonds.
Zwitterionic protomers, denoted as \textbf{P4} and \textbf{P5} in Fig. \ref{fig:protonationsketch}, do not appear in the gas-phase.

Inspecting the map in Fig. \ref{fig:Arg_Sketchmap}(a), it is clear that low-energy conformers are almost exclusively present in the upper hemisphere of the plot. 
This can be rationalized in terms of the structural motifs that occupy these two halves of conformational space:
In Fig. \ref{fig:Arg_Sketchmap}(b), we color-code the dots in terms of the longest extension of the conformers. 
While the upper hemisphere features compact structures, the lower hemisphere of the map is populated by extended conformers (with longest extensions between 7.5 \AA~ and 10.0 \AA).
Many of them do not contain any H-bond or contain only one H-bond between the carboxyl and amino group.
Extended conformers of Arg are energetically unfavoured in the gas-phase as the formation of strong H-bonds is crucial for the stabilization of Arg in isolation. 
Comparing the different plots in Figure \ref{fig:Arg_Sketchmap}, we see that all low-energy structures with $\Delta E_{\text{rel}} < 0.1$ eV are indeed compact with one or two H-bonds. 

In Fig. \ref{fig:Arg_Sketchmap}(c), we identify in total 13 different families with respect to the number and character of H-bonds in the molecule, with $\Delta E_{\text{rel}} < 0.5$ eV. Representative structures of all families are shown in panel (a). 
This family classification  helps us understand why in Fig. \ref{fig:Arg_Sketchmap}(a) there are structures of higher energies at similar regions as structures with lower energies. Even though these structures are typically in the same protomeric state and have a similar arrangement of the heavy-atoms, the carboxyl group can rotate, giving rise to different H-bond patterns.
These different arrangements can give rise to energy differences of up to 0.2 eV, as exemplified in Fig. S4 in the SI. Including hydrogens in the SOAP descriptors used to build the 2D map could provide a better energy separation, but it would prevent us from comparing different protonation states, as shown in the next section.

\subsection{Adding a proton: Arg-H$^+$ in isolation \label{sec:Proton}}

\begin{figure}[htbp]
\center
\includegraphics[width=0.9\linewidth]{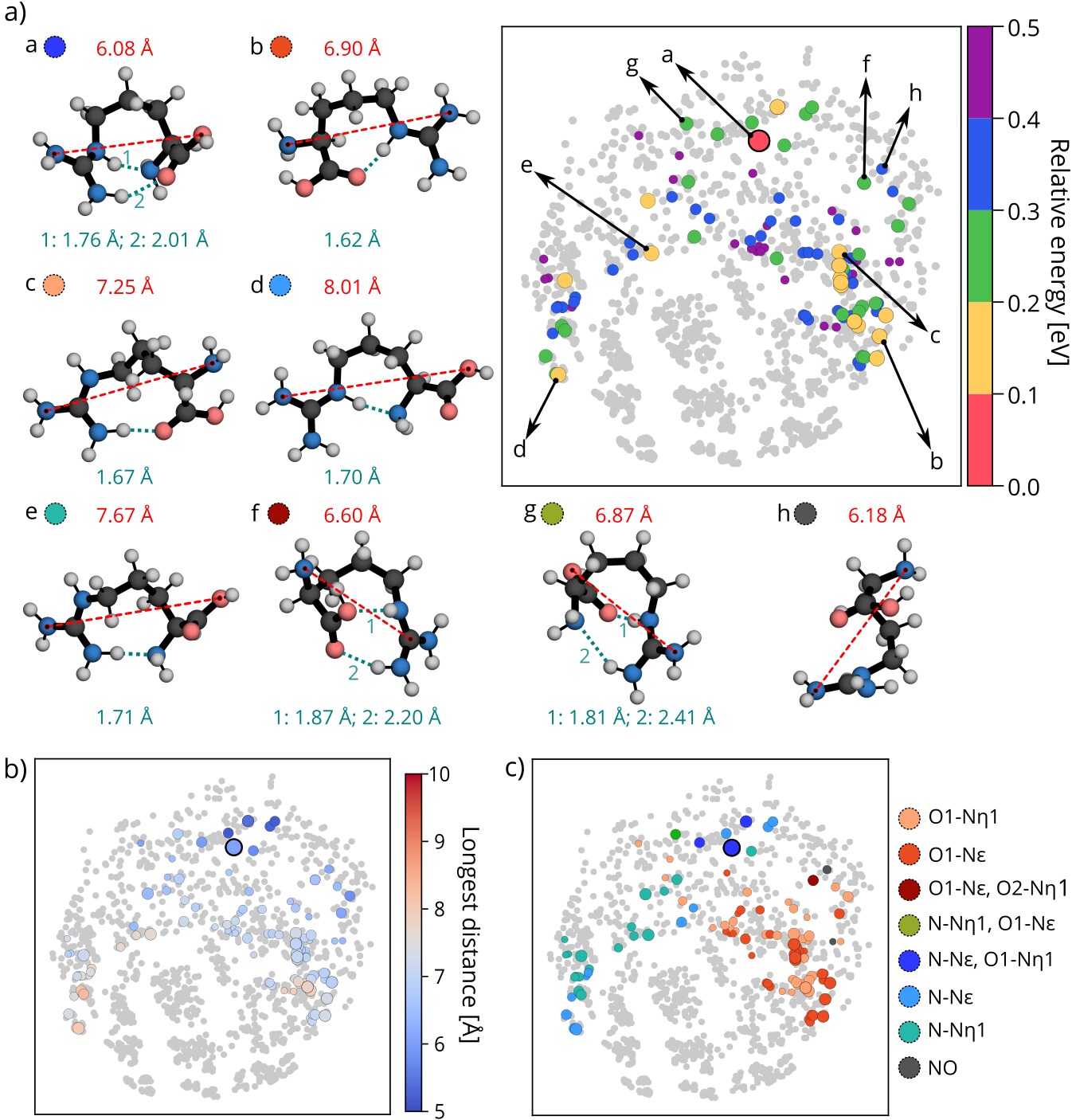}
\caption{Representative conformers of the populated structure families within 0.5 eV of the global minimum of isolated Arg-H$^+$ and low-dimensional projections of all populated conformers onto the Arg map. Grey dots represent all structures from the original map of isolated Arg in Figure 3, and serve as a guide to the eye. The maps are colored with respect to a) relative energy, b) longest distance within the molecule, and c) H-bond pattern.}
\label{fig:arghmaps}
\end{figure}

Arg-H$^+$ is of particular interest as it is the most abundant form of this amino acid under physiological pH conditions \cite{BolgerBook1995}, and we thus investigate changes of the conformational space introduced by the addition of a proton to the Arg amino-acid. 
To that end, we plot a projection of all stationary points of the Arg-H$^+$ PES with $\Delta E_{\text{rel}} < 0.5$ eV (referenced to its own global minimum) onto the map that was previously created for Arg. 
In Fig. \ref{fig:arghmaps}(a), we color the dots in the map according to $\Delta E_{\text{rel}}$, in Fig. \ref{fig:arghmaps}(b) according to the longest distance between heavy atoms in the molecule, and in Fig. \ref{fig:arghmaps}(c) according to the H-bond pattern. The grey dots in the maps represent all points in the Arg map of Figure \ref{fig:Arg_Sketchmap} and are shown for ease of comparison.

The unique conformation types of Arg-H$^+$ can be grouped into 8 different families in this energy range, which are represented in Fig. \ref{fig:arghmaps}(a). Most families only have one H-bond and there are no zwitterionic protomers. 
This means that in isolation only the protomer \textbf{P6} is populated.
It is worth noting that under physiological conditions (in solution), the zwitterionic protomer \textbf{P7} is preferred.

There are only two very similar structures (same family) with $\Delta E_{\text{rel}} < 0.1$ eV in this case. The global minimum, labeled \textit{a} in Fig. \ref{fig:arghmaps}(a), contains two H-bonds within the molecule, between atoms N-N$\mathrm{\varepsilon}$ and O1-N$\eta$ (see Fig. \ref{fig:protonationsketch}). 
This particular structure resembles the lowest-energy structure of Arg with a proton added to the carboxyl group.
This protonation results in an extension of the molecule by around 1 \AA. 
That correlates with the location of the lowest-energy structure being slightly shifted on the map towards the region containing more extended structures.

\begin{figure}[ht]
\centering
\includegraphics[width=0.75\linewidth]{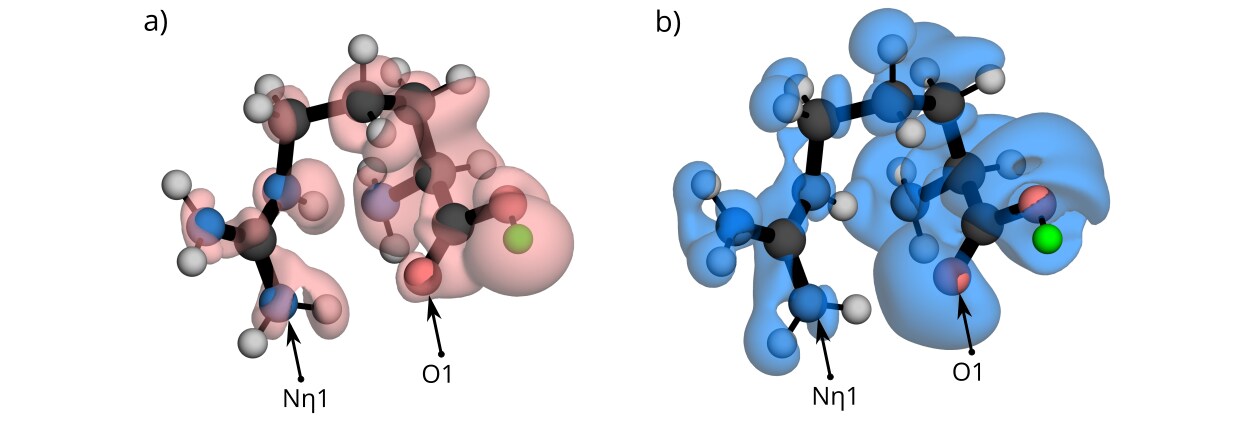}
\caption{Electron density difference between Arg-H$^+$ and Arg calculated by neutralizing the charge and removing the hydrogen connected to the carboxyl group (marked in green) from the lowest energy structure of Arg-H$^+$. The isosurfaces of electron density with value $\pm0.005$ e/Bohr$^3$ corresponding to the a) regions of electron accumulation on Arg-H$^+$ and b) where the electron depletion on Arg-H$^+$, both compared to Arg.}
\label{fig:ArghArg_electrondensitydifferencce}
\end{figure}

We note that the structure space of Arg-H$^+$ is contained within the conformational space of Arg and also drastically reduced in numbers if compared to Arg: 
There are only 108 structures with $\Delta E_{\text{rel}} < 0.5$ eV, compared to 1179 structures in the Arg case. 
In this energy range, regions of the map with very compact and very extended structures are not populated in this protonation state. 
This can be traced to the constraint imposed by the addition of the proton, that make extended structures less stable due to the strong driving force to neutralize the charge imbalance created by the proton on the guanidino group.
To rationalize why the most compact conformers are also less populated, we show in Fig. \ref{fig:ArghArg_electrondensitydifferencce} the electron-density differences between the lowest energy Arg-H$^+$ conformer and an Arg conformer created by fixing the same Arg-H$^+$ structure, but neutralizing the charge and removing the hydrogen connected to the carboxyl group. This modification yields the same covalent connectivity observed in the global minimum of Arg. 
We show isosurfaces corresponding to electron accumulation in Arg-H$^+$ in red and electron depletion in Arg-H$^+$ (accumulation in Arg) in blue. 
We observe a density surplus between the O1 and N$\eta$ atoms in Arg, favoring the formation of a stronger H-bond leading to a more compact structure.

\subsection{Adsorption of Arg on Cu, Ag, Au (111) surfaces}

\begin{figure}[htbp]
\centering
\includegraphics[width=0.85\linewidth]{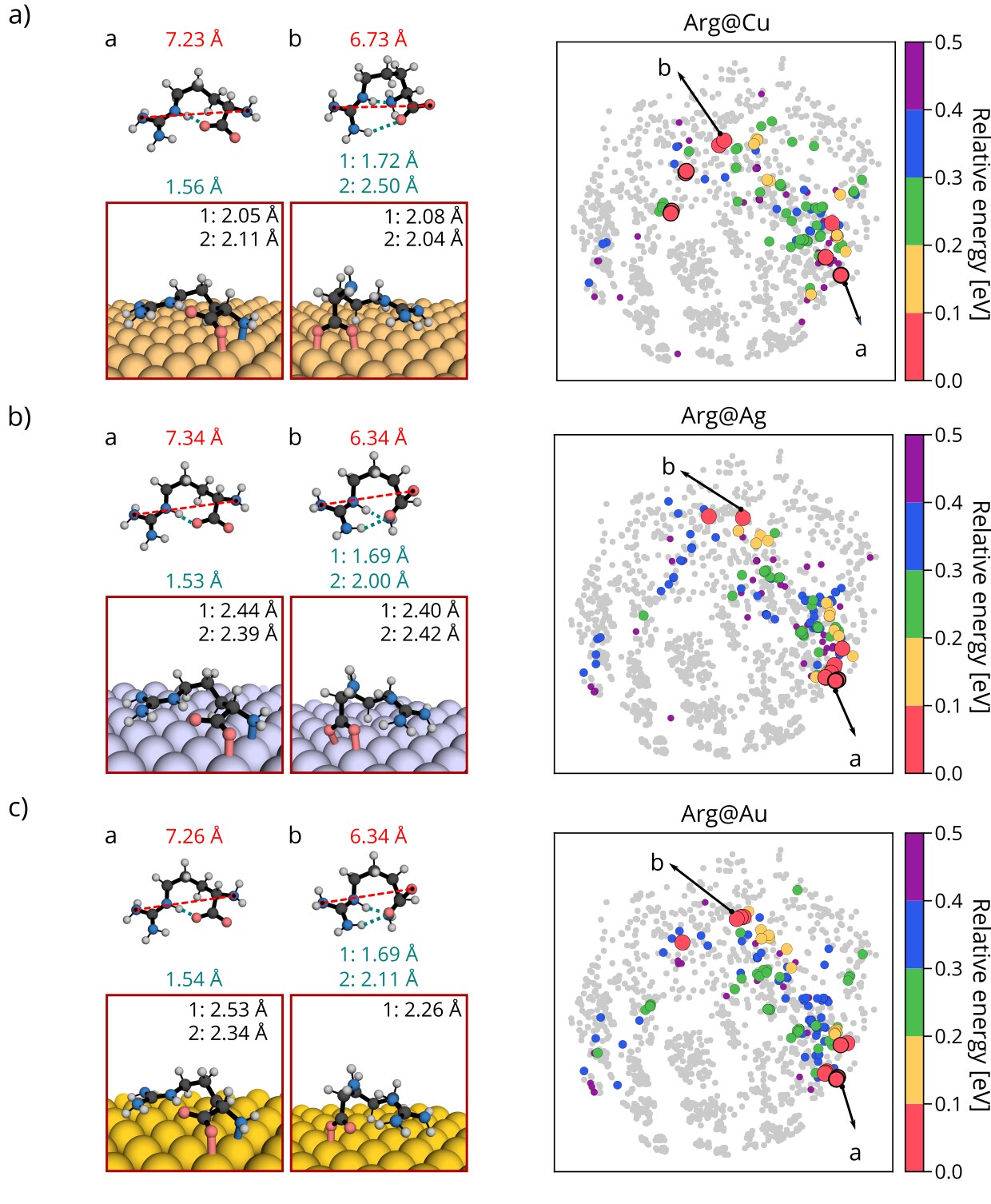}
\caption{
Low-dimensional projections of conformers of Arg adsorbed on a) Cu(111), b) Ag(111), and c) Au(111), onto the gas-phase Arg map of Fig. 3. Only conformers within 0.5 eV from the respective global minimum are colored. Grey dots represent all structures from the original map of gas-phase Arg, and serve as a guide to the eye.
In each panel, representative structures are shown from two perspectives: a side view where molecule and surface are shown (bottom), and the corresponding top view (top) where only the molecule is shown. The longest distance within each visualized conformer is reported in red and H-bond lengths are reported in turquoise.}
\label{fig:argsurf_sketchmap}
\end{figure}

We now turn to the analysis of the conformational space of Arg when in contact with metal surfaces, namely Cu(111), Ag(111), and Au(111). 
In Figure \ref{fig:argsurf_sketchmap}, we show map-projections of the stationary points with $\Delta E_{\text{rel}} < 0.5$ eV (referenced to each respective global minimum) of Arg adsorbed on the three surfaces.
The conformational space of Arg upon adsorption is reduced and the conformers occupy similar regions of the map as the ones from Arg-H$^+$. 
We will learn in the following that this is mainly due to the formation of strong bonds with the surface that result in steric constraints of the space and also partially due to electron donation from the molecule to the metallic surfaces.

The lowest energy structure lies on the same part of the map on all surfaces, which is different from the area where the gas-phase global minimum of Arg was located. These conformers, labeled \textit{a} in Figure \ref{fig:argsurf_sketchmap}(a), (b) and (c) form a strong H-bond between atoms O1 and N$\epsilon$. The longest distance within molecule lies between 7.20-7.35 \AA. This structure binds most strongly to all three surfaces through its amino and carboxyl groups. 

Other low-energy structures in all surfaces form strong bonds to the surfaces only through the carboxyl group, as exemplified by the structure labeled \textit{b} in all panels of Fig. \ref{fig:argsurf_sketchmap}. 
These bonds are formed most favorably on \textit{top} positions, i.e. vertically on top of a surface metal atom.
In particular for Cu(111), the atomic spacing of the Cu atoms on the surface favors both oxygens to bind on \textit{top} positions simultaneously.

The favorable formation of these bonds is connected with the fact that all conformers with $\Delta E_{\text{rel}} < 0.2$ eV are in the protomeric state \textbf{P3}, in which the carboxyl group is deprotonated.
The bonds to the surface and a favorable vdW attraction effectively flatten the molecular conformation, thus energetically favoring more elongated structures. 
Protomers of type \textbf{P1}, which was dominant in the gas-phase, only appear  with $\Delta E_{\text{rel}} > 0.3$ eV on Cu and Ag, and with $\Delta E_{\text{rel}} > 0.2$ eV on Au.
Protomers \textbf{P4} and \textbf{P5} are again not observed. 
Regarding the intramolecular H-bond patterns, within 0.5 eV from the global minimum
we can identify 7 different families on Cu(111), 
and 6 families on both Ag(111) and Au(111). 
These families contain H-bonds where the carboxyl group predominantly participates. All families are represented in Table S4 in the SI.

\subsection{Adsorption of Arg-H$^+$ on Cu, Ag, Au (111) surfaces}

\begin{figure}[htbp]
\centering
\includegraphics[width=0.83\linewidth]{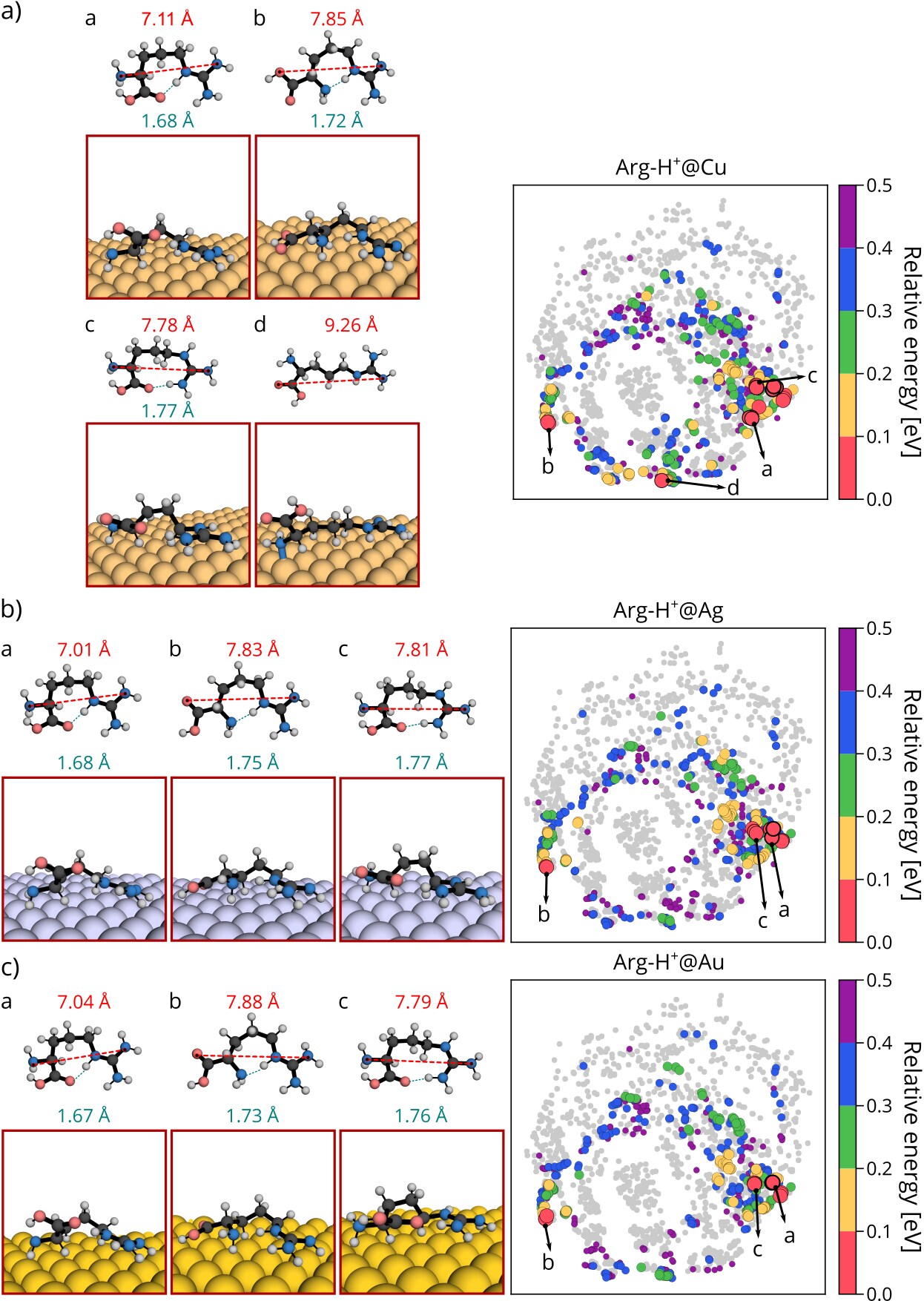}
\caption{Low-dimensional projections of conformers of Arg-H$^+$ adsorbed on a) Cu(111), b) Ag(111), and c) Au(111), onto the gas-phase Arg map of Fig. 3. Only conformers within 0.5 eV from the respective global minimum are colored. Grey dots represent all structures from the original map of gas-phase Arg, and serve as a guide to the eye.
In each panel, representative structures are shown from two perspectives: a side view where molecule and surface are shown (bottom), and the corresponding top view (top) where only the molecule is shown. The longest distance within each visualized conformer is reported in red and H-bond lengths are reported in turquoise.}
\label{fig:arghsurf_map}
\end{figure}

Finally, we characterize the conformational-space changes arising from the simultaneous addition of a proton and the adsorption onto metallic surfaces.
In Figure \ref{fig:arghsurf_map}, we show the projection of Arg-H$^+$ conformers adsorbed on Cu, Ag, and Au(111) onto the map of isolated Arg. 
These projections, in particular the comparison of the plots in Figs. \ref{fig:argsurf_sketchmap} and \ref{fig:arghsurf_map}, reveal that 
the conformational space of adsorbed Arg-H$^+$ is larger than the one of adsorbed Arg. 
While Arg-H$^+$ features more than 500 conformers within $ \Delta E_{\text{rel}} < 0.5$ eV, Arg only counts about 150 conformers in the same energy range.
Interestingly, the adsorption of Arg-H$^+$ to a metal surface also results in an increase of the occupied structure space in comparison to isolated Arg-H$^+$ (108 structures with $ \Delta E_{\text{rel}} < 0.5$ eV), shown in Fig. \ref{fig:arghmaps}. 
In fact, the structures occupy similar regions of the map as the ones occupied by Arg-H$^+$, with the addition of extended structures that are located in the bottom of the map.


We identify 4 different families on Cu(111) and 3 on Ag(111) and Au(111) with $ \Delta E_{\text{rel}} < 0.1$ eV. 
Representative conformers of these families are shown in Fig. \ref{fig:arghsurf_map}. 
The lowest energy conformer, labeled \textit{a} in Fig. \ref{fig:arghsurf_map}(a)-(c), appears on all surfaces at the same region of the map as for adsorbed Arg.  The largest distance within the molecule lies around 7 \AA~ and it also has a strong H-bond linking the carboxyl-O and the N$\varepsilon$ atoms. 
The structure, however, does not present the same orientation to the surface as compared to the lowest energy conformer of Arg, and does not form strong bonds with the surface.
With the exception of the extended structure on Cu(111), labeled \textit{d} in  Fig. \ref{fig:arghsurf_map}(a), all conformers with $ \Delta E_{\text{rel}} < 0.1$ eV on all surfaces contain one intramolecular H-bond involving either 
carboxyl-O and N$\varepsilon$ atom (labeled a), 
backbone N and N$\varepsilon$ atoms (labeled b) or 
carboxyl O and a N$\eta$ atom (labeled c). 
Compared to adsorbed Arg, adsorbed Arg-H$^+$ structures become on average 1.0 \AA~ more extended (see SI). The guanidino and carboxyl groups often lie parallel to the surface. 
The protomer \textbf{P6}, the only one present in the gas-phase, is dominantly populated also on the surfaces. 
However, we do observe very few conformers in the zwitterionic \textbf{P7} state. These structures are at least 0.2 eV higher in energy than than the global minimum. 

With respect to the number of bonds that Arg-H$^+$ forms with the surface, the picture is very different from adsorbed Arg. 
Within the lower 0.15 eV, we do not observe short (strong) bonds of O or N atoms to the surfaces. 
This lack of constraint by the surface contributes to the increased structure space of adsorbed Arg-H$^+$. 
In addition, the molecule accepts electrons from the surface, becoming less positively charged, as we discuss in detail in the next section. 
We conclude that Arg-H$^+$ interacts with the metallic surfaces mostly through van der Waals and electrostatic interactions.

\subsection{Electronic structure and trends across surfaces}


In the previous sections we focused on structural aspects of the adsorbed molecules and the most prominent bonds the molecules make with the metallic surfaces. In the following, we will discuss different aspects of the molecule-surface interactions with the goal of identifying trends across these systems.

The binding energies for all surfaces were calculated as discussed in Section \ref{sec:BindingEnergies}. 
The larger negative values in Fig. \ref{fig:relrel}(d) correspond to stronger binding of the molecule to the surface. 
In the case of adsorbed Arg, many conformers bind to Cu more strongly than to Ag and Au.
As mentioned previously, Arg forms strong bonds to the surfaces, but the binding of the deprotonated carboxyl group of Arg to the Cu(111) surface is geometrically favored as discussed above.
In the case of adsorbed Arg-H$^+$, there is no pronounced difference in binding strengths to the surfaces and the values are comparable to binding energies obtained for Arg adsorbed on Cu(111). 
This correlates with the observation that the interaction of Arg-H$^+$ to the surfaces happens mostly through dispersion and electrostatic interactions. Despite the strong binding to the surface, it is also visible in Fig. \ref{fig:relrel}(a) that the interaction of Arg-H$^+$ with the surface does not strongly template the conformations of this molecule, implying a low corrugation (i.e. homogeneity) of the molecule-surface interaction and allowing for a larger variety of conformers with similiar energies. This is in contrast to the molecule-surface interation of Arg, that is more inhomogeneous due to the formation of bonds through specific chemical groups.
Additionally, we have estimated harmonic vibrational free energies for all conformers with $\Delta E_{\text{rel}} < 0.1$ eV in each surface. In contrast to what has been reported for longer helical peptides \cite{Rossi_2013,Schubert_2015}, the global minimum remains the same in all cases, as reported in Fig. S6 in the SI. For Arg-H$^+$ we observe relative energy rearrangements of up to 50 meV at 300 K, which changes the relative energy hierarchy of conformers less stable than the global minimum.

\begin{figure}[htbp]
\centering
\includegraphics[width=0.95\textwidth]{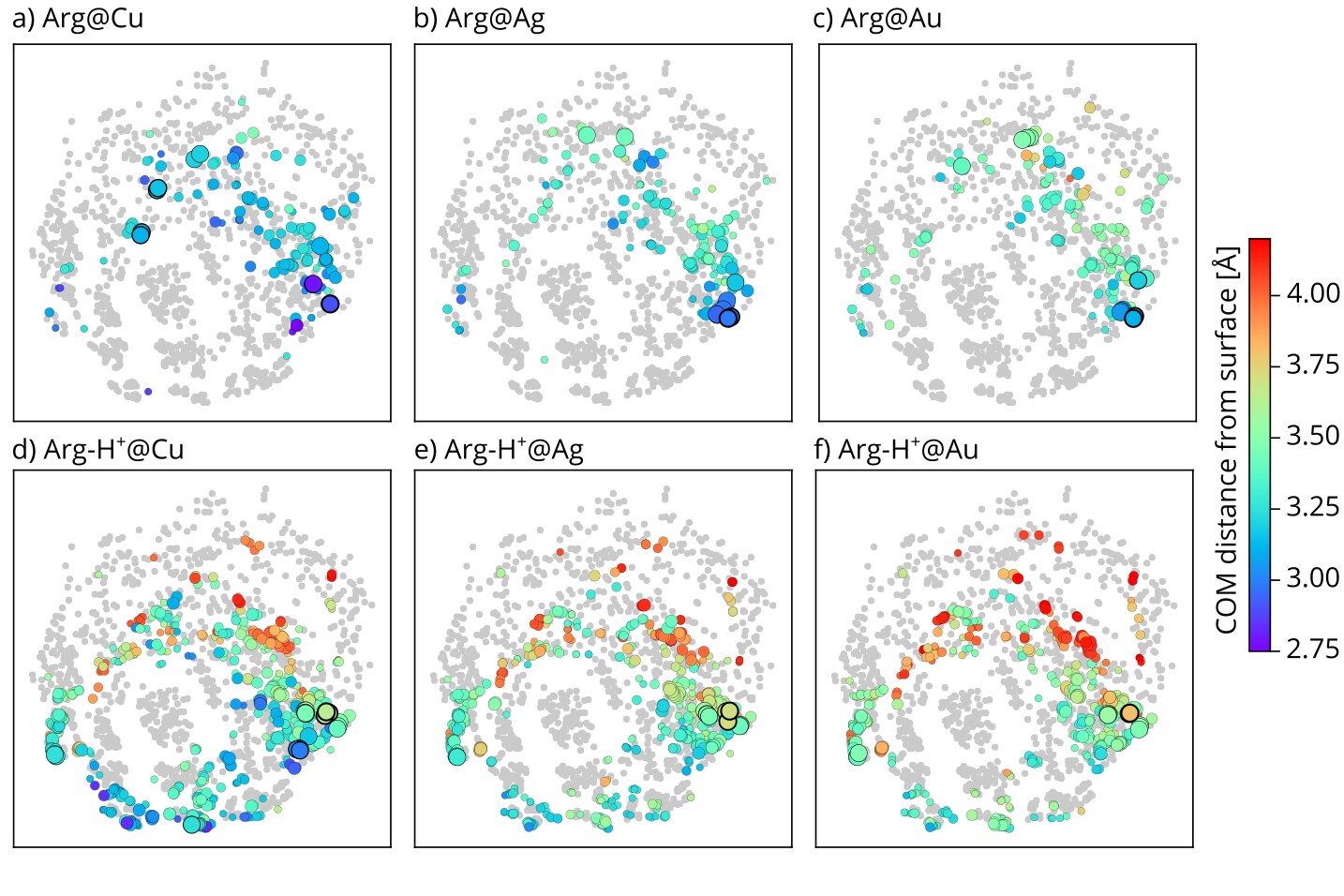}
\caption{Low dimensional projections of adsorbed Arg and Arg-H$^+$ on Cu(111), Ag(111) and Au(111) color-coded with respect to the distance of the center of mass of the molecule with respect to the surface. Grey dots represent all structures from the original map of isolated Arg where the projection was made, and serve as a guide to the eye. }
\label{fig:Closest_map}
\end{figure}

We then focus on the distance between the molecule and the surfaces. We define this quantity by measuring the distance of the center of mass (COM) of the molecule with respect to the  
surface plane defined by the top layer of surface atoms.
These distances are collected in Fig. \ref{fig:Closest_map}. 
The COM is closer to Cu(111) than to Ag(111) and Au(111) for both Arg and Arg-H$^+$. 
In addition, in all surfaces, Arg lies closer than Arg-H$^+$, in agreement with the observation that Arg forms stronger bonds to the surface. 
The extended structures of Arg-H$^+$, at the bottom of the maps, tend to be closer to the surface than those that have H-bonds within the molecule, likely due to the stronger vdW attraction to the surface by extended conformations.

\begin{figure}[ht!]
\centering
\includegraphics[width=0.9\textwidth]{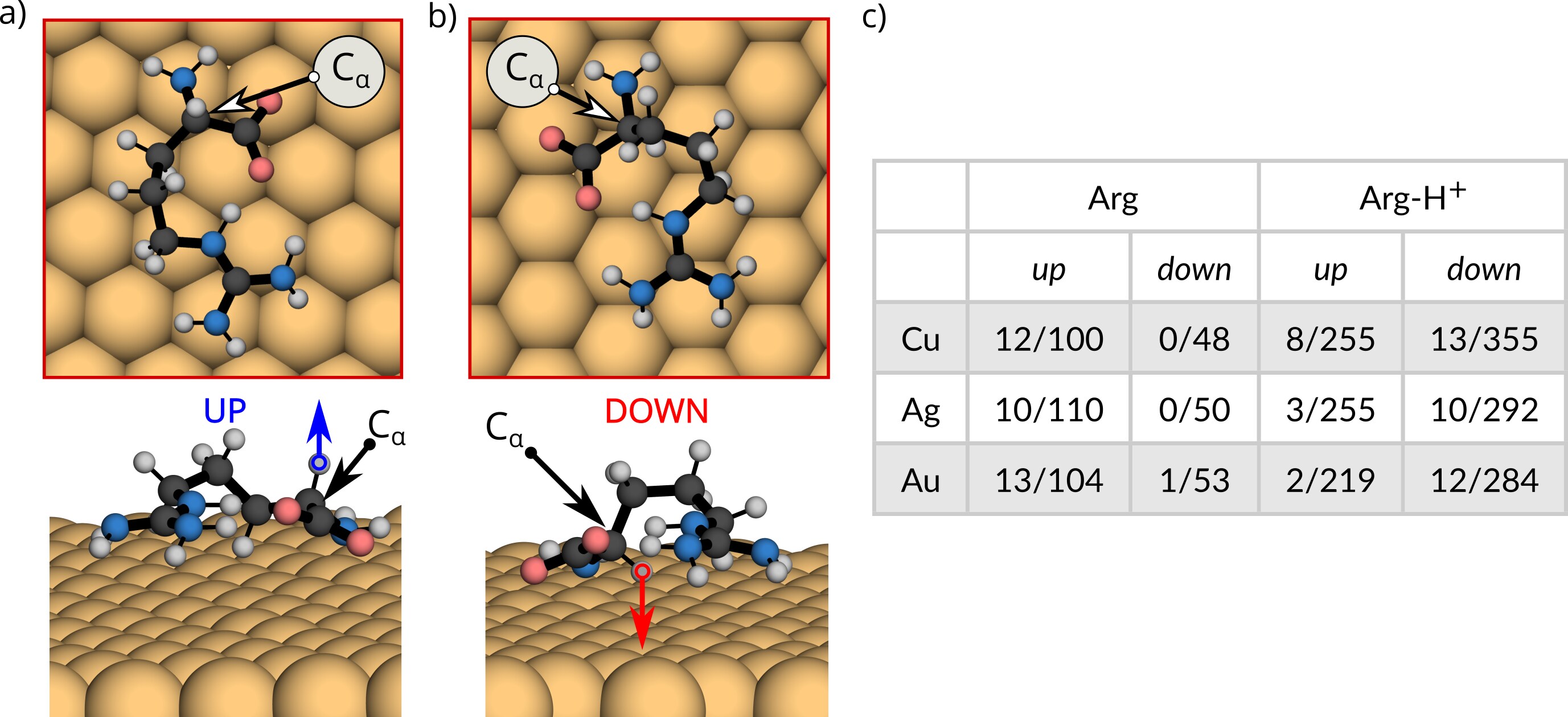}
\caption{Orientation of the C$_\alpha$H group in a) \textit{up} orientation (hydrogen pointing towards vacuum) and b) \textit{down} orientation (hydrogen pointing towards the surfaces). c) The amount of structures with \textit{up} and \textit{down} orientation within 0.1/0.5 eV from the global minimum of each surface.}
\label{fig:UpDown}
\end{figure}


The differences in COM distance to the surfaces between Arg and Arg-H$^+$ are apparently related to the preferred orientation of the chiral center of the molecule to the surface. 
The chiral C$_\alpha$ carbon can point its bonded hydrogen towards the surface (labeled \textit{down} in the following), or towards the vacuum region (labeled \textit{up} in the following).
Examples of different molecular orientation are shown in Fig. \ref{fig:UpDown}(a).
The dominant orientation with respect to the surface is different in the cases of Arg and Arg-H$^+$, as evidenced by the numbers presented in Fig. \ref{fig:UpDown}(b). 
The lower energy structures are mostly in the \textit{up} orientation for Arg and mostly in the \textit{down} orientation for Arg-H$^+$ (see also map in Fig. S7).
As discussed in the previous sections, despite showing different orientations of their C$_\alpha$H groups, the lowest energy structures for both molecules adsorbed on each surface have very similar conformations. 
Since the addition or removal of a proton can apparently alter the preference of the chiral-center orientation, we propose that it could template different chiralities of  self-assembled super-structures on the surface \cite{LingenfelderThesis}.

\begin{figure}[hb!]
\includegraphics[width=\textwidth]{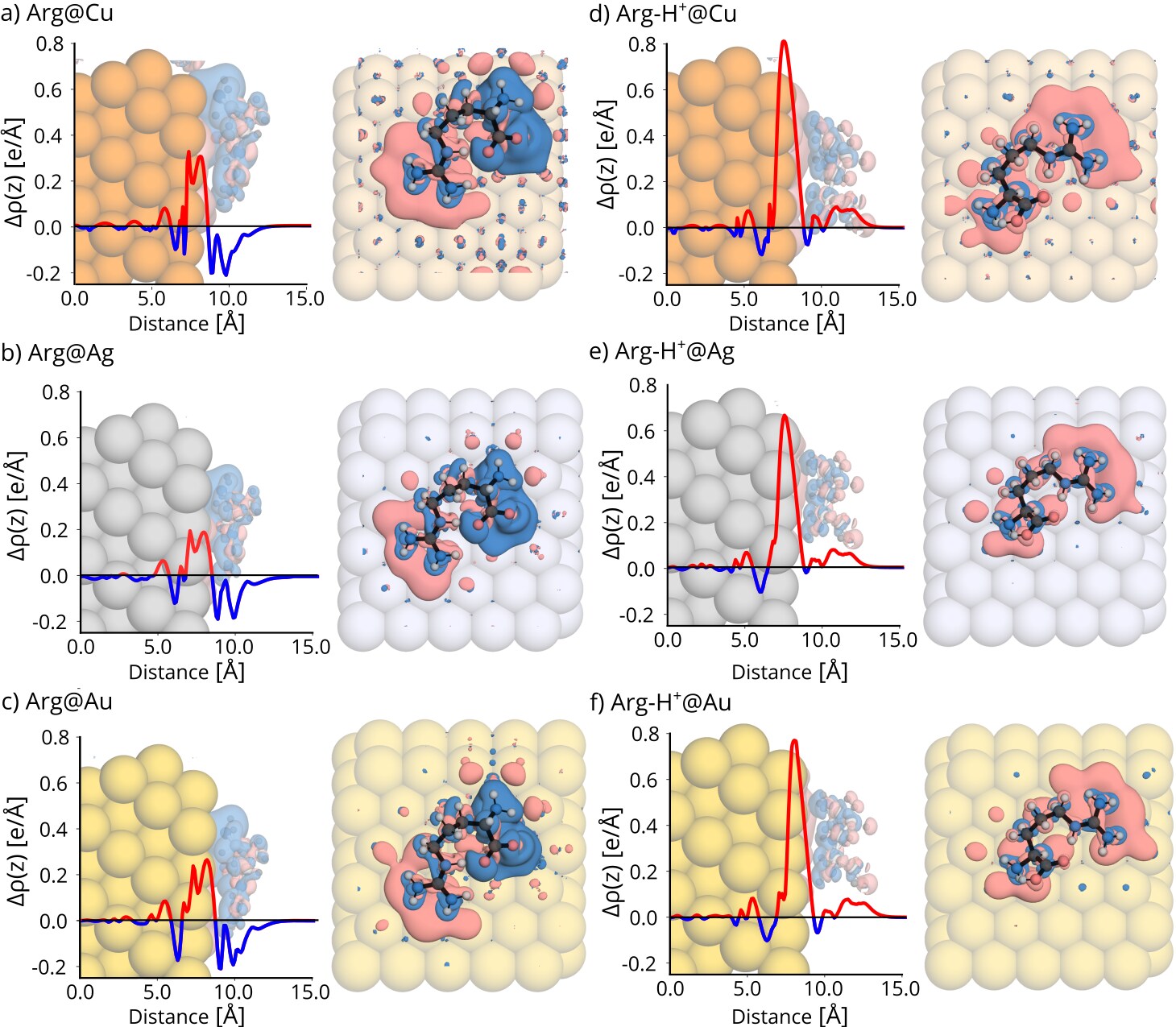}
\caption{Electronic-density difference averaged over the directions parallel to the surface for the lowest energy conformers of Arg adsorbed on Cu(111) (a), Ag(111) (b), and Au(111) (c), as well as of Arg-H$^+$ adsorbed on Cu(111) (d), Ag(111) (e), and Au(111) (f). Positive values (red) correspond to electron density accumulation and negative values (blue) correspond to electron density depletion.
In each panel, we also show a side and top view of the 3D electronic density rearrangement.  Blue isosurfaces correspond to an electron density of +0.05 e/Bohr$^3$ and red isosurfaces to -0.05 e/Bohr$^3$. \label{fig:charge-re}}
\end{figure}

We then investigated the rearrangement of the electronic density upon binding of the molecules to the different surfaces. 
In Fig. \ref{fig:charge-re} we show the electronic density rearrangement created by the lowest energy conformer at each surface, integrated over the axis parallel to the surface, overlaid on the side-view of the 3D density rearrangement. In addition, we show a top view of the density rearrangement in each case. Examples of further conformers are summarized in the SI, Figs. S8-S13. The data shows that Arg donates electrons to the surface, while Arg-H$^+$ accepts electrons from the surface. We have checked this propensity for selected conformers by integration of the electronic density rearrangement around the molecule and by calculating the Hirshfeld charge remaining on the molecule for the full database (see SI, Table S6). When comparing Hirshfeld charges on the molecule and those obtained from the electronic density rearrangement, we observe that Hirshfeld charges are always 0.3-0.5 e underestimated. 
In addition, we observe that the depletion and accumulation of charge is not uniform through the lateral extension of the molecule. 
This behavior is consistent with the level alignment predicted by the PBE Kohn-Sham energy levels, as shown in Fig. S14 in the SI. However, we note that quantitative values of charge transfer are often inaccurate at this level of theory, as characterized in Refs. \cite{Egger_2015, Liu_2017}. Optimally tuned range-separated hybrid functionals would yield more accurate values, but their computational cost is prohibitive for the use in this whole database. Nevertheless, hybrid-functional calculations of selected conformers (see SI, Fig. S15) confirm the qualitative trend. Therefore, we conclude that the protonation state again critically impacts these systems, in this case by qualitatively changing the redistribution of electronic charge.   

\begin{figure}[ht!]
 \centering
 \includegraphics[width=0.65\textwidth]{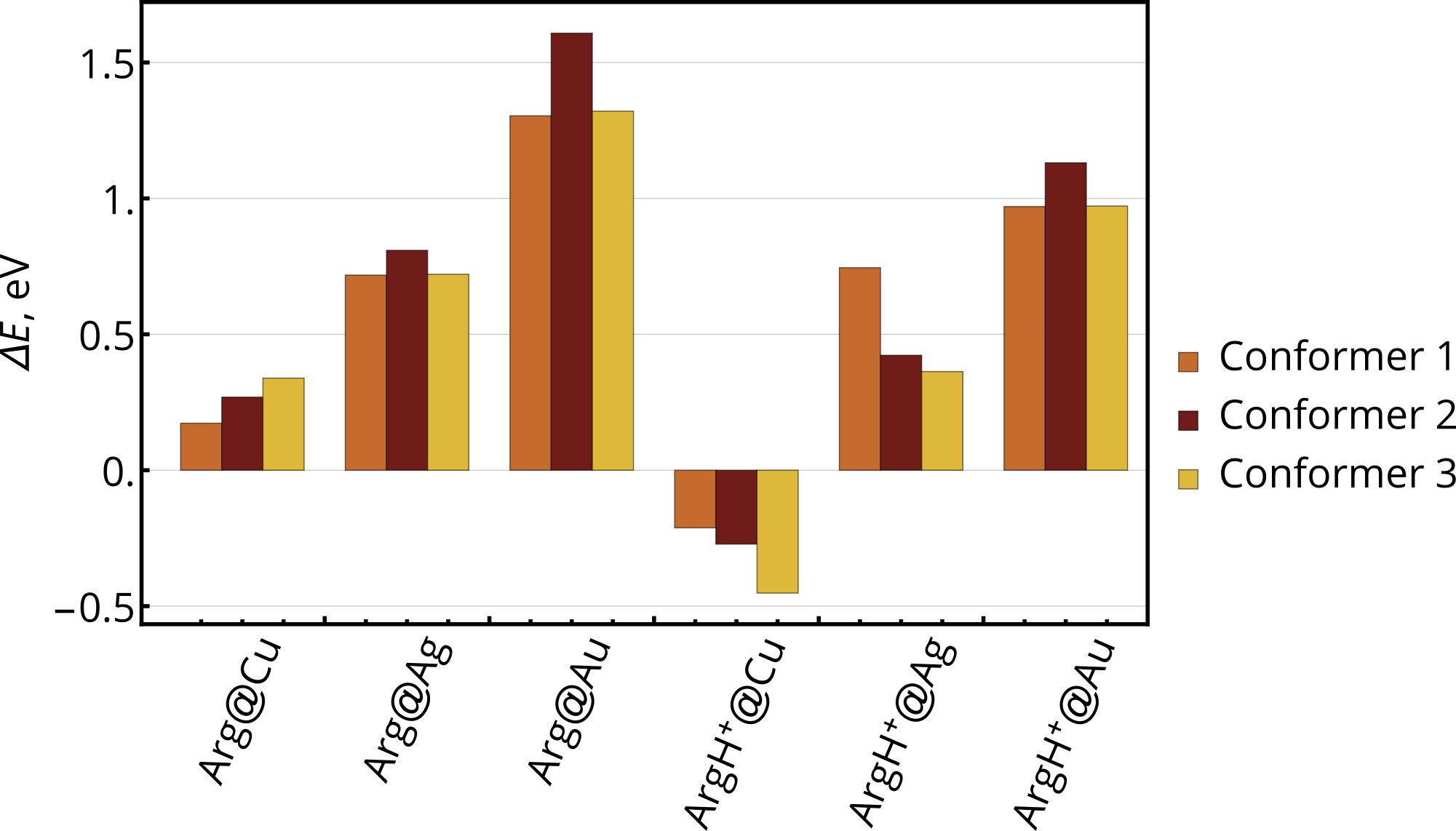}
 \caption{Energy differences upon hydrogen dissociation for selected conformers (see text and SI) of Arg and Arg-H$^+$ on all metallic surfaces. $\Delta E = E_{\textbf{dep}}-E$, where $E_{\textbf{dep}}$ is the total energy of the dissociated structure after optimization (including the adsorbed hydrogen) and $E$ the energy of the optimized intact structure. A negative $\Delta E$ indicates that deprotonation is favored.}
 \label{fig:deprotonation}
 \end{figure}

It was observed experimentally that amino acids can undergo deprotonation on reactive surfaces \cite{BARLOW1998322, BarlowRaval2004, METHIVIER201588, MATEOMARTI2002191, MARTI200485, EralpTrolle2010}. 
Here we also investigated whether deprotonation of Arg and Arg-H$^+$  was favorable on any of the surfaces studied here, the details of the procedure can be found in the SI. The results are summarized in Fig. \ref{fig:deprotonation}. They show that deprotonation of Arg-H$^+$ is favorable in Cu(111), such that Arg-H$^+$ would be predominantely deprotonated. The population of deprotonated Arg on Cu(111) could reach about 0.1\% of the molecules at room temperature, from a simple Arrhenius estimation. Given that we have not observed any spontaneous dissociation upon optimization of Arg-H$^+$ on Cu(111), we conclude that, although favorable, this dissociation of H does not occur without a barrier. In all other surfaces, the barrier for dissociation would be rather high for both molecules.

\section{Conclusions}

In this paper, we have characterized the conformational space of the arginine amino acid in its neutral and protonated form in different non-biological environments, i.e. in isolation and in contact with metallic surfaces. In particular, we have analyzed how and why different parts of the conformational space become accessible or excluded depending on the protonation state and the environment, showing the importance of bond formation and charge rearrangement in these systems. 

This study included the construction of a database based on thousands of structures optimized by density-functional theory including dispersion interactions. The construction of this database is a result in itself. We found, for example, that it is advantageous to start from a comprehensive sampling of the conformational space of the least-constrained molecular form, which in our case was the neutral Arg amino acid in the gas-phase. This is evidenced by the fact that in our low-dimensional projections, all low-energy conformers we observe on the surfaces, for both Arg and Arg-H$^+$ lie among structural conformations that were already present in the gas-phase sampling of Arg, albeit often with high relative energies. In addition, we find that within Cu, Ag, and Au surfaces, relative energies between different conformers are largely preserved when changing the substrate.

In recent years much effort has been invested in developing and improving classical force fields (FF) for simulation of organic-inorganic interfaces, including peptides on surfaces \cite{Iori2009, Wright2013, Heinz2013}. However, as we discuss in the SI, the existing force fields are not accurate for these scenarios where amino acids are in contact with metal surfaces and under vacuum conditions. Nevertheless, these situations are extremely relevant for scanning tunneling microscopy experiments \cite{Rauschenbach2017, Rauschenbach2016, Abb2016, KoslowskiThesis} and other technological applications \cite{WangGiovanni2013, GuoCahen2016}. As we illustrate in the SI, even though these force-fields can sample the relevant areas of conformational space, they are not able to capture consistent energy hierarchies. Additionally, the molecular chemical groups show a preference to adsorb on different surface sites, which could have considerable impact on self-assembly studies. Databases such as this one will serve as an important source of data for further parametrization and improvement of these potentials. 

Regarding the structural space of Arg and Arg-H$^+$ adsorbed on (111) surfaces of Cu, Ag and Au, we have learned the following:
The adsorption of Arg leads to the formation of strong bonds to the surface that involve mostly the carboxyl and amino groups. This stabilizes the protomer that we label \textbf{P3} in this work, where the carboxyl group is deprotonated and the side chain is protonated. This is different to the dominant protomer in the gas phase \textbf{P1}. The bonds to the surface sterically constrain the conformations of this molecule, thus decreasing the amount of structures 
with respect to the numbers observed in the gas phase. 
The molecule also donates electrons to the surface, becoming slightly positively charged. 
We do not observe fully extended structures lying on the surface and most conformers exhibit intramolecular H-bonds. The majority of conformers of Arg in the low-energy region adsorbs with the C$_{\alpha}$H chiral center pointing the hydrogen atom away from the surfaces.

Arginine in its protonated form, i.e. Arg-H$^+$ is the most abundant form of this amino-acid in biological environments, where it typically adopts the zwitterionic protomer \textbf{P7}. 
In the gas-phase, we observe that the non-zwitterionic state \textbf{P6} is dominant and that the addition of the proton decreases the number of allowed conformations with respect to isolated Arg due to the added electrostatic interactions, and the passivation of the carboxyl group that would otherwise be involved in intramolecular H-bonds. 
Upon adsorption to the metallic surfaces, we observe that the protomer \textbf{P6} is still dominant and that there are no strong bonds formed to the surface. In addition, this molecule receives electrons from the surface, thus becoming less positively charged. Both effects conspire to yield a homogeneous (flat) molecule-surface interaction and a relatively high population of different structures in the low-energy range. Contrary to Arg, most low-energy conformers of Arg-H$^+$ adsorb with the C$_{\alpha}$-H chiral center pointing the hydrogen atom towards to the surfaces. Finally, through the calculation of dissociation energies, we also conclude that the deprotonation of Arg-H$^+$ is energetically favorable only on Cu(111). 

Our observations regarding the preferred protomers and deprotonation propensities discussed above are consistent with the observations in the literature that the adsorption of amino acids in their anionic and deprotonated form is common on reactive metals like Cu(111) \cite{Costa:2015:IMPORTANT}. 
One pronounced difference that we find among surfaces is the average adsorption height of the molecules: They follow the trend Cu(111) $<$ Ag(111) $<$ Au(111), and Arg is always closer than Arg-H$^+$ to the same respective surface.

The set of electronic-structure calculations presented here show that a flexible amino-acid like Arginine presents a rich conformational space involving different protomeric states and molecule orientations with respect to the surface, allied to a complex charge rearrangement. Going forward, it is clear that the likes of this study based solely on DFT cannot become a routine method due to the elevated computational cost. Addressing the whole breadth of amino acids as well as self assembly of these structures on surfaces will profit from this study as a benchmark and a means to develop models, possibly based on different machine-learning techniques \cite{Todorovic2019, Oganov2009, Dieterich2010, fafoom, genarris, HORMANN2019, Olsson8265}, that can bypass the cost of thousands of DFT structure optimizations.


\section*{Availability of the data}
The data presented here is has been uploaded to the NOMAD repository \cite{DOI_of_the_data_set_at_NOMAD}.

\section*{Acknowledgements}
 This work was supported by the Max Planck-EPFL Center for Molecular Nanoscience and Technology. We acknowledge Michele Ceriotti for insightful discussions and critical comments to the manuscript. We also acknowledge fruitful discussions with Yair Litman, Stephan Rauschenbach and Sabine Abb. MR also acknowledges support from the Max Planck Research Network on Big-Data-Driven Science (BigMAX).

\bibliography{bibliography}

\begin{thebibliography}{69}%
\makeatletter
\providecommand \@ifxundefined [1]{%
 \@ifx{#1\undefined}
}%
\providecommand \@ifnum [1]{%
 \ifnum #1\expandafter \@firstoftwo
 \else \expandafter \@secondoftwo
 \fi
}%
\providecommand \@ifx [1]{%
 \ifx #1\expandafter \@firstoftwo
 \else \expandafter \@secondoftwo
 \fi
}%
\providecommand \natexlab [1]{#1}%
\providecommand \enquote  [1]{``#1''}%
\providecommand \bibnamefont  [1]{#1}%
\providecommand \bibfnamefont [1]{#1}%
\providecommand \citenamefont [1]{#1}%
\providecommand \href@noop [0]{\@secondoftwo}%
\providecommand \href [0]{\begingroup \@sanitize@url \@href}%
\providecommand \@href[1]{\@@startlink{#1}\@@href}%
\providecommand \@@href[1]{\endgroup#1\@@endlink}%
\providecommand \@sanitize@url [0]{\catcode `\\12\catcode `\$12\catcode
  `\&12\catcode `\#12\catcode `\^12\catcode `\_12\catcode `\%12\relax}%
\providecommand \@@startlink[1]{}%
\providecommand \@@endlink[0]{}%
\providecommand \url  [0]{\begingroup\@sanitize@url \@url }%
\providecommand \@url [1]{\endgroup\@href {#1}{\urlprefix }}%
\providecommand \urlprefix  [0]{URL }%
\providecommand \Eprint [0]{\href }%
\providecommand \doibase [0]{http://dx.doi.org/}%
\providecommand \selectlanguage [0]{\@gobble}%
\providecommand \bibinfo  [0]{\@secondoftwo}%
\providecommand \bibfield  [0]{\@secondoftwo}%
\providecommand \translation [1]{[#1]}%
\providecommand \BibitemOpen [0]{}%
\providecommand \bibitemStop [0]{}%
\providecommand \bibitemNoStop [0]{.\EOS\space}%
\providecommand \EOS [0]{\spacefactor3000\relax}%
\providecommand \BibitemShut  [1]{\csname bibitem#1\endcsname}%
\let\auto@bib@innerbib\@empty
\bibitem [{\citenamefont {Wang}\ \emph {et~al.}(2013)\citenamefont {Wang},
  \citenamefont {Lingenfelder}, \citenamefont {Fabris}, \citenamefont
  {Fratesi}, \citenamefont {Ferrando}, \citenamefont {Classen}, \citenamefont
  {Kern},\ and\ \citenamefont {Costantini}}]{WangGiovanni2013}%
  \BibitemOpen
  \bibfield  {author} {\bibinfo {author} {\bibfnamefont {Y.}~\bibnamefont
  {Wang}}, \bibinfo {author} {\bibfnamefont {M.}~\bibnamefont {Lingenfelder}},
  \bibinfo {author} {\bibfnamefont {S.}~\bibnamefont {Fabris}}, \bibinfo
  {author} {\bibfnamefont {G.}~\bibnamefont {Fratesi}}, \bibinfo {author}
  {\bibfnamefont {R.}~\bibnamefont {Ferrando}}, \bibinfo {author}
  {\bibfnamefont {T.}~\bibnamefont {Classen}}, \bibinfo {author} {\bibfnamefont
  {K.}~\bibnamefont {Kern}}, \ and\ \bibinfo {author} {\bibfnamefont
  {G.}~\bibnamefont {Costantini}},\ }\href {\doibase 10.1021/jp309566s}
  {\bibfield  {journal} {\bibinfo  {journal} {The Journal of Physical Chemistry
  C}\ }\textbf {\bibinfo {volume} {117}},\ \bibinfo {pages} {3440} (\bibinfo
  {year} {2013})}\BibitemShut {NoStop}%
\bibitem [{\citenamefont {Guo}\ \emph {et~al.}(2016)\citenamefont {Guo},
  \citenamefont {Yu}, \citenamefont {Refaely-Abramson}, \citenamefont
  {Sepunaru}, \citenamefont {Bendikov}, \citenamefont {Pecht}, \citenamefont
  {Kronik}, \citenamefont {Vilan}, \citenamefont {Sheves},\ and\ \citenamefont
  {Cahen}}]{GuoCahen2016}%
  \BibitemOpen
  \bibfield  {author} {\bibinfo {author} {\bibfnamefont {C.}~\bibnamefont
  {Guo}}, \bibinfo {author} {\bibfnamefont {X.}~\bibnamefont {Yu}}, \bibinfo
  {author} {\bibfnamefont {S.}~\bibnamefont {Refaely-Abramson}}, \bibinfo
  {author} {\bibfnamefont {L.}~\bibnamefont {Sepunaru}}, \bibinfo {author}
  {\bibfnamefont {T.}~\bibnamefont {Bendikov}}, \bibinfo {author}
  {\bibfnamefont {I.}~\bibnamefont {Pecht}}, \bibinfo {author} {\bibfnamefont
  {L.}~\bibnamefont {Kronik}}, \bibinfo {author} {\bibfnamefont
  {A.}~\bibnamefont {Vilan}}, \bibinfo {author} {\bibfnamefont
  {M.}~\bibnamefont {Sheves}}, \ and\ \bibinfo {author} {\bibfnamefont
  {D.}~\bibnamefont {Cahen}},\ }\href {\doibase 10.1073/pnas.1606779113}
  {\bibfield  {journal} {\bibinfo  {journal} {Proceedings of the National
  Academy of Sciences}\ }\textbf {\bibinfo {volume} {113}},\ \bibinfo {pages}
  {10785} (\bibinfo {year} {2016})}\BibitemShut {NoStop}%
\bibitem [{\citenamefont {Khatayevich}\ \emph {et~al.}(2014)\citenamefont
  {Khatayevich}, \citenamefont {Page}, \citenamefont {Gresswell}, \citenamefont
  {Hayamizu}, \citenamefont {Grady},\ and\ \citenamefont
  {Sarikaya}}]{Khatayevich}%
  \BibitemOpen
  \bibfield  {author} {\bibinfo {author} {\bibfnamefont {D.}~\bibnamefont
  {Khatayevich}}, \bibinfo {author} {\bibfnamefont {T.}~\bibnamefont {Page}},
  \bibinfo {author} {\bibfnamefont {C.}~\bibnamefont {Gresswell}}, \bibinfo
  {author} {\bibfnamefont {Y.}~\bibnamefont {Hayamizu}}, \bibinfo {author}
  {\bibfnamefont {W.}~\bibnamefont {Grady}}, \ and\ \bibinfo {author}
  {\bibfnamefont {M.}~\bibnamefont {Sarikaya}},\ }\href {\doibase
  10.1002/smll.201302188} {\bibfield  {journal} {\bibinfo  {journal} {Small}\
  }\textbf {\bibinfo {volume} {10}},\ \bibinfo {pages} {1505} (\bibinfo {year}
  {2014})}\BibitemShut {NoStop}%
\bibitem [{\citenamefont {Mannoor}\ \emph {et~al.}(2012)\citenamefont
  {Mannoor}, \citenamefont {Tao}, \citenamefont {Clayton}, \citenamefont
  {Sengupta}, \citenamefont {Kaplan}, \citenamefont {Naik}, \citenamefont
  {Verma}, \citenamefont {Omenetto},\ and\ \citenamefont {McAlpine}}]{Mannoor}%
  \BibitemOpen
  \bibfield  {author} {\bibinfo {author} {\bibfnamefont {M.~S.}\ \bibnamefont
  {Mannoor}}, \bibinfo {author} {\bibfnamefont {H.}~\bibnamefont {Tao}},
  \bibinfo {author} {\bibfnamefont {J.~D.}\ \bibnamefont {Clayton}}, \bibinfo
  {author} {\bibfnamefont {A.}~\bibnamefont {Sengupta}}, \bibinfo {author}
  {\bibfnamefont {D.~L.}\ \bibnamefont {Kaplan}}, \bibinfo {author}
  {\bibfnamefont {R.~R.}\ \bibnamefont {Naik}}, \bibinfo {author}
  {\bibfnamefont {N.}~\bibnamefont {Verma}}, \bibinfo {author} {\bibfnamefont
  {F.~G.}\ \bibnamefont {Omenetto}}, \ and\ \bibinfo {author} {\bibfnamefont
  {M.~C.}\ \bibnamefont {McAlpine}},\ }\href {\doibase 10.1038/ncomms1767}
  {\bibfield  {journal} {\bibinfo  {journal} {Nature Communications}\ }\textbf
  {\bibinfo {volume} {3}},\ \bibinfo {pages} {763} (\bibinfo {year}
  {2012})}\BibitemShut {NoStop}%
\bibitem [{\citenamefont {Guy}\ \emph {et~al.}(2012)\citenamefont {Guy},
  \citenamefont {Burwell}, \citenamefont {Tehrani}, \citenamefont {Castaing},
  \citenamefont {Walker},\ and\ \citenamefont {Doak}}]{Guy}%
  \BibitemOpen
  \bibfield  {author} {\bibinfo {author} {\bibfnamefont {O.~J.}\ \bibnamefont
  {Guy}}, \bibinfo {author} {\bibfnamefont {G.}~\bibnamefont {Burwell}},
  \bibinfo {author} {\bibfnamefont {Z.}~\bibnamefont {Tehrani}}, \bibinfo
  {author} {\bibfnamefont {A.}~\bibnamefont {Castaing}}, \bibinfo {author}
  {\bibfnamefont {K.~A.}\ \bibnamefont {Walker}}, \ and\ \bibinfo {author}
  {\bibfnamefont {S.}~\bibnamefont {Doak}},\ }\href {\doibase
  10.4028/www.scientific.net/MSF.711.246} {\bibfield  {journal} {\bibinfo
  {journal} {Materials Science Forum}\ }\textbf {\bibinfo {volume} {711}},\
  \bibinfo {pages} {246} (\bibinfo {year} {2012})}\BibitemShut {NoStop}%
\bibitem [{\citenamefont {Zhao}\ \emph {et~al.}(2008)\citenamefont {Zhao},
  \citenamefont {Pan},\ and\ \citenamefont {Lu}}]{Zhao}%
  \BibitemOpen
  \bibfield  {author} {\bibinfo {author} {\bibfnamefont {X.}~\bibnamefont
  {Zhao}}, \bibinfo {author} {\bibfnamefont {F.}~\bibnamefont {Pan}}, \ and\
  \bibinfo {author} {\bibfnamefont {J.~R.}\ \bibnamefont {Lu}},\ }\href
  {\doibase 10.1016/j.pnsc.2008.01.012} {\bibfield  {journal} {\bibinfo
  {journal} {Progress in Natural Science}\ }\textbf {\bibinfo {volume} {18}},\
  \bibinfo {pages} {653} (\bibinfo {year} {2008})}\BibitemShut {NoStop}%
\bibitem [{\citenamefont {Sarikaya}\ \emph {et~al.}(2003)\citenamefont
  {Sarikaya}, \citenamefont {Tamerler}, \citenamefont {Jen}, \citenamefont
  {Schulten},\ and\ \citenamefont {Baneyx}}]{Sarikaya:2003-biomim}%
  \BibitemOpen
  \bibfield  {author} {\bibinfo {author} {\bibfnamefont {M.}~\bibnamefont
  {Sarikaya}}, \bibinfo {author} {\bibfnamefont {C.}~\bibnamefont {Tamerler}},
  \bibinfo {author} {\bibfnamefont {A.~K.}\ \bibnamefont {Jen}}, \bibinfo
  {author} {\bibfnamefont {K.}~\bibnamefont {Schulten}}, \ and\ \bibinfo
  {author} {\bibfnamefont {F.}~\bibnamefont {Baneyx}},\ }\href@noop {}
  {\bibfield  {journal} {\bibinfo  {journal} {Nature Materials}\ }\textbf
  {\bibinfo {volume} {2}},\ \bibinfo {pages} {577} (\bibinfo {year}
  {2003})}\BibitemShut {NoStop}%
\bibitem [{\citenamefont {Costa}\ \emph {et~al.}(2015)\citenamefont {Costa},
  \citenamefont {Pradier}, \citenamefont {Tielens},\ and\ \citenamefont
  {Savio}}]{Costa:2015:IMPORTANT}%
  \BibitemOpen
  \bibfield  {author} {\bibinfo {author} {\bibfnamefont {D.}~\bibnamefont
  {Costa}}, \bibinfo {author} {\bibfnamefont {C.-M.}\ \bibnamefont {Pradier}},
  \bibinfo {author} {\bibfnamefont {F.}~\bibnamefont {Tielens}}, \ and\
  \bibinfo {author} {\bibfnamefont {L.}~\bibnamefont {Savio}},\ }\href@noop {}
  {\bibfield  {journal} {\bibinfo  {journal} {Surface Science Reports}\
  }\textbf {\bibinfo {volume} {70}},\ \bibinfo {pages} {449} (\bibinfo {year}
  {2015})}\BibitemShut {NoStop}%
\bibitem [{\citenamefont {Heinz}\ and\ \citenamefont
  {Ramezani-Dakhel}(2016)}]{Heinz:2016:Review}%
  \BibitemOpen
  \bibfield  {author} {\bibinfo {author} {\bibfnamefont {H.}~\bibnamefont
  {Heinz}}\ and\ \bibinfo {author} {\bibfnamefont {H.}~\bibnamefont
  {Ramezani-Dakhel}},\ }\href {\doibase 10.1039/C5CS00890E} {\bibfield
  {journal} {\bibinfo  {journal} {Chem. Soc. Rev.}\ }\textbf {\bibinfo {volume}
  {45}},\ \bibinfo {pages} {412} (\bibinfo {year} {2016})}\BibitemShut
  {NoStop}%
\bibitem [{\citenamefont {Walsh}\ and\ \citenamefont
  {Knecht}(2017)}]{Walsh:2017ci}%
  \BibitemOpen
  \bibfield  {author} {\bibinfo {author} {\bibfnamefont {T.~R.}\ \bibnamefont
  {Walsh}}\ and\ \bibinfo {author} {\bibfnamefont {M.~R.}\ \bibnamefont
  {Knecht}},\ }\href@noop {} {\bibfield  {journal} {\bibinfo  {journal}
  {Chemical Reviews}\ }\textbf {\bibinfo {volume} {117}},\ \bibinfo {pages}
  {12641} (\bibinfo {year} {2017})}\BibitemShut {NoStop}%
\bibitem [{\citenamefont {Rauschenbach}\ \emph {et~al.}(2017)\citenamefont
  {Rauschenbach}, \citenamefont {Rinke}, \citenamefont {Gutzler}, \citenamefont
  {Abb}, \citenamefont {Albarghash}, \citenamefont {Le}, \citenamefont
  {Rahman}, \citenamefont {Duy}, \citenamefont {Harnau},\ and\ \citenamefont
  {Kern}}]{Rauschenbach2017}%
  \BibitemOpen
  \bibfield  {author} {\bibinfo {author} {\bibfnamefont {S.}~\bibnamefont
  {Rauschenbach}}, \bibinfo {author} {\bibfnamefont {G.}~\bibnamefont {Rinke}},
  \bibinfo {author} {\bibfnamefont {R.}~\bibnamefont {Gutzler}}, \bibinfo
  {author} {\bibfnamefont {S.}~\bibnamefont {Abb}}, \bibinfo {author}
  {\bibfnamefont {A.}~\bibnamefont {Albarghash}}, \bibinfo {author}
  {\bibfnamefont {D.}~\bibnamefont {Le}}, \bibinfo {author} {\bibfnamefont
  {T.~S.}\ \bibnamefont {Rahman}}, \bibinfo {author} {\bibfnamefont
  {M.}~\bibnamefont {Duy}}, \bibinfo {author} {\bibfnamefont {L.}~\bibnamefont
  {Harnau}}, \ and\ \bibinfo {author} {\bibfnamefont {K.}~\bibnamefont
  {Kern}},\ }\href {\doibase 10.1021/acsnano.6b06145} {\bibfield  {journal}
  {\bibinfo  {journal} {ACS Nano}\ }\textbf {\bibinfo {volume} {11}},\ \bibinfo
  {pages} {2420} (\bibinfo {year} {2017})}\BibitemShut {NoStop}%
\bibitem [{\citenamefont {Rauschenbach}\ \emph {et~al.}(2016)\citenamefont
  {Rauschenbach}, \citenamefont {Ternes}, \citenamefont {Harnau},\ and\
  \citenamefont {Kern}}]{Rauschenbach2016}%
  \BibitemOpen
  \bibfield  {author} {\bibinfo {author} {\bibfnamefont {S.}~\bibnamefont
  {Rauschenbach}}, \bibinfo {author} {\bibfnamefont {M.}~\bibnamefont
  {Ternes}}, \bibinfo {author} {\bibfnamefont {L.}~\bibnamefont {Harnau}}, \
  and\ \bibinfo {author} {\bibfnamefont {K.}~\bibnamefont {Kern}},\ }\href
  {\doibase 10.1146/annurev-anchem-071015-041633} {\bibfield  {journal}
  {\bibinfo  {journal} {Annual Review of Analytical Chemistry}\ }\textbf
  {\bibinfo {volume} {9}},\ \bibinfo {pages} {473} (\bibinfo {year}
  {2016})}\BibitemShut {NoStop}%
\bibitem [{\citenamefont {Abb}\ \emph {et~al.}(2016)\citenamefont {Abb},
  \citenamefont {Harnau}, \citenamefont {Gutzler}, \citenamefont
  {Rauschenbach},\ and\ \citenamefont {Kern}}]{Abb2016}%
  \BibitemOpen
  \bibfield  {author} {\bibinfo {author} {\bibfnamefont {S.}~\bibnamefont
  {Abb}}, \bibinfo {author} {\bibfnamefont {L.}~\bibnamefont {Harnau}},
  \bibinfo {author} {\bibfnamefont {R.}~\bibnamefont {Gutzler}}, \bibinfo
  {author} {\bibfnamefont {S.}~\bibnamefont {Rauschenbach}}, \ and\ \bibinfo
  {author} {\bibfnamefont {K.}~\bibnamefont {Kern}},\ }\href@noop {} {\bibfield
   {journal} {\bibinfo  {journal} {Nature Communications}\ }\textbf {\bibinfo
  {volume} {7}},\ \bibinfo {pages} {10335} (\bibinfo {year}
  {2016})}\BibitemShut {NoStop}%
\bibitem [{\citenamefont {M\'ethivier}\ \emph {et~al.}(2016)\citenamefont
  {M\'ethivier}, \citenamefont {Cruguel}, \citenamefont {Costa}, \citenamefont
  {Pradier},\ and\ \citenamefont {Humblot}}]{Humblot2016}%
  \BibitemOpen
  \bibfield  {author} {\bibinfo {author} {\bibfnamefont {C.}~\bibnamefont
  {M\'ethivier}}, \bibinfo {author} {\bibfnamefont {H.}~\bibnamefont
  {Cruguel}}, \bibinfo {author} {\bibfnamefont {D.}~\bibnamefont {Costa}},
  \bibinfo {author} {\bibfnamefont {C.-M.}\ \bibnamefont {Pradier}}, \ and\
  \bibinfo {author} {\bibfnamefont {V.}~\bibnamefont {Humblot}},\ }\href
  {\doibase 10.1021/acs.langmuir.6b03553} {\bibfield  {journal} {\bibinfo
  {journal} {Langmuir}\ }\textbf {\bibinfo {volume} {32}},\ \bibinfo {pages}
  {13759} (\bibinfo {year} {2016})}\BibitemShut {NoStop}%
\bibitem [{\citenamefont {{Di Felice}}\ \emph {et~al.}(2003)\citenamefont {{Di
  Felice}}, \citenamefont {Selloni},\ and\ \citenamefont
  {Molinari}}]{DiFelice2003}%
  \BibitemOpen
  \bibfield  {author} {\bibinfo {author} {\bibfnamefont {R.}~\bibnamefont {{Di
  Felice}}}, \bibinfo {author} {\bibfnamefont {A.}~\bibnamefont {Selloni}}, \
  and\ \bibinfo {author} {\bibfnamefont {E.}~\bibnamefont {Molinari}},\ }\href
  {\doibase 10.1021/jp0272421} {\bibfield  {journal} {\bibinfo  {journal} {The
  Journal of Physical Chemistry B}\ }\textbf {\bibinfo {volume} {107}},\
  \bibinfo {pages} {1151} (\bibinfo {year} {2003})}\BibitemShut {NoStop}%
\bibitem [{\citenamefont {{Di Felice}}\ and\ \citenamefont
  {Selloni}(2004)}]{DiFelice2004}%
  \BibitemOpen
  \bibfield  {author} {\bibinfo {author} {\bibfnamefont {R.}~\bibnamefont {{Di
  Felice}}}\ and\ \bibinfo {author} {\bibfnamefont {A.}~\bibnamefont
  {Selloni}},\ }\href {\doibase 10.1063/1.1645789} {\bibfield  {journal}
  {\bibinfo  {journal} {The Journal of Chemical Physics}\ }\textbf {\bibinfo
  {volume} {120}},\ \bibinfo {pages} {4906} (\bibinfo {year}
  {2004})}\BibitemShut {NoStop}%
\bibitem [{\citenamefont {Ghiringhelli}\ \emph {et~al.}(2006)\citenamefont
  {Ghiringhelli}, \citenamefont {Schravendijk},\ and\ \citenamefont {{Delle
  Site}}}]{Ghiringhelli2006}%
  \BibitemOpen
  \bibfield  {author} {\bibinfo {author} {\bibfnamefont {L.~M.}\ \bibnamefont
  {Ghiringhelli}}, \bibinfo {author} {\bibfnamefont {P.}~\bibnamefont
  {Schravendijk}}, \ and\ \bibinfo {author} {\bibfnamefont {L.}~\bibnamefont
  {{Delle Site}}},\ }\href {\doibase 10.1103/PhysRevB.74.035437} {\bibfield
  {journal} {\bibinfo  {journal} {Phys. Rev. B}\ }\textbf {\bibinfo {volume}
  {74}},\ \bibinfo {pages} {35437} (\bibinfo {year} {2006})}\BibitemShut
  {NoStop}%
\bibitem [{\citenamefont {Arrouvel}\ \emph {et~al.}(2007)\citenamefont
  {Arrouvel}, \citenamefont {Diawara}, \citenamefont {Costa},\ and\
  \citenamefont {Marcus}}]{Arrouvel2007}%
  \BibitemOpen
  \bibfield  {author} {\bibinfo {author} {\bibfnamefont {C.}~\bibnamefont
  {Arrouvel}}, \bibinfo {author} {\bibfnamefont {B.}~\bibnamefont {Diawara}},
  \bibinfo {author} {\bibfnamefont {D.}~\bibnamefont {Costa}}, \ and\ \bibinfo
  {author} {\bibfnamefont {P.}~\bibnamefont {Marcus}},\ }\href {\doibase
  10.1021/jp0741408} {\bibfield  {journal} {\bibinfo  {journal} {The Journal of
  Physical Chemistry C}\ }\textbf {\bibinfo {volume} {111}},\ \bibinfo {pages}
  {18164} (\bibinfo {year} {2007})}\BibitemShut {NoStop}%
\bibitem [{\citenamefont {Iori}\ \emph {et~al.}(2008)\citenamefont {Iori},
  \citenamefont {Corni},\ and\ \citenamefont {{Di Felice}}}]{Iori2008}%
  \BibitemOpen
  \bibfield  {author} {\bibinfo {author} {\bibfnamefont {F.}~\bibnamefont
  {Iori}}, \bibinfo {author} {\bibfnamefont {S.}~\bibnamefont {Corni}}, \ and\
  \bibinfo {author} {\bibfnamefont {R.}~\bibnamefont {{Di Felice}}},\ }\href
  {\doibase 10.1021/jp801542s} {\bibfield  {journal} {\bibinfo  {journal} {The
  Journal of Physical Chemistry C}\ }\textbf {\bibinfo {volume} {112}},\
  \bibinfo {pages} {13540} (\bibinfo {year} {2008})}\BibitemShut {NoStop}%
\bibitem [{\citenamefont {Hong}\ \emph {et~al.}(2009)\citenamefont {Hong},
  \citenamefont {Heinz}, \citenamefont {Naik}, \citenamefont {Farmer},\ and\
  \citenamefont {Pachter}}]{Hong2009}%
  \BibitemOpen
  \bibfield  {author} {\bibinfo {author} {\bibfnamefont {G.}~\bibnamefont
  {Hong}}, \bibinfo {author} {\bibfnamefont {H.}~\bibnamefont {Heinz}},
  \bibinfo {author} {\bibfnamefont {R.~R.}\ \bibnamefont {Naik}}, \bibinfo
  {author} {\bibfnamefont {B.~L.}\ \bibnamefont {Farmer}}, \ and\ \bibinfo
  {author} {\bibfnamefont {R.}~\bibnamefont {Pachter}},\ }\href {\doibase
  10.1021/am800099z} {\bibfield  {journal} {\bibinfo  {journal} {ACS Applied
  Materials {\&} Interfaces}\ }\textbf {\bibinfo {volume} {1}},\ \bibinfo
  {pages} {388} (\bibinfo {year} {2009})}\BibitemShut {NoStop}%
\bibitem [{\citenamefont {Ropo}\ \emph
  {et~al.}(2016{\natexlab{a}})\citenamefont {Ropo}, \citenamefont {Blum},\ and\
  \citenamefont {Baldauf}}]{ropo2016trends}%
  \BibitemOpen
  \bibfield  {author} {\bibinfo {author} {\bibfnamefont {M.}~\bibnamefont
  {Ropo}}, \bibinfo {author} {\bibfnamefont {V.}~\bibnamefont {Blum}}, \ and\
  \bibinfo {author} {\bibfnamefont {C.}~\bibnamefont {Baldauf}},\ }\href@noop
  {} {\bibfield  {journal} {\bibinfo  {journal} {Scientific Reports}\ }\textbf
  {\bibinfo {volume} {6}},\ \bibinfo {pages} {35772} (\bibinfo {year}
  {2016}{\natexlab{a}})}\BibitemShut {NoStop}%
\bibitem [{\citenamefont {Rossi}\ \emph {et~al.}(2013)\citenamefont {Rossi},
  \citenamefont {Scheffler},\ and\ \citenamefont {Blum}}]{Rossi_2013}%
  \BibitemOpen
  \bibfield  {author} {\bibinfo {author} {\bibfnamefont {M.}~\bibnamefont
  {Rossi}}, \bibinfo {author} {\bibfnamefont {M.}~\bibnamefont {Scheffler}}, \
  and\ \bibinfo {author} {\bibfnamefont {V.}~\bibnamefont {Blum}},\ }\href@noop
  {} {\bibfield  {journal} {\bibinfo  {journal} {Journal Of Physical Chemistry
  B}\ }\textbf {\bibinfo {volume} {117}},\ \bibinfo {pages} {5574} (\bibinfo
  {year} {2013})}\BibitemShut {NoStop}%
\bibitem [{\citenamefont {Rossi}\ \emph {et~al.}(2014)\citenamefont {Rossi},
  \citenamefont {Chutia}, \citenamefont {Scheffler},\ and\ \citenamefont
  {Blum}}]{Rossi_2014}%
  \BibitemOpen
  \bibfield  {author} {\bibinfo {author} {\bibfnamefont {M.}~\bibnamefont
  {Rossi}}, \bibinfo {author} {\bibfnamefont {S.}~\bibnamefont {Chutia}},
  \bibinfo {author} {\bibfnamefont {M.}~\bibnamefont {Scheffler}}, \ and\
  \bibinfo {author} {\bibfnamefont {V.}~\bibnamefont {Blum}},\ }\href@noop {}
  {\bibfield  {journal} {\bibinfo  {journal} {Journal Of Physical Chemistry A}\
  }\textbf {\bibinfo {volume} {118}},\ \bibinfo {pages} {7349} (\bibinfo {year}
  {2014})}\BibitemShut {NoStop}%
\bibitem [{\citenamefont {Schubert}\ \emph {et~al.}(2015)\citenamefont
  {Schubert}, \citenamefont {Rossi}, \citenamefont {Baldauf}, \citenamefont
  {Pagel}, \citenamefont {Warnke}, \citenamefont {von Helden}, \citenamefont
  {Filsinger}, \citenamefont {Kupser}, \citenamefont {Meijer}, \citenamefont
  {Salwiczek}, \citenamefont {Koksch}, \citenamefont {Scheffler},\ and\
  \citenamefont {Blum}}]{Schubert_2015}%
  \BibitemOpen
  \bibfield  {author} {\bibinfo {author} {\bibfnamefont {F.}~\bibnamefont
  {Schubert}}, \bibinfo {author} {\bibfnamefont {M.}~\bibnamefont {Rossi}},
  \bibinfo {author} {\bibfnamefont {C.}~\bibnamefont {Baldauf}}, \bibinfo
  {author} {\bibfnamefont {K.}~\bibnamefont {Pagel}}, \bibinfo {author}
  {\bibfnamefont {S.}~\bibnamefont {Warnke}}, \bibinfo {author} {\bibfnamefont
  {G.}~\bibnamefont {von Helden}}, \bibinfo {author} {\bibfnamefont
  {F.}~\bibnamefont {Filsinger}}, \bibinfo {author} {\bibfnamefont
  {P.}~\bibnamefont {Kupser}}, \bibinfo {author} {\bibfnamefont
  {G.}~\bibnamefont {Meijer}}, \bibinfo {author} {\bibfnamefont
  {M.}~\bibnamefont {Salwiczek}}, \bibinfo {author} {\bibfnamefont
  {B.}~\bibnamefont {Koksch}}, \bibinfo {author} {\bibfnamefont
  {M.}~\bibnamefont {Scheffler}}, \ and\ \bibinfo {author} {\bibfnamefont
  {V.}~\bibnamefont {Blum}},\ }\href@noop {} {\bibfield  {journal} {\bibinfo
  {journal} {Physical Chemistry Chemical Physics}\ }\textbf {\bibinfo {volume}
  {17}},\ \bibinfo {pages} {7373} (\bibinfo {year} {2015})}\BibitemShut
  {NoStop}%
\bibitem [{\citenamefont {Baldauf}\ and\ \citenamefont
  {Rossi}(2015)}]{Baldauf_2015}%
  \BibitemOpen
  \bibfield  {author} {\bibinfo {author} {\bibfnamefont {C.}~\bibnamefont
  {Baldauf}}\ and\ \bibinfo {author} {\bibfnamefont {M.}~\bibnamefont
  {Rossi}},\ }\href@noop {} {\bibfield  {journal} {\bibinfo  {journal} {Journal
  of Physics: Condensed Matter}\ }\textbf {\bibinfo {volume} {27}},\ \bibinfo
  {pages} {493002} (\bibinfo {year} {2015})}\BibitemShut {NoStop}%
\bibitem [{\citenamefont {Ceriotti}\ \emph {et~al.}(2011)\citenamefont
  {Ceriotti}, \citenamefont {Tribello},\ and\ \citenamefont
  {Parrinello}}]{Ceriotti2011}%
  \BibitemOpen
  \bibfield  {author} {\bibinfo {author} {\bibfnamefont {M.}~\bibnamefont
  {Ceriotti}}, \bibinfo {author} {\bibfnamefont {G.~A.}\ \bibnamefont
  {Tribello}}, \ and\ \bibinfo {author} {\bibfnamefont {M.}~\bibnamefont
  {Parrinello}},\ }\href {\doibase 10.1073/pnas.1108486108} {\bibfield
  {journal} {\bibinfo  {journal} {Proceedings of the National Academy of
  Sciences}\ }\textbf {\bibinfo {volume} {108}},\ \bibinfo {pages} {13023}
  (\bibinfo {year} {2011})}\BibitemShut {NoStop}%
\bibitem [{\citenamefont {Tribello}\ \emph {et~al.}(2012)\citenamefont
  {Tribello}, \citenamefont {Ceriotti},\ and\ \citenamefont
  {Parrinello}}]{Tribello2012}%
  \BibitemOpen
  \bibfield  {author} {\bibinfo {author} {\bibfnamefont {G.~A.}\ \bibnamefont
  {Tribello}}, \bibinfo {author} {\bibfnamefont {M.}~\bibnamefont {Ceriotti}},
  \ and\ \bibinfo {author} {\bibfnamefont {M.}~\bibnamefont {Parrinello}},\
  }\href {\doibase 10.1073/pnas.1201152109} {\bibfield  {journal} {\bibinfo
  {journal} {Proceedings of the National Academy of Sciences of the United
  States of America}\ }\textbf {\bibinfo {volume} {109}},\ \bibinfo {pages}
  {5196} (\bibinfo {year} {2012})}\BibitemShut {NoStop}%
\bibitem [{\citenamefont {Ceriotti}\ \emph {et~al.}(2013)\citenamefont
  {Ceriotti}, \citenamefont {Tribello},\ and\ \citenamefont
  {Parrinello}}]{Ceriotti2013}%
  \BibitemOpen
  \bibfield  {author} {\bibinfo {author} {\bibfnamefont {M.}~\bibnamefont
  {Ceriotti}}, \bibinfo {author} {\bibfnamefont {G.~A.}\ \bibnamefont
  {Tribello}}, \ and\ \bibinfo {author} {\bibfnamefont {M.}~\bibnamefont
  {Parrinello}},\ }\href {\doibase 10.1021/ct3010563} {\bibfield  {journal}
  {\bibinfo  {journal} {Journal of Chemical Theory and Computation}\ }\textbf
  {\bibinfo {volume} {9}},\ \bibinfo {pages} {1521} (\bibinfo {year}
  {2013})}\BibitemShut {NoStop}%
\bibitem [{\citenamefont {De}\ \emph {et~al.}(2016)\citenamefont {De},
  \citenamefont {Bartók}, \citenamefont {Csányi},\ and\ \citenamefont
  {Ceriotti}}]{De2016}%
  \BibitemOpen
  \bibfield  {author} {\bibinfo {author} {\bibfnamefont {S.}~\bibnamefont
  {De}}, \bibinfo {author} {\bibfnamefont {A.~P.}\ \bibnamefont {Bartók}},
  \bibinfo {author} {\bibfnamefont {G.}~\bibnamefont {Csányi}}, \ and\
  \bibinfo {author} {\bibfnamefont {M.}~\bibnamefont {Ceriotti}},\ }\href
  {\doibase 10.1039/C6CP00415F} {\bibfield  {journal} {\bibinfo  {journal}
  {Phys. Chem. Chem. Phys.}\ }\textbf {\bibinfo {volume} {18}},\ \bibinfo
  {pages} {13754} (\bibinfo {year} {2016})}\BibitemShut {NoStop}%
\bibitem [{\citenamefont {Bart{\'o}k}\ \emph {et~al.}(2017)\citenamefont
  {Bart{\'o}k}, \citenamefont {De}, \citenamefont {Poelking}, \citenamefont
  {Bernstein}, \citenamefont {Kermode}, \citenamefont {Cs{\'a}nyi},\ and\
  \citenamefont {Ceriotti}}]{Bartok2017}%
  \BibitemOpen
  \bibfield  {author} {\bibinfo {author} {\bibfnamefont {A.~P.}\ \bibnamefont
  {Bart{\'o}k}}, \bibinfo {author} {\bibfnamefont {S.}~\bibnamefont {De}},
  \bibinfo {author} {\bibfnamefont {C.}~\bibnamefont {Poelking}}, \bibinfo
  {author} {\bibfnamefont {N.}~\bibnamefont {Bernstein}}, \bibinfo {author}
  {\bibfnamefont {J.~R.}\ \bibnamefont {Kermode}}, \bibinfo {author}
  {\bibfnamefont {G.}~\bibnamefont {Cs{\'a}nyi}}, \ and\ \bibinfo {author}
  {\bibfnamefont {M.}~\bibnamefont {Ceriotti}},\ }\href {\doibase
  10.1126/sciadv.1701816} {\bibfield  {journal} {\bibinfo  {journal} {Science
  Advances}\ }\textbf {\bibinfo {volume} {3}} (\bibinfo {year} {2017}),\
  10.1126/sciadv.1701816}\BibitemShut {NoStop}%
\bibitem [{\citenamefont {De}\ \emph {et~al.}(2017)\citenamefont {De},
  \citenamefont {Musil}, \citenamefont {Ingram}, \citenamefont {Baldauf},\ and\
  \citenamefont {Ceriotti}}]{De2017}%
  \BibitemOpen
  \bibfield  {author} {\bibinfo {author} {\bibfnamefont {S.}~\bibnamefont
  {De}}, \bibinfo {author} {\bibfnamefont {F.}~\bibnamefont {Musil}}, \bibinfo
  {author} {\bibfnamefont {T.}~\bibnamefont {Ingram}}, \bibinfo {author}
  {\bibfnamefont {C.}~\bibnamefont {Baldauf}}, \ and\ \bibinfo {author}
  {\bibfnamefont {M.}~\bibnamefont {Ceriotti}},\ }\href {\doibase
  10.1186/s13321-017-0192-4} {\bibfield  {journal} {\bibinfo  {journal}
  {Journal of Cheminformatics}\ }\textbf {\bibinfo {volume} {9}},\ \bibinfo
  {pages} {6} (\bibinfo {year} {2017})}\BibitemShut {NoStop}%
\bibitem [{\citenamefont {Maksimov}\ \emph {et~al.}()\citenamefont {Maksimov},
  \citenamefont {Baldauf},\ and\ \citenamefont
  {Rossi}}]{DOI_of_the_data_set_at_NOMAD}%
  \BibitemOpen
  \bibfield  {author} {\bibinfo {author} {\bibfnamefont {D.}~\bibnamefont
  {Maksimov}}, \bibinfo {author} {\bibfnamefont {C.}~\bibnamefont {Baldauf}}, \
  and\ \bibinfo {author} {\bibfnamefont {M.}~\bibnamefont {Rossi}},\ }\href
  {\doibase https://dx.doi.org/10.17172/NOMAD/2020.03.24-1} {\enquote {\bibinfo
  {title} {Database of arg and arg-h$^+$ adsorbed on cu(111), ag(111) and
  au(111) in the {NOMAD} repository.}}\ }\BibitemShut {NoStop}%
\bibitem [{\citenamefont {Blum}\ \emph {et~al.}(2009)\citenamefont {Blum},
  \citenamefont {Gehrke}, \citenamefont {Hanke}, \citenamefont {Havu},
  \citenamefont {Havu}, \citenamefont {Ren}, \citenamefont {Reuter},\ and\
  \citenamefont {Scheffler}}]{Blum2009}%
  \BibitemOpen
  \bibfield  {author} {\bibinfo {author} {\bibfnamefont {V.}~\bibnamefont
  {Blum}}, \bibinfo {author} {\bibfnamefont {R.}~\bibnamefont {Gehrke}},
  \bibinfo {author} {\bibfnamefont {F.}~\bibnamefont {Hanke}}, \bibinfo
  {author} {\bibfnamefont {P.}~\bibnamefont {Havu}}, \bibinfo {author}
  {\bibfnamefont {V.}~\bibnamefont {Havu}}, \bibinfo {author} {\bibfnamefont
  {X.}~\bibnamefont {Ren}}, \bibinfo {author} {\bibfnamefont {K.}~\bibnamefont
  {Reuter}}, \ and\ \bibinfo {author} {\bibfnamefont {M.}~\bibnamefont
  {Scheffler}},\ }\href {\doibase 10.1016/j.cpc.2009.06.022} {\bibfield
  {journal} {\bibinfo  {journal} {Computer Physics Communications}\ }\textbf
  {\bibinfo {volume} {180}},\ \bibinfo {pages} {2175} (\bibinfo {year}
  {2009})}\BibitemShut {NoStop}%
\bibitem [{\citenamefont {Havu}\ \emph {et~al.}(2009)\citenamefont {Havu},
  \citenamefont {Blum}, \citenamefont {Havu},\ and\ \citenamefont
  {Scheffler}}]{Havu2009}%
  \BibitemOpen
  \bibfield  {author} {\bibinfo {author} {\bibfnamefont {V.}~\bibnamefont
  {Havu}}, \bibinfo {author} {\bibfnamefont {V.}~\bibnamefont {Blum}}, \bibinfo
  {author} {\bibfnamefont {P.}~\bibnamefont {Havu}}, \ and\ \bibinfo {author}
  {\bibfnamefont {M.}~\bibnamefont {Scheffler}},\ }\href {\doibase
  https://doi.org/10.1016/j.jcp.2009.08.008} {\bibfield  {journal} {\bibinfo
  {journal} {Journal of Computational Physics}\ }\textbf {\bibinfo {volume}
  {228}},\ \bibinfo {pages} {8367} (\bibinfo {year} {2009})}\BibitemShut
  {NoStop}%
\bibitem [{\citenamefont {{Hermann}}\ and\ \citenamefont
  {{Tkatchenko}}(2019)}]{Hermann2019}%
  \BibitemOpen
  \bibfield  {author} {\bibinfo {author} {\bibfnamefont {J.}~\bibnamefont
  {{Hermann}}}\ and\ \bibinfo {author} {\bibfnamefont {A.}~\bibnamefont
  {{Tkatchenko}}},\ }\href@noop {} {\bibfield  {journal} {\bibinfo  {journal}
  {arXiv e-prints}\ ,\ \bibinfo {eid} {arXiv:1910.03073}} (\bibinfo {year}
  {2019})}\BibitemShut {NoStop}%
\bibitem [{\citenamefont {Ruiz}\ \emph {et~al.}(2016)\citenamefont {Ruiz},
  \citenamefont {Liu},\ and\ \citenamefont {Tkatchenko}}]{Ruiz2016}%
  \BibitemOpen
  \bibfield  {author} {\bibinfo {author} {\bibfnamefont {V.~G.}\ \bibnamefont
  {Ruiz}}, \bibinfo {author} {\bibfnamefont {W.}~\bibnamefont {Liu}}, \ and\
  \bibinfo {author} {\bibfnamefont {A.}~\bibnamefont {Tkatchenko}},\ }\href
  {\doibase 10.1103/PhysRevB.93.035118} {\bibfield  {journal} {\bibinfo
  {journal} {Physical Review B}\ }\textbf {\bibinfo {volume} {93}},\ \bibinfo
  {pages} {035118} (\bibinfo {year} {2016})}\BibitemShut {NoStop}%
\bibitem [{\citenamefont {Ruiz~Lopez}(2016)}]{Ruiz-thesis}%
  \BibitemOpen
  \bibfield  {author} {\bibinfo {author} {\bibfnamefont {V.~G.}\ \bibnamefont
  {Ruiz~Lopez}},\ }\emph {\bibinfo {title} {Adsorption of atoms and molecules
  on surfaces : density-functional theory with screened van der Waals
  interactions}},\ \href {\doibase 10.14279/depositonce-5478} {\bibinfo {type}
  {Doctoral thesis}},\ \bibinfo  {school} {Technische Universitaet Berlin},
  \bibinfo {address} {Berlin} (\bibinfo {year} {2016})\BibitemShut {NoStop}%
\bibitem [{\citenamefont {Liu}\ \emph {et~al.}(2013)\citenamefont {Liu},
  \citenamefont {Ruiz}, \citenamefont {Zhang}, \citenamefont {Santra},
  \citenamefont {Ren}, \citenamefont {Scheffler},\ and\ \citenamefont
  {Tkatchenko}}]{Liu2013}%
  \BibitemOpen
  \bibfield  {author} {\bibinfo {author} {\bibfnamefont {W.}~\bibnamefont
  {Liu}}, \bibinfo {author} {\bibfnamefont {V.~G.}\ \bibnamefont {Ruiz}},
  \bibinfo {author} {\bibfnamefont {G.-X.}\ \bibnamefont {Zhang}}, \bibinfo
  {author} {\bibfnamefont {B.}~\bibnamefont {Santra}}, \bibinfo {author}
  {\bibfnamefont {X.}~\bibnamefont {Ren}}, \bibinfo {author} {\bibfnamefont
  {M.}~\bibnamefont {Scheffler}}, \ and\ \bibinfo {author} {\bibfnamefont
  {A.}~\bibnamefont {Tkatchenko}},\ }\href {\doibase
  10.1088/1367-2630/15/5/053046} {\bibfield  {journal} {\bibinfo  {journal}
  {New Journal of Physics}\ }\textbf {\bibinfo {volume} {15}},\ \bibinfo
  {pages} {53046} (\bibinfo {year} {2013})}\BibitemShut {NoStop}%
\bibitem [{\citenamefont {Liu}\ \emph {et~al.}(2012)\citenamefont {Liu},
  \citenamefont {Carrasco}, \citenamefont {Santra}, \citenamefont
  {Michaelides}, \citenamefont {Scheffler},\ and\ \citenamefont
  {Tkatchenko}}]{Liu2012}%
  \BibitemOpen
  \bibfield  {author} {\bibinfo {author} {\bibfnamefont {W.}~\bibnamefont
  {Liu}}, \bibinfo {author} {\bibfnamefont {J.}~\bibnamefont {Carrasco}},
  \bibinfo {author} {\bibfnamefont {B.}~\bibnamefont {Santra}}, \bibinfo
  {author} {\bibfnamefont {A.}~\bibnamefont {Michaelides}}, \bibinfo {author}
  {\bibfnamefont {M.}~\bibnamefont {Scheffler}}, \ and\ \bibinfo {author}
  {\bibfnamefont {A.}~\bibnamefont {Tkatchenko}},\ }\href {\doibase
  10.1103/PhysRevB.86.245405} {\bibfield  {journal} {\bibinfo  {journal} {Phys.
  Rev. B}\ }\textbf {\bibinfo {volume} {86}},\ \bibinfo {pages} {245405}
  (\bibinfo {year} {2012})}\BibitemShut {NoStop}%
\bibitem [{\citenamefont {Al-Saidi}\ \emph {et~al.}(2012)\citenamefont
  {Al-Saidi}, \citenamefont {Feng},\ and\ \citenamefont
  {Fichthorn}}]{Al-Saidi2012}%
  \BibitemOpen
  \bibfield  {author} {\bibinfo {author} {\bibfnamefont {W.~A.}\ \bibnamefont
  {Al-Saidi}}, \bibinfo {author} {\bibfnamefont {H.}~\bibnamefont {Feng}}, \
  and\ \bibinfo {author} {\bibfnamefont {K.~A.}\ \bibnamefont {Fichthorn}},\
  }\href {\doibase 10.1021/nl2041113} {\bibfield  {journal} {\bibinfo
  {journal} {Nano Letters}\ }\textbf {\bibinfo {volume} {12}},\ \bibinfo
  {pages} {997} (\bibinfo {year} {2012})}\BibitemShut {NoStop}%
\bibitem [{\citenamefont {van Ruitenbeek}(2012)}]{VanRuitenbeek2012}%
  \BibitemOpen
  \bibfield  {author} {\bibinfo {author} {\bibfnamefont {J.}~\bibnamefont {van
  Ruitenbeek}},\ }\href {\doibase 10.1038/nmat3436} {\bibfield  {journal}
  {\bibinfo  {journal} {Nature Materials}\ }\textbf {\bibinfo {volume} {11}},\
  \bibinfo {pages} {834} (\bibinfo {year} {2012})}\BibitemShut {NoStop}%
\bibitem [{\citenamefont {Wagner}\ \emph {et~al.}(2012)\citenamefont {Wagner},
  \citenamefont {Fournier}, \citenamefont {Tautz},\ and\ \citenamefont
  {Temirov}}]{Wagner2012}%
  \BibitemOpen
  \bibfield  {author} {\bibinfo {author} {\bibfnamefont {C.}~\bibnamefont
  {Wagner}}, \bibinfo {author} {\bibfnamefont {N.}~\bibnamefont {Fournier}},
  \bibinfo {author} {\bibfnamefont {F.~S.}\ \bibnamefont {Tautz}}, \ and\
  \bibinfo {author} {\bibfnamefont {R.}~\bibnamefont {Temirov}},\ }\href
  {\doibase 10.1103/PhysRevLett.109.076102} {\bibfield  {journal} {\bibinfo
  {journal} {Phys. Rev. Lett.}\ }\textbf {\bibinfo {volume} {109}},\ \bibinfo
  {pages} {076102} (\bibinfo {year} {2012})}\BibitemShut {NoStop}%
\bibitem [{\citenamefont {Carrasco}\ \emph {et~al.}(2014)\citenamefont
  {Carrasco}, \citenamefont {Liu}, \citenamefont {Michaelides},\ and\
  \citenamefont {Tkatchenko}}]{Carrasco2014}%
  \BibitemOpen
  \bibfield  {author} {\bibinfo {author} {\bibfnamefont {J.}~\bibnamefont
  {Carrasco}}, \bibinfo {author} {\bibfnamefont {W.}~\bibnamefont {Liu}},
  \bibinfo {author} {\bibfnamefont {A.}~\bibnamefont {Michaelides}}, \ and\
  \bibinfo {author} {\bibfnamefont {A.}~\bibnamefont {Tkatchenko}},\ }\href
  {\doibase 10.1063/1.4866175} {\bibfield  {journal} {\bibinfo  {journal} {The
  Journal of Chemical Physics}\ }\textbf {\bibinfo {volume} {140}},\ \bibinfo
  {pages} {84704} (\bibinfo {year} {2014})}\BibitemShut {NoStop}%
\bibitem [{\citenamefont {Ropo}\ \emph
  {et~al.}(2016{\natexlab{b}})\citenamefont {Ropo}, \citenamefont {Schneider},
  \citenamefont {Baldauf},\ and\ \citenamefont {Blum}}]{Ropo2016}%
  \BibitemOpen
  \bibfield  {author} {\bibinfo {author} {\bibfnamefont {M.}~\bibnamefont
  {Ropo}}, \bibinfo {author} {\bibfnamefont {M.}~\bibnamefont {Schneider}},
  \bibinfo {author} {\bibfnamefont {C.}~\bibnamefont {Baldauf}}, \ and\
  \bibinfo {author} {\bibfnamefont {V.}~\bibnamefont {Blum}},\ }\href {\doibase
  10.1038/sdata.2016.9} {\bibfield  {journal} {\bibinfo  {journal} {Scientific
  Data}\ }\textbf {\bibinfo {volume} {3}},\ \bibinfo {pages} {160009} (\bibinfo
  {year} {2016}{\natexlab{b}})}\BibitemShut {NoStop}%
\bibitem [{\citenamefont {Ceriotti}\ \emph {et~al.}()\citenamefont {Ceriotti},
  \citenamefont {De},\ and\ \citenamefont {Musil}}]{GLOSIM}%
  \BibitemOpen
  \bibfield  {author} {\bibinfo {author} {\bibfnamefont {M.}~\bibnamefont
  {Ceriotti}}, \bibinfo {author} {\bibfnamefont {S.}~\bibnamefont {De}}, \ and\
  \bibinfo {author} {\bibfnamefont {F.}~\bibnamefont {Musil}},\ }\href
  {https://github.com/cosmo-epfl/glosim} {\enquote {\bibinfo {title} {Glosim
  package},}\ }\bibinfo {note} {Code assessed in 2020-01-01}\BibitemShut
  {NoStop}%
\bibitem [{\citenamefont {Pedregosa}\ \emph {et~al.}(2011)\citenamefont
  {Pedregosa}, \citenamefont {Varoquaux}, \citenamefont {Gramfort},
  \citenamefont {Michel}, \citenamefont {Thirion}, \citenamefont {Grisel},
  \citenamefont {Blondel}, \citenamefont {Prettenhofer}, \citenamefont {Weiss},
  \citenamefont {Dubourg}, \citenamefont {Vanderplas}, \citenamefont {Passos},
  \citenamefont {Cournapeau}, \citenamefont {Brucher}, \citenamefont {Perrot},\
  and\ \citenamefont {Duchesnay}}]{scikit-learn}%
  \BibitemOpen
  \bibfield  {author} {\bibinfo {author} {\bibfnamefont {F.}~\bibnamefont
  {Pedregosa}}, \bibinfo {author} {\bibfnamefont {G.}~\bibnamefont
  {Varoquaux}}, \bibinfo {author} {\bibfnamefont {A.}~\bibnamefont {Gramfort}},
  \bibinfo {author} {\bibfnamefont {V.}~\bibnamefont {Michel}}, \bibinfo
  {author} {\bibfnamefont {B.}~\bibnamefont {Thirion}}, \bibinfo {author}
  {\bibfnamefont {O.}~\bibnamefont {Grisel}}, \bibinfo {author} {\bibfnamefont
  {M.}~\bibnamefont {Blondel}}, \bibinfo {author} {\bibfnamefont
  {P.}~\bibnamefont {Prettenhofer}}, \bibinfo {author} {\bibfnamefont
  {R.}~\bibnamefont {Weiss}}, \bibinfo {author} {\bibfnamefont
  {V.}~\bibnamefont {Dubourg}}, \bibinfo {author} {\bibfnamefont
  {J.}~\bibnamefont {Vanderplas}}, \bibinfo {author} {\bibfnamefont
  {A.}~\bibnamefont {Passos}}, \bibinfo {author} {\bibfnamefont
  {D.}~\bibnamefont {Cournapeau}}, \bibinfo {author} {\bibfnamefont
  {M.}~\bibnamefont {Brucher}}, \bibinfo {author} {\bibfnamefont
  {M.}~\bibnamefont {Perrot}}, \ and\ \bibinfo {author} {\bibfnamefont
  {E.}~\bibnamefont {Duchesnay}},\ }\href@noop {} {\bibfield  {journal}
  {\bibinfo  {journal} {Journal of Machine Learning Research}\ }\textbf
  {\bibinfo {volume} {12}},\ \bibinfo {pages} {2825} (\bibinfo {year}
  {2011})}\BibitemShut {NoStop}%
\bibitem [{\citenamefont {McQuarrie}(2000)}]{McQuarrie}%
  \BibitemOpen
  \bibfield  {author} {\bibinfo {author} {\bibfnamefont {D.~A.}\ \bibnamefont
  {McQuarrie}},\ }\href@noop {} {\emph {\bibinfo {title} {{Statistical
  Mechanics}}}}\ (\bibinfo  {publisher} {University Science Books},\ \bibinfo
  {year} {2000})\BibitemShut {NoStop}%
\bibitem [{\citenamefont {Fultz}(2010)}]{Fultz2010}%
  \BibitemOpen
  \bibfield  {author} {\bibinfo {author} {\bibfnamefont {B.}~\bibnamefont
  {Fultz}},\ }\href {\doibase https://doi.org/10.1016/j.pmatsci.2009.05.002}
  {\bibfield  {journal} {\bibinfo  {journal} {Progress in Materials Science}\
  }\textbf {\bibinfo {volume} {55}},\ \bibinfo {pages} {247} (\bibinfo {year}
  {2010})}\BibitemShut {NoStop}%
\bibitem [{\citenamefont {Bolger}(1995)}]{BolgerBook1995}%
  \BibitemOpen
  \bibfield  {author} {\bibinfo {author} {\bibfnamefont {M.~B.}\ \bibnamefont
  {Bolger}},\ }in\ \href {\doibase
  https://doi.org/10.1016/B978-012286230-4/50010-9} {\emph {\bibinfo
  {booktitle} {Introduction to Biophysical Methods for Protein and Nucleic Acid
  Research}}},\ \bibinfo {editor} {edited by\ \bibinfo {editor} {\bibfnamefont
  {J.~A.}\ \bibnamefont {Glasel}}, \bibinfo {editor} {\bibfnamefont {M.~P.}\
  \bibnamefont {Deutscher}}, \ and\ \bibinfo {editor} {\bibfnamefont {M.~P.}\
  \bibnamefont {Deutscher}}}\ (\bibinfo  {publisher} {Academic Press},\
  \bibinfo {address} {San Diego},\ \bibinfo {year} {1995})\ pp.\ \bibinfo
  {pages} {433--490}\BibitemShut {NoStop}%
\bibitem [{\citenamefont {Lingenfelder}(2008)}]{LingenfelderThesis}%
  \BibitemOpen
  \bibfield  {author} {\bibinfo {author} {\bibfnamefont {M.}~\bibnamefont
  {Lingenfelder}},\ }\emph {\bibinfo {title} {Chiral recognition and
  supramolecular self-assembly of adsorbed amino acids and dipeptides at the
  submolecular level}},\ \href {\doibase 10.5075/epfl-thesis-3920} {Ph.D.
  thesis},\ \bibinfo  {school} {\'Ecole Polytechnique F\'ed\`erale de Lausanne}
  (\bibinfo {year} {2008})\BibitemShut {NoStop}%
\bibitem [{\citenamefont {Egger}\ \emph {et~al.}(2015)\citenamefont {Egger},
  \citenamefont {Liu}, \citenamefont {Neaton},\ and\ \citenamefont
  {Kronik}}]{Egger_2015}%
  \BibitemOpen
  \bibfield  {author} {\bibinfo {author} {\bibfnamefont {D.~A.}\ \bibnamefont
  {Egger}}, \bibinfo {author} {\bibfnamefont {Z.-F.}\ \bibnamefont {Liu}},
  \bibinfo {author} {\bibfnamefont {J.~B.}\ \bibnamefont {Neaton}}, \ and\
  \bibinfo {author} {\bibfnamefont {L.}~\bibnamefont {Kronik}},\ }\href@noop {}
  {\bibfield  {journal} {\bibinfo  {journal} {Nano Letters}\ }\textbf {\bibinfo
  {volume} {15}},\ \bibinfo {pages} {2448} (\bibinfo {year}
  {2015})}\BibitemShut {NoStop}%
\bibitem [{\citenamefont {Liu}\ \emph {et~al.}(2017)\citenamefont {Liu},
  \citenamefont {Egger}, \citenamefont {Refaely-Abramson}, \citenamefont
  {Kronik},\ and\ \citenamefont {Neaton}}]{Liu_2017}%
  \BibitemOpen
  \bibfield  {author} {\bibinfo {author} {\bibfnamefont {Z.-F.}\ \bibnamefont
  {Liu}}, \bibinfo {author} {\bibfnamefont {D.~A.}\ \bibnamefont {Egger}},
  \bibinfo {author} {\bibfnamefont {S.}~\bibnamefont {Refaely-Abramson}},
  \bibinfo {author} {\bibfnamefont {L.}~\bibnamefont {Kronik}}, \ and\ \bibinfo
  {author} {\bibfnamefont {J.~B.}\ \bibnamefont {Neaton}},\ }\href@noop {}
  {\bibfield  {journal} {\bibinfo  {journal} {The Journal of Chemical Physics}\
  }\textbf {\bibinfo {volume} {146}},\ \bibinfo {pages} {092326} (\bibinfo
  {year} {2017})}\BibitemShut {NoStop}%
\bibitem [{\citenamefont {Barlow}\ \emph {et~al.}(1998)\citenamefont {Barlow},
  \citenamefont {Kitching}, \citenamefont {Haq},\ and\ \citenamefont
  {Richardson}}]{BARLOW1998322}%
  \BibitemOpen
  \bibfield  {author} {\bibinfo {author} {\bibfnamefont {S.}~\bibnamefont
  {Barlow}}, \bibinfo {author} {\bibfnamefont {K.}~\bibnamefont {Kitching}},
  \bibinfo {author} {\bibfnamefont {S.}~\bibnamefont {Haq}}, \ and\ \bibinfo
  {author} {\bibfnamefont {N.}~\bibnamefont {Richardson}},\ }\href {\doibase
  https://doi.org/10.1016/S0039-6028(97)01086-8} {\bibfield  {journal}
  {\bibinfo  {journal} {Surface Science}\ }\textbf {\bibinfo {volume} {401}},\
  \bibinfo {pages} {322} (\bibinfo {year} {1998})}\BibitemShut {NoStop}%
\bibitem [{\citenamefont {Barlow}\ \emph {et~al.}(2004)\citenamefont {Barlow},
  \citenamefont {Louafi}, \citenamefont {Le~Roux}, \citenamefont {Williams},
  \citenamefont {Muryn}, \citenamefont {Haq},\ and\ \citenamefont
  {Raval}}]{BarlowRaval2004}%
  \BibitemOpen
  \bibfield  {author} {\bibinfo {author} {\bibfnamefont {S.~M.}\ \bibnamefont
  {Barlow}}, \bibinfo {author} {\bibfnamefont {S.}~\bibnamefont {Louafi}},
  \bibinfo {author} {\bibfnamefont {D.}~\bibnamefont {Le~Roux}}, \bibinfo
  {author} {\bibfnamefont {J.}~\bibnamefont {Williams}}, \bibinfo {author}
  {\bibfnamefont {C.}~\bibnamefont {Muryn}}, \bibinfo {author} {\bibfnamefont
  {S.}~\bibnamefont {Haq}}, \ and\ \bibinfo {author} {\bibfnamefont
  {R.}~\bibnamefont {Raval}},\ }\href {\doibase 10.1021/la049391b} {\bibfield
  {journal} {\bibinfo  {journal} {Langmuir}\ }\textbf {\bibinfo {volume}
  {20}},\ \bibinfo {pages} {7171} (\bibinfo {year} {2004})}\BibitemShut
  {NoStop}%
\bibitem [{\citenamefont {Méthivier}\ \emph {et~al.}(2015)\citenamefont
  {Méthivier}, \citenamefont {Humblot},\ and\ \citenamefont
  {Pradier}}]{METHIVIER201588}%
  \BibitemOpen
  \bibfield  {author} {\bibinfo {author} {\bibfnamefont {C.}~\bibnamefont
  {Méthivier}}, \bibinfo {author} {\bibfnamefont {V.}~\bibnamefont {Humblot}},
  \ and\ \bibinfo {author} {\bibfnamefont {C.-M.}\ \bibnamefont {Pradier}},\
  }\href {\doibase https://doi.org/10.1016/j.susc.2014.09.007} {\bibfield
  {journal} {\bibinfo  {journal} {Surface Science}\ }\textbf {\bibinfo {volume}
  {632}},\ \bibinfo {pages} {88} (\bibinfo {year} {2015})}\BibitemShut
  {NoStop}%
\bibitem [{\citenamefont {Marti}\ \emph {et~al.}(2002)\citenamefont {Marti},
  \citenamefont {Barlow}, \citenamefont {Haq},\ and\ \citenamefont
  {Raval}}]{MATEOMARTI2002191}%
  \BibitemOpen
  \bibfield  {author} {\bibinfo {author} {\bibfnamefont {E.~M.}\ \bibnamefont
  {Marti}}, \bibinfo {author} {\bibfnamefont {S.}~\bibnamefont {Barlow}},
  \bibinfo {author} {\bibfnamefont {S.}~\bibnamefont {Haq}}, \ and\ \bibinfo
  {author} {\bibfnamefont {R.}~\bibnamefont {Raval}},\ }\href {\doibase
  https://doi.org/10.1016/S0039-6028(01)02026-X} {\bibfield  {journal}
  {\bibinfo  {journal} {Surface Science}\ }\textbf {\bibinfo {volume} {501}},\
  \bibinfo {pages} {191} (\bibinfo {year} {2002})}\BibitemShut {NoStop}%
\bibitem [{\citenamefont {Marti}\ \emph {et~al.}(2004)\citenamefont {Marti},
  \citenamefont {Quash}, \citenamefont {Methivier}, \citenamefont {Dubot},\
  and\ \citenamefont {Pradier}}]{MARTI200485}%
  \BibitemOpen
  \bibfield  {author} {\bibinfo {author} {\bibfnamefont {E.~M.}\ \bibnamefont
  {Marti}}, \bibinfo {author} {\bibfnamefont {A.}~\bibnamefont {Quash}},
  \bibinfo {author} {\bibfnamefont {C.}~\bibnamefont {Methivier}}, \bibinfo
  {author} {\bibfnamefont {P.}~\bibnamefont {Dubot}}, \ and\ \bibinfo {author}
  {\bibfnamefont {C.}~\bibnamefont {Pradier}},\ }\href {\doibase
  https://doi.org/10.1016/j.colsurfa.2004.08.055} {\bibfield  {journal}
  {\bibinfo  {journal} {Colloids and Surfaces A: Physicochemical and
  Engineering Aspects}\ }\textbf {\bibinfo {volume} {249}},\ \bibinfo {pages}
  {85} (\bibinfo {year} {2004})}\BibitemShut {NoStop}%
\bibitem [{\citenamefont {Eralp}\ \emph {et~al.}(2010)\citenamefont {Eralp},
  \citenamefont {Shavorskiy}, \citenamefont {Zheleva}, \citenamefont {Held},
  \citenamefont {Kalashnyk}, \citenamefont {Ning},\ and\ \citenamefont
  {Linderoth}}]{EralpTrolle2010}%
  \BibitemOpen
  \bibfield  {author} {\bibinfo {author} {\bibfnamefont {T.}~\bibnamefont
  {Eralp}}, \bibinfo {author} {\bibfnamefont {A.}~\bibnamefont {Shavorskiy}},
  \bibinfo {author} {\bibfnamefont {Z.~V.}\ \bibnamefont {Zheleva}}, \bibinfo
  {author} {\bibfnamefont {G.}~\bibnamefont {Held}}, \bibinfo {author}
  {\bibfnamefont {N.}~\bibnamefont {Kalashnyk}}, \bibinfo {author}
  {\bibfnamefont {Y.}~\bibnamefont {Ning}}, \ and\ \bibinfo {author}
  {\bibfnamefont {T.~R.}\ \bibnamefont {Linderoth}},\ }\href {\doibase
  10.1021/la1036772} {\bibfield  {journal} {\bibinfo  {journal} {Langmuir}\
  }\textbf {\bibinfo {volume} {26}},\ \bibinfo {pages} {18841} (\bibinfo {year}
  {2010})}\BibitemShut {NoStop}%
\bibitem [{\citenamefont {Iori}\ \emph {et~al.}(2009)\citenamefont {Iori},
  \citenamefont {{Di Felice}}, \citenamefont {Molinari},\ and\ \citenamefont
  {Corni}}]{Iori2009}%
  \BibitemOpen
  \bibfield  {author} {\bibinfo {author} {\bibfnamefont {F.}~\bibnamefont
  {Iori}}, \bibinfo {author} {\bibfnamefont {R.}~\bibnamefont {{Di Felice}}},
  \bibinfo {author} {\bibfnamefont {E.}~\bibnamefont {Molinari}}, \ and\
  \bibinfo {author} {\bibfnamefont {S.}~\bibnamefont {Corni}},\ }\href
  {\doibase 10.1002/jcc.21165} {\bibfield  {journal} {\bibinfo  {journal}
  {Journal of Computational Chemistry}\ }\textbf {\bibinfo {volume} {30}},\
  \bibinfo {pages} {1465} (\bibinfo {year} {2009})}\BibitemShut {NoStop}%
\bibitem [{\citenamefont {Wright}\ \emph {et~al.}(2013)\citenamefont {Wright},
  \citenamefont {Rodger}, \citenamefont {Corni},\ and\ \citenamefont
  {Walsh}}]{Wright2013}%
  \BibitemOpen
  \bibfield  {author} {\bibinfo {author} {\bibfnamefont {L.~B.}\ \bibnamefont
  {Wright}}, \bibinfo {author} {\bibfnamefont {P.~M.}\ \bibnamefont {Rodger}},
  \bibinfo {author} {\bibfnamefont {S.}~\bibnamefont {Corni}}, \ and\ \bibinfo
  {author} {\bibfnamefont {T.~R.}\ \bibnamefont {Walsh}},\ }\href {\doibase
  10.1021/ct301018m} {\bibfield  {journal} {\bibinfo  {journal} {Journal of
  Chemical Theory and Computation}\ }\textbf {\bibinfo {volume} {9}},\ \bibinfo
  {pages} {1616} (\bibinfo {year} {2013})}\BibitemShut {NoStop}%
\bibitem [{\citenamefont {Heinz}\ \emph {et~al.}(2013)\citenamefont {Heinz},
  \citenamefont {Lin}, \citenamefont {{Kishore Mishra}},\ and\ \citenamefont
  {Emami}}]{Heinz2013}%
  \BibitemOpen
  \bibfield  {author} {\bibinfo {author} {\bibfnamefont {H.}~\bibnamefont
  {Heinz}}, \bibinfo {author} {\bibfnamefont {T.-J.}\ \bibnamefont {Lin}},
  \bibinfo {author} {\bibfnamefont {R.}~\bibnamefont {{Kishore Mishra}}}, \
  and\ \bibinfo {author} {\bibfnamefont {F.~S.}\ \bibnamefont {Emami}},\ }\href
  {\doibase 10.1021/la3038846} {\bibfield  {journal} {\bibinfo  {journal}
  {Langmuir}\ }\textbf {\bibinfo {volume} {29}},\ \bibinfo {pages} {1754}
  (\bibinfo {year} {2013})}\BibitemShut {NoStop}%
\bibitem [{\citenamefont {Koslowski}(2017)}]{KoslowskiThesis}%
  \BibitemOpen
  \bibfield  {author} {\bibinfo {author} {\bibfnamefont {S.}~\bibnamefont
  {Koslowski}},\ }\emph {\bibinfo {title} {Scanning tunneling microscopy on
  large bio-molecular systems on surfaces}},\ \href {\doibase
  10.5075/epfl-thesis-8159} {Ph.D. thesis},\ \bibinfo  {school} {\'Ecole
  Polytechnique F\'ed\`erale de Lausanne} (\bibinfo {year} {2017})\BibitemShut
  {NoStop}%
\bibitem [{\citenamefont {Todorovi{\'{c}}}\ \emph {et~al.}(2019)\citenamefont
  {Todorovi{\'{c}}}, \citenamefont {Gutmann}, \citenamefont {Corander},\ and\
  \citenamefont {Rinke}}]{Todorovic2019}%
  \BibitemOpen
  \bibfield  {author} {\bibinfo {author} {\bibfnamefont {M.}~\bibnamefont
  {Todorovi{\'{c}}}}, \bibinfo {author} {\bibfnamefont {M.~U.}\ \bibnamefont
  {Gutmann}}, \bibinfo {author} {\bibfnamefont {J.}~\bibnamefont {Corander}}, \
  and\ \bibinfo {author} {\bibfnamefont {P.}~\bibnamefont {Rinke}},\ }\href
  {\doibase 10.1038/s41524-019-0175-2} {\bibfield  {journal} {\bibinfo
  {journal} {npj Computational Materials}\ }\textbf {\bibinfo {volume} {5}},\
  \bibinfo {pages} {35} (\bibinfo {year} {2019})}\BibitemShut {NoStop}%
\bibitem [{\citenamefont {Sch{\"{o}}nborn}\ \emph {et~al.}(2009)\citenamefont
  {Sch{\"{o}}nborn}, \citenamefont {Goedecker}, \citenamefont {Roy},\ and\
  \citenamefont {Oganov}}]{Oganov2009}%
  \BibitemOpen
  \bibfield  {author} {\bibinfo {author} {\bibfnamefont {S.~E.}\ \bibnamefont
  {Sch{\"{o}}nborn}}, \bibinfo {author} {\bibfnamefont {S.}~\bibnamefont
  {Goedecker}}, \bibinfo {author} {\bibfnamefont {S.}~\bibnamefont {Roy}}, \
  and\ \bibinfo {author} {\bibfnamefont {A.~R.}\ \bibnamefont {Oganov}},\
  }\href {\doibase 10.1063/1.3097197} {\bibfield  {journal} {\bibinfo
  {journal} {The Journal of Chemical Physics}\ }\textbf {\bibinfo {volume}
  {130}},\ \bibinfo {pages} {144108} (\bibinfo {year} {2009})}\BibitemShut
  {NoStop}%
\bibitem [{\citenamefont {Dieterich}\ and\ \citenamefont
  {Hartke}(2010)}]{Dieterich2010}%
  \BibitemOpen
  \bibfield  {author} {\bibinfo {author} {\bibfnamefont {J.~M.}\ \bibnamefont
  {Dieterich}}\ and\ \bibinfo {author} {\bibfnamefont {B.}~\bibnamefont
  {Hartke}},\ }\href {\doibase 10.1080/00268970903446756} {\bibfield  {journal}
  {\bibinfo  {journal} {Molecular Physics}\ }\textbf {\bibinfo {volume}
  {108}},\ \bibinfo {pages} {279} (\bibinfo {year} {2010})}\BibitemShut
  {NoStop}%
\bibitem [{\citenamefont {Supady}\ \emph {et~al.}(2015)\citenamefont {Supady},
  \citenamefont {Blum},\ and\ \citenamefont {Baldauf}}]{fafoom}%
  \BibitemOpen
  \bibfield  {author} {\bibinfo {author} {\bibfnamefont {A.}~\bibnamefont
  {Supady}}, \bibinfo {author} {\bibfnamefont {V.}~\bibnamefont {Blum}}, \ and\
  \bibinfo {author} {\bibfnamefont {C.}~\bibnamefont {Baldauf}},\ }\href
  {\doibase 10.1021/acs.jcim.5b00243} {\bibfield  {journal} {\bibinfo
  {journal} {Journal of Chemical Information and Modeling}\ }\textbf {\bibinfo
  {volume} {55}},\ \bibinfo {pages} {2338} (\bibinfo {year}
  {2015})}\BibitemShut {NoStop}%
\bibitem [{\citenamefont {Tom}\ \emph {et~al.}(2020)\citenamefont {Tom},
  \citenamefont {Rose}, \citenamefont {Bier}, \citenamefont {O’Brien},
  \citenamefont {Álvaro Vázquez-Mayagoitia},\ and\ \citenamefont
  {Marom}}]{genarris}%
  \BibitemOpen
  \bibfield  {author} {\bibinfo {author} {\bibfnamefont {R.}~\bibnamefont
  {Tom}}, \bibinfo {author} {\bibfnamefont {T.}~\bibnamefont {Rose}}, \bibinfo
  {author} {\bibfnamefont {I.}~\bibnamefont {Bier}}, \bibinfo {author}
  {\bibfnamefont {H.}~\bibnamefont {O’Brien}}, \bibinfo {author}
  {\bibnamefont {Álvaro Vázquez-Mayagoitia}}, \ and\ \bibinfo {author}
  {\bibfnamefont {N.}~\bibnamefont {Marom}},\ }\href {\doibase
  https://doi.org/10.1016/j.cpc.2020.107170} {\bibfield  {journal} {\bibinfo
  {journal} {Computer Physics Communications}\ }\textbf {\bibinfo {volume}
  {250}},\ \bibinfo {pages} {107170} (\bibinfo {year} {2020})}\BibitemShut
  {NoStop}%
\bibitem [{\citenamefont {Hörmann}\ \emph {et~al.}(2019)\citenamefont
  {Hörmann}, \citenamefont {Jeindl}, \citenamefont {Egger}, \citenamefont
  {Scherbela},\ and\ \citenamefont {Hofmann}}]{HORMANN2019}%
  \BibitemOpen
  \bibfield  {author} {\bibinfo {author} {\bibfnamefont {L.}~\bibnamefont
  {Hörmann}}, \bibinfo {author} {\bibfnamefont {A.}~\bibnamefont {Jeindl}},
  \bibinfo {author} {\bibfnamefont {A.~T.}\ \bibnamefont {Egger}}, \bibinfo
  {author} {\bibfnamefont {M.}~\bibnamefont {Scherbela}}, \ and\ \bibinfo
  {author} {\bibfnamefont {O.~T.}\ \bibnamefont {Hofmann}},\ }\href {\doibase
  https://doi.org/10.1016/j.cpc.2019.06.010} {\bibfield  {journal} {\bibinfo
  {journal} {Computer Physics Communications}\ }\textbf {\bibinfo {volume}
  {244}},\ \bibinfo {pages} {143} (\bibinfo {year} {2019})}\BibitemShut
  {NoStop}%
\bibitem [{\citenamefont {Olsson}\ \emph {et~al.}(2017)\citenamefont {Olsson},
  \citenamefont {Wu}, \citenamefont {Paul}, \citenamefont {Clementi},\ and\
  \citenamefont {No{\'e}}}]{Olsson8265}%
  \BibitemOpen
  \bibfield  {author} {\bibinfo {author} {\bibfnamefont {S.}~\bibnamefont
  {Olsson}}, \bibinfo {author} {\bibfnamefont {H.}~\bibnamefont {Wu}}, \bibinfo
  {author} {\bibfnamefont {F.}~\bibnamefont {Paul}}, \bibinfo {author}
  {\bibfnamefont {C.}~\bibnamefont {Clementi}}, \ and\ \bibinfo {author}
  {\bibfnamefont {F.}~\bibnamefont {No{\'e}}},\ }\href {\doibase
  10.1073/pnas.1704803114} {\bibfield  {journal} {\bibinfo  {journal}
  {Proceedings of the National Academy of Sciences}\ }\textbf {\bibinfo
  {volume} {114}},\ \bibinfo {pages} {8265} (\bibinfo {year}
  {2017})}\BibitemShut {NoStop}%
\end{thebibliography}%


\begin{thebibliography}{14}%
\makeatletter
\providecommand \@ifxundefined [1]{%
 \@ifx{#1\undefined}
}%
\providecommand \@ifnum [1]{%
 \ifnum #1\expandafter \@firstoftwo
 \else \expandafter \@secondoftwo
 \fi
}%
\providecommand \@ifx [1]{%
 \ifx #1\expandafter \@firstoftwo
 \else \expandafter \@secondoftwo
 \fi
}%
\providecommand \natexlab [1]{#1}%
\providecommand \enquote  [1]{``#1''}%
\providecommand \bibnamefont  [1]{#1}%
\providecommand \bibfnamefont [1]{#1}%
\providecommand \citenamefont [1]{#1}%
\providecommand \href@noop [0]{\@secondoftwo}%
\providecommand \href [0]{\begingroup \@sanitize@url \@href}%
\providecommand \@href[1]{\@@startlink{#1}\@@href}%
\providecommand \@@href[1]{\endgroup#1\@@endlink}%
\providecommand \@sanitize@url [0]{\catcode `\\12\catcode `\$12\catcode
  `\&12\catcode `\#12\catcode `\^12\catcode `\_12\catcode `\%12\relax}%
\providecommand \@@startlink[1]{}%
\providecommand \@@endlink[0]{}%
\providecommand \url  [0]{\begingroup\@sanitize@url \@url }%
\providecommand \@url [1]{\endgroup\@href {#1}{\urlprefix }}%
\providecommand \urlprefix  [0]{URL }%
\providecommand \Eprint [0]{\href }%
\providecommand \doibase [0]{http://dx.doi.org/}%
\providecommand \selectlanguage [0]{\@gobble}%
\providecommand \bibinfo  [0]{\@secondoftwo}%
\providecommand \bibfield  [0]{\@secondoftwo}%
\providecommand \translation [1]{[#1]}%
\providecommand \BibitemOpen [0]{}%
\providecommand \bibitemStop [0]{}%
\providecommand \bibitemNoStop [0]{.\EOS\space}%
\providecommand \EOS [0]{\spacefactor3000\relax}%
\providecommand \BibitemShut  [1]{\csname bibitem#1\endcsname}%
\let\auto@bib@innerbib\@empty
\bibitem [{\citenamefont {van Lenthe}\ \emph {et~al.}(2000)\citenamefont {van
  Lenthe}, \citenamefont {Faas},\ and\ \citenamefont
  {Snijders}}]{VanLenthe2000}%
  \BibitemOpen
  \bibfield  {author} {\bibinfo {author} {\bibfnamefont {J.}~\bibnamefont {van
  Lenthe}}, \bibinfo {author} {\bibfnamefont {S.}~\bibnamefont {Faas}}, \ and\
  \bibinfo {author} {\bibfnamefont {J.}~\bibnamefont {Snijders}},\ }\href
  {\doibase https://doi.org/10.1016/S0009-2614(00)00832-0} {\bibfield
  {journal} {\bibinfo  {journal} {Chemical Physics Letters}\ }\textbf {\bibinfo
  {volume} {328}},\ \bibinfo {pages} {107} (\bibinfo {year}
  {2000})}\BibitemShut {NoStop}%
\bibitem [{\citenamefont {van Wüllen}(1998)}]{VanWullen1998}%
  \BibitemOpen
  \bibfield  {author} {\bibinfo {author} {\bibfnamefont {C.}~\bibnamefont {van
  Wüllen}},\ }\href {\doibase 10.1063/1.476576} {\bibfield  {journal}
  {\bibinfo  {journal} {The Journal of Chemical Physics}\ }\textbf {\bibinfo
  {volume} {109}},\ \bibinfo {pages} {392} (\bibinfo {year}
  {1998})}\BibitemShut {NoStop}%
\bibitem [{\citenamefont {Perdew}\ \emph {et~al.}(1997)\citenamefont {Perdew},
  \citenamefont {Burke},\ and\ \citenamefont {Ernzerhof}}]{Perdew1997}%
  \BibitemOpen
  \bibfield  {author} {\bibinfo {author} {\bibfnamefont {J.~P.}\ \bibnamefont
  {Perdew}}, \bibinfo {author} {\bibfnamefont {K.}~\bibnamefont {Burke}}, \
  and\ \bibinfo {author} {\bibfnamefont {M.}~\bibnamefont {Ernzerhof}},\ }\href
  {\doibase 10.1103/PhysRevLett.78.1396} {\bibfield  {journal} {\bibinfo
  {journal} {Physical Review Letters}\ }\textbf {\bibinfo {volume} {78}},\
  \bibinfo {pages} {1396} (\bibinfo {year} {1997})}\BibitemShut {NoStop}%
\bibitem [{\citenamefont {Liu}\ \emph {et~al.}(2013)\citenamefont {Liu},
  \citenamefont {Ruiz}, \citenamefont {Zhang}, \citenamefont {Santra},
  \citenamefont {Ren}, \citenamefont {Scheffler},\ and\ \citenamefont
  {Tkatchenko}}]{Liu2013}%
  \BibitemOpen
  \bibfield  {author} {\bibinfo {author} {\bibfnamefont {W.}~\bibnamefont
  {Liu}}, \bibinfo {author} {\bibfnamefont {V.~G.}\ \bibnamefont {Ruiz}},
  \bibinfo {author} {\bibfnamefont {G.-X.}\ \bibnamefont {Zhang}}, \bibinfo
  {author} {\bibfnamefont {B.}~\bibnamefont {Santra}}, \bibinfo {author}
  {\bibfnamefont {X.}~\bibnamefont {Ren}}, \bibinfo {author} {\bibfnamefont
  {M.}~\bibnamefont {Scheffler}}, \ and\ \bibinfo {author} {\bibfnamefont
  {A.}~\bibnamefont {Tkatchenko}},\ }\href {\doibase
  10.1088/1367-2630/15/5/053046} {\bibfield  {journal} {\bibinfo  {journal}
  {New Journal of Physics}\ }\textbf {\bibinfo {volume} {15}},\ \bibinfo
  {pages} {53046} (\bibinfo {year} {2013})}\BibitemShut {NoStop}%
\bibitem [{\citenamefont {Haas}\ \emph {et~al.}(2009)\citenamefont {Haas},
  \citenamefont {Tran},\ and\ \citenamefont {Blaha}}]{Haas2009}%
  \BibitemOpen
  \bibfield  {author} {\bibinfo {author} {\bibfnamefont {P.}~\bibnamefont
  {Haas}}, \bibinfo {author} {\bibfnamefont {F.}~\bibnamefont {Tran}}, \ and\
  \bibinfo {author} {\bibfnamefont {P.}~\bibnamefont {Blaha}},\ }\href
  {\doibase 10.1103/PhysRevB.79.085104} {\bibfield  {journal} {\bibinfo
  {journal} {Phys. Rev. B}\ }\textbf {\bibinfo {volume} {79}},\ \bibinfo
  {pages} {85104} (\bibinfo {year} {2009})}\BibitemShut {NoStop}%
\bibitem [{\citenamefont {Tkatchenko}\ and\ \citenamefont
  {Scheffler}(2009)}]{Tkatchenko2009}%
  \BibitemOpen
  \bibfield  {author} {\bibinfo {author} {\bibfnamefont {A.}~\bibnamefont
  {Tkatchenko}}\ and\ \bibinfo {author} {\bibfnamefont {M.}~\bibnamefont
  {Scheffler}},\ }\href {\doibase 10.1103/PhysRevLett.102.073005} {\bibfield
  {journal} {\bibinfo  {journal} {Physical Review Letters}\ }\textbf {\bibinfo
  {volume} {102}},\ \bibinfo {pages} {073005} (\bibinfo {year}
  {2009})}\BibitemShut {NoStop}%
\bibitem [{\citenamefont {Ruiz}\ \emph {et~al.}(2016)\citenamefont {Ruiz},
  \citenamefont {Liu},\ and\ \citenamefont {Tkatchenko}}]{Ruiz2016}%
  \BibitemOpen
  \bibfield  {author} {\bibinfo {author} {\bibfnamefont {V.~G.}\ \bibnamefont
  {Ruiz}}, \bibinfo {author} {\bibfnamefont {W.}~\bibnamefont {Liu}}, \ and\
  \bibinfo {author} {\bibfnamefont {A.}~\bibnamefont {Tkatchenko}},\ }\href
  {\doibase 10.1103/PhysRevB.93.035118} {\bibfield  {journal} {\bibinfo
  {journal} {Physical Review B}\ }\textbf {\bibinfo {volume} {93}},\ \bibinfo
  {pages} {035118} (\bibinfo {year} {2016})}\BibitemShut {NoStop}%
\bibitem [{Note1()}]{Note1}%
  \BibitemOpen
  \bibinfo {note} {We here used the original parameters published in Ref. \cite
  {Ruiz2016}.}\BibitemShut {Stop}%
\bibitem [{\citenamefont {McQuarrie}(2000)}]{McQuarrie}%
  \BibitemOpen
  \bibfield  {author} {\bibinfo {author} {\bibfnamefont {D.~A.}\ \bibnamefont
  {McQuarrie}},\ }\href@noop {} {\emph {\bibinfo {title} {{Statistical
  Mechanics}}}}\ (\bibinfo  {publisher} {University Science Books},\ \bibinfo
  {year} {2000})\BibitemShut {NoStop}%
\bibitem [{\citenamefont {Fultz}(2010)}]{Fultz2010}%
  \BibitemOpen
  \bibfield  {author} {\bibinfo {author} {\bibfnamefont {B.}~\bibnamefont
  {Fultz}},\ }\href {\doibase https://doi.org/10.1016/j.pmatsci.2009.05.002}
  {\bibfield  {journal} {\bibinfo  {journal} {Progress in Materials Science}\
  }\textbf {\bibinfo {volume} {55}},\ \bibinfo {pages} {247} (\bibinfo {year}
  {2010})}\BibitemShut {NoStop}%
\bibitem [{\citenamefont {Togo}\ and\ \citenamefont {Tanaka}(2015)}]{phonopy}%
  \BibitemOpen
  \bibfield  {author} {\bibinfo {author} {\bibfnamefont {A.}~\bibnamefont
  {Togo}}\ and\ \bibinfo {author} {\bibfnamefont {I.}~\bibnamefont {Tanaka}},\
  }\href@noop {} {\bibfield  {journal} {\bibinfo  {journal} {Scr. Mater.}\
  }\textbf {\bibinfo {volume} {108}},\ \bibinfo {pages} {1} (\bibinfo {year}
  {2015})}\BibitemShut {NoStop}%
\bibitem [{\citenamefont {Fidanyan}()}]{phonopy-fidaynyan}%
  \BibitemOpen
  \bibfield  {author} {\bibinfo {author} {\bibfnamefont {K.}~\bibnamefont
  {Fidanyan}},\ }\href {https://github.com/fidanyan/phonopy} {\enquote
  {\bibinfo {title} {Development version of phonopy},}\ }\bibinfo {note}
  {Accessed: 2019-03-01}\BibitemShut {NoStop}%
\bibitem [{\citenamefont {Heinz}\ and\ \citenamefont
  {Ramezani-Dakhel}(2016)}]{Heinz:2016:Review}%
  \BibitemOpen
  \bibfield  {author} {\bibinfo {author} {\bibfnamefont {H.}~\bibnamefont
  {Heinz}}\ and\ \bibinfo {author} {\bibfnamefont {H.}~\bibnamefont
  {Ramezani-Dakhel}},\ }\href {\doibase 10.1039/C5CS00890E} {\bibfield
  {journal} {\bibinfo  {journal} {Chem. Soc. Rev.}\ }\textbf {\bibinfo {volume}
  {45}},\ \bibinfo {pages} {412} (\bibinfo {year} {2016})}\BibitemShut
  {NoStop}%
\bibitem [{\citenamefont {Phillips}\ \emph {et~al.}(2005)\citenamefont
  {Phillips}, \citenamefont {Braun}, \citenamefont {Wang}, \citenamefont
  {Gumbart}, \citenamefont {Tajkhorshid}, \citenamefont {Villa}, \citenamefont
  {Chipot}, \citenamefont {Skeel}, \citenamefont {Kalé},\ and\ \citenamefont
  {Schulten}}]{Phillips2005}%
  \BibitemOpen
  \bibfield  {author} {\bibinfo {author} {\bibfnamefont {J.~C.}\ \bibnamefont
  {Phillips}}, \bibinfo {author} {\bibfnamefont {R.}~\bibnamefont {Braun}},
  \bibinfo {author} {\bibfnamefont {W.}~\bibnamefont {Wang}}, \bibinfo {author}
  {\bibfnamefont {J.}~\bibnamefont {Gumbart}}, \bibinfo {author} {\bibfnamefont
  {E.}~\bibnamefont {Tajkhorshid}}, \bibinfo {author} {\bibfnamefont
  {E.}~\bibnamefont {Villa}}, \bibinfo {author} {\bibfnamefont
  {C.}~\bibnamefont {Chipot}}, \bibinfo {author} {\bibfnamefont {R.~D.}\
  \bibnamefont {Skeel}}, \bibinfo {author} {\bibfnamefont {L.}~\bibnamefont
  {Kalé}}, \ and\ \bibinfo {author} {\bibfnamefont {K.}~\bibnamefont
  {Schulten}},\ }\href {\doibase 10.1002/jcc.20289} {\bibfield  {journal}
  {\bibinfo  {journal} {Journal of Computational Chemistry}\ }\textbf {\bibinfo
  {volume} {26}},\ \bibinfo {pages} {1781} (\bibinfo {year}
  {2005})}\BibitemShut {NoStop}%
\end{thebibliography}%

\end{document}


\title{Supporting Information: The Conformational Space of a Flexible Amino Acid at Metallic Surfaces}

\author{Dmitrii Maksimov}
\email{maksimov@fhi-berlin.mpg.de}
\affiliation{Fritz-Haber-Institut der Max-Planck-Gesellschaft, Faradayweg 4-6, 14195 Berlin, Germany}
\affiliation{Max Planck Institute for Structure and Dynamics of Matter, Luruper Chaussee 149, 22761 Hamburg, Germany}

\author{Carsten Baldauf}
\email{baldauf@fhi-berlin.mpg.de}
\affiliation{Fritz-Haber-Institut der Max-Planck-Gesellschaft, Faradayweg 4-6, 14195 Berlin, Germany}

\author{Mariana Rossi}
\email{mariana.rossi@mpsd.mpg.de}
\affiliation{Fritz-Haber-Institut der Max-Planck-Gesellschaft, Faradayweg 4-6, 14195 Berlin, Germany}
\affiliation{Max Planck Institute for Structure and Dynamics of Matter, Luruper Chaussee 149, 22761 Hamburg, Germany}

\keywords{Arginine, adsorption, metal, conformational space, electronic structure}
\maketitle

\section{Details of the calculations}

For Cu, Ag and Au, the bulk lattice constants were determined by optimizing the fcc unit cell. The convergence criteria were set to 0.001 eV/\AA~ for the final forces, 10$^{-4}$ e/Bohr$^{3}$ for the charge density, and 10$^{-5}$ eV for the total energy of the system. A 30$\times$30$\times$30 k-grid mesh  was used for the sampling of the Brillouin zone. Relativistic effects were considered by the zeroth order regular approximation (ZORA) \cite{VanLenthe2000, VanWullen1998}. The values obtained with the PBE functional\cite{Perdew1997} are in good agreement with previous works \cite{Liu2013, Haas2009} and are shown in Table \ref{tbl:Bulk_constants}. In that Table, we compare these values with lattice constants obtained when including pairwise van der Waals dispersion from the original Tkatchenko-Scheffler scheme (+vdW)\cite{Tkatchenko2009} and from the one that includes an effective electronic screening optimized for metallic surfaces (+vdW$^{\text{surf}}$)\cite{Ruiz2016}\footnote{We here used the original parameters published in Ref. \cite{Ruiz2016}.}. 

\begin{table}[h!]
\center
 \caption{Lattice constants (in \AA) of bulk metals determined with the PBE, PBE+vdW and PBE+vdW$^\text{surf}$ functionals (\textit{light} settings). }
 \label{tbl:Bulk_constants}
\begin{tabular}{|r|c|c|c|}
\hline
Method     & Cu & Ag & Au \\
\hline
PBE                   & 3.633     & 4.156   & 4.157   \\
\hline
PBE+vdW               & 3.545     & 4.077     & 4.114               \\
\hline
PBE+vdW$^{\text{surf}}$ & 3.604     & 4.022   & 4.173    \\
\hline
Exp              & 3.603      & 4.069     & 4.065      \\
\hline
\end{tabular}
\end{table}

\begin{table}
\center
\caption{Fermi energies calculated with the PBE functional for the 4-layer slabs with (111) surface orientation used in our calculations of the binding energies of Arg-H$^+$ to the different surfaces. All values in eV.}
\label{tbl:fermi}
\begin{tabular}{|l|c|c|c|c|}
\hline
& Cu & Ag  & Au \\ \hline
Slab $E_f$ & -4.73     & -4.30     & -5.02   \\ \hline
\end{tabular}
\end{table}

\begin{table}[h!]
\center
\caption{Relative binding energies (in eV) of relaxed Arg@Cu for different surface unit cell sizes with a 8$\times$8$\times$1 k-grid for the cell sizes less than 10$\times$12 and 4$\times$4$\times$1 for the 10$\times$12 unit cell. All numbers are reported with respect to the binding energy for the structure A modelled with a $5\times6$ surface unit cell.}
\label{tbl:surf_unit_cell_ArgCu}
\begin{tabular}{|r|c|c|c|}
\hline
size     & A & B & C \\
\hline
5$\times$6                   & 0.000     &  0.011   & 0.216  \\
\hline
6$\times$6                   & -0.011     &   -0.013    & 0.190              \\
\hline
6$\times$7                   & -0.021     &  -0.030     & 0.174    \\
\hline
10$\times$12                 &   -0.048     &  -0.053  &       0.151  \\
\hline
\end{tabular}
\end{table}

\begin{figure}[h!]
\centering
\includegraphics[width=0.75\linewidth]{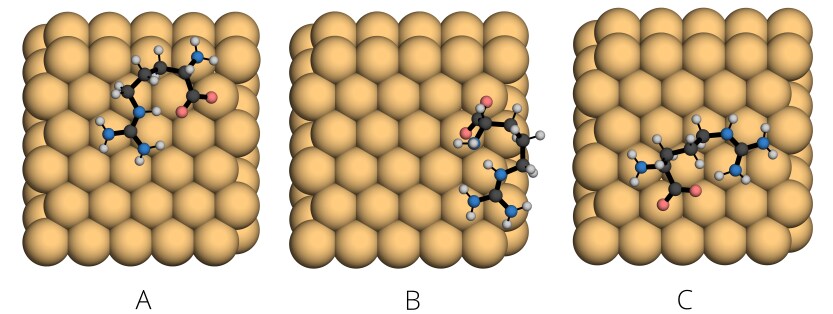}
\caption{Structures that were used for the surface unit cell size convergence test of Arg@Cu. Image unit cell size is $5 \times 6$.}
\label{fig:surfaceunitcellArgCu}
\end{figure}

\begin{table}[h!]
\center
 \caption{Relative binding energies (in eV) of relaxed Arg-H$^+$@Cu for different surface unit cell sizes with a 8$\times$8$\times$1 k-grid for the cell sizes less than 10$\times$12 and 4$\times$4$\times$1 for the 10$\times$12 unit cell. All numbers are reported with respect to the binding energy for the structure A modelled with a $5\times6$ surface unit cell.}
\label{tbl:surf_unit_cell_ArgHCu}
\begin{tabular}{|r|c|c|c|}
\hline
size     & A & B & C \\
\hline
5$\times$6                   & 0.000     &  0.080   & 0.035  \\
\hline
6$\times$6                   & -0.050     &   0.041    & -0.017              \\
\hline
6$\times$7                   & -0.055     &  0.029     & -0.033    \\
\hline
10$\times$12                 & -0.044      & -0.007  &   -0.057     \\
\hline
\end{tabular}
\end{table}

\begin{figure}[h!]
\centering
\includegraphics[width=0.75\linewidth]{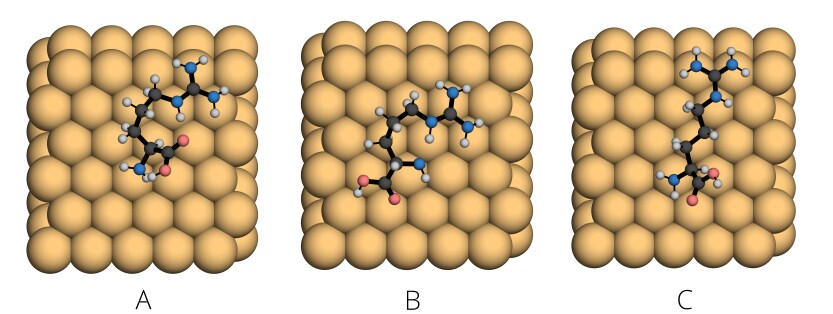}
\caption{Structures that were used for surface unit cell size convergence test for Arg-H$^+$@Cu. Image unit cell size is $5 \times 6$.}
\label{fig:surfaceunitcellArgHCu}
\end{figure}
\clearpage

\section{Family Classification According to Hydrogen Bond Patterns}

\begin{figure}[htb]
\centering
\includegraphics[width=1.0\linewidth]{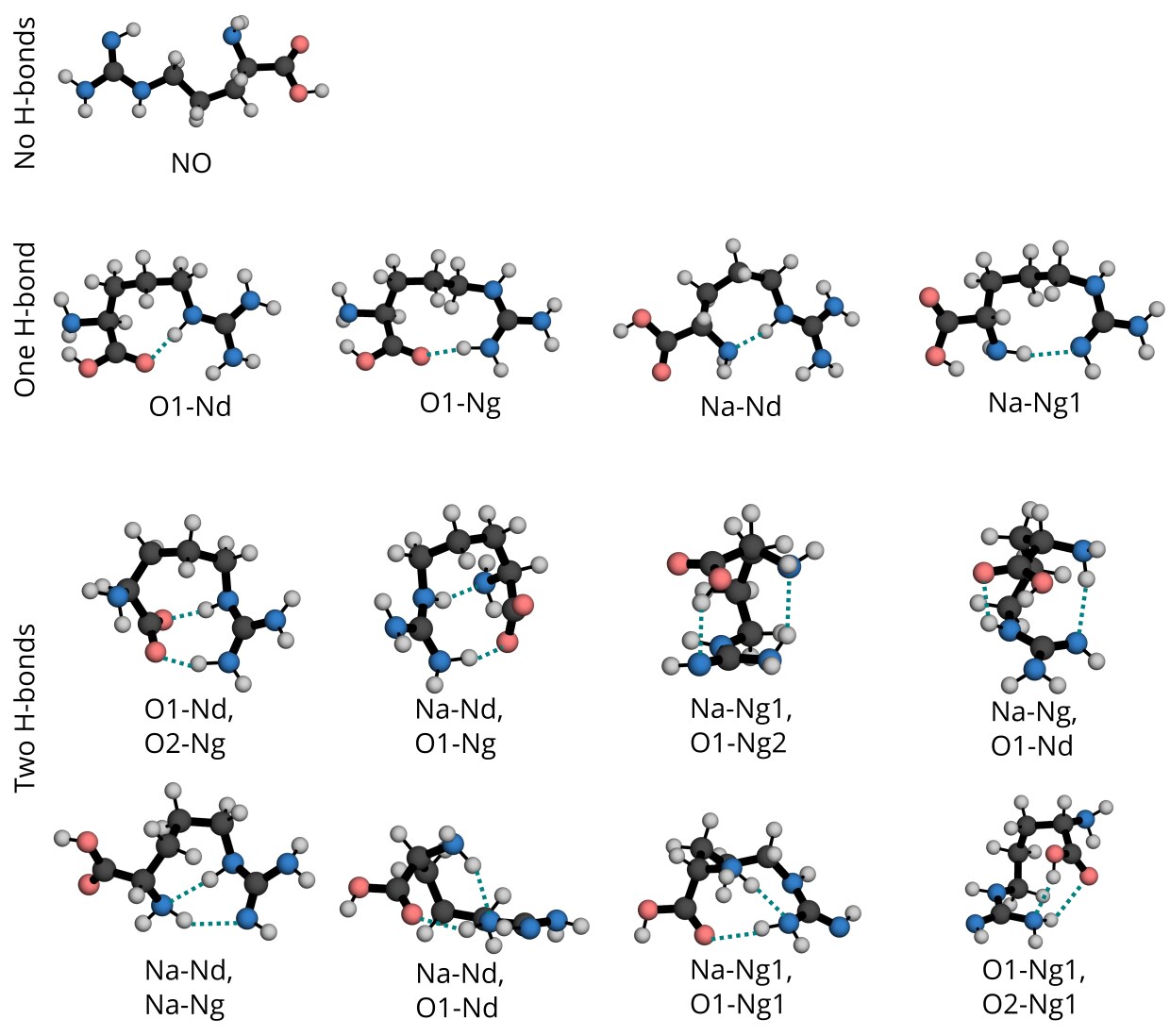}
\caption{Labeling of all H-bond patterns considered in this manuscript.}
\label{fig:hbonds}
\end{figure}
\clearpage
\section{Structure space representation}
\begin{figure}[h!]
\centering
\includegraphics[width=0.99\linewidth]{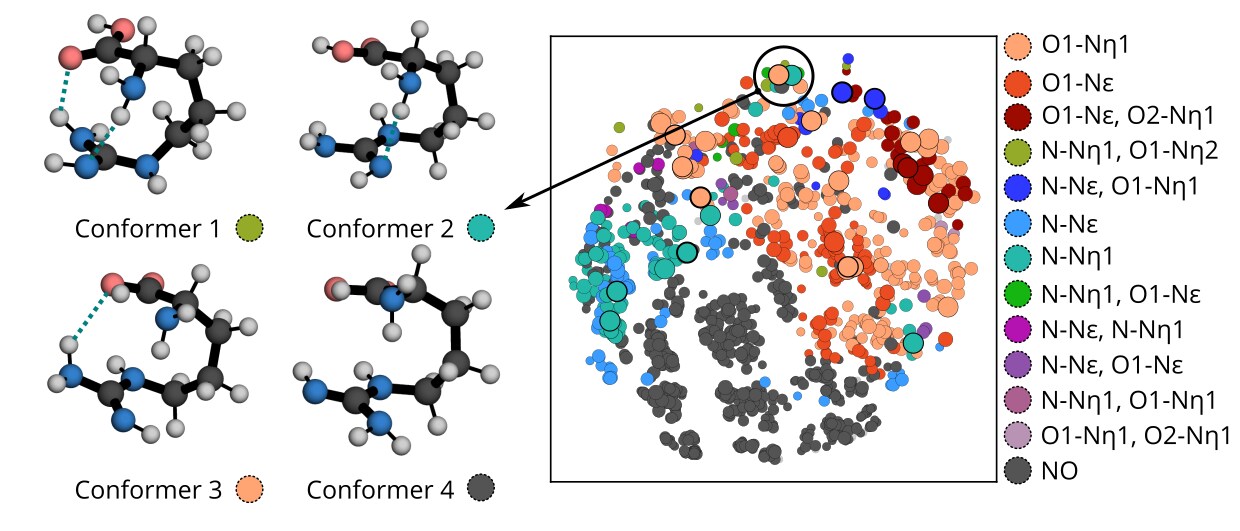}
\caption{Representative conformers with similar backbone structure but different H-bonds within the molecule. The different H-bond pattern can cause energy differences of up to 0.2 eV for similar structures, as discussed in the main text.}
\label{fig:hbonds}
\end{figure}

\begin{figure}[h!]
\centering
\includegraphics[width=0.99\linewidth]{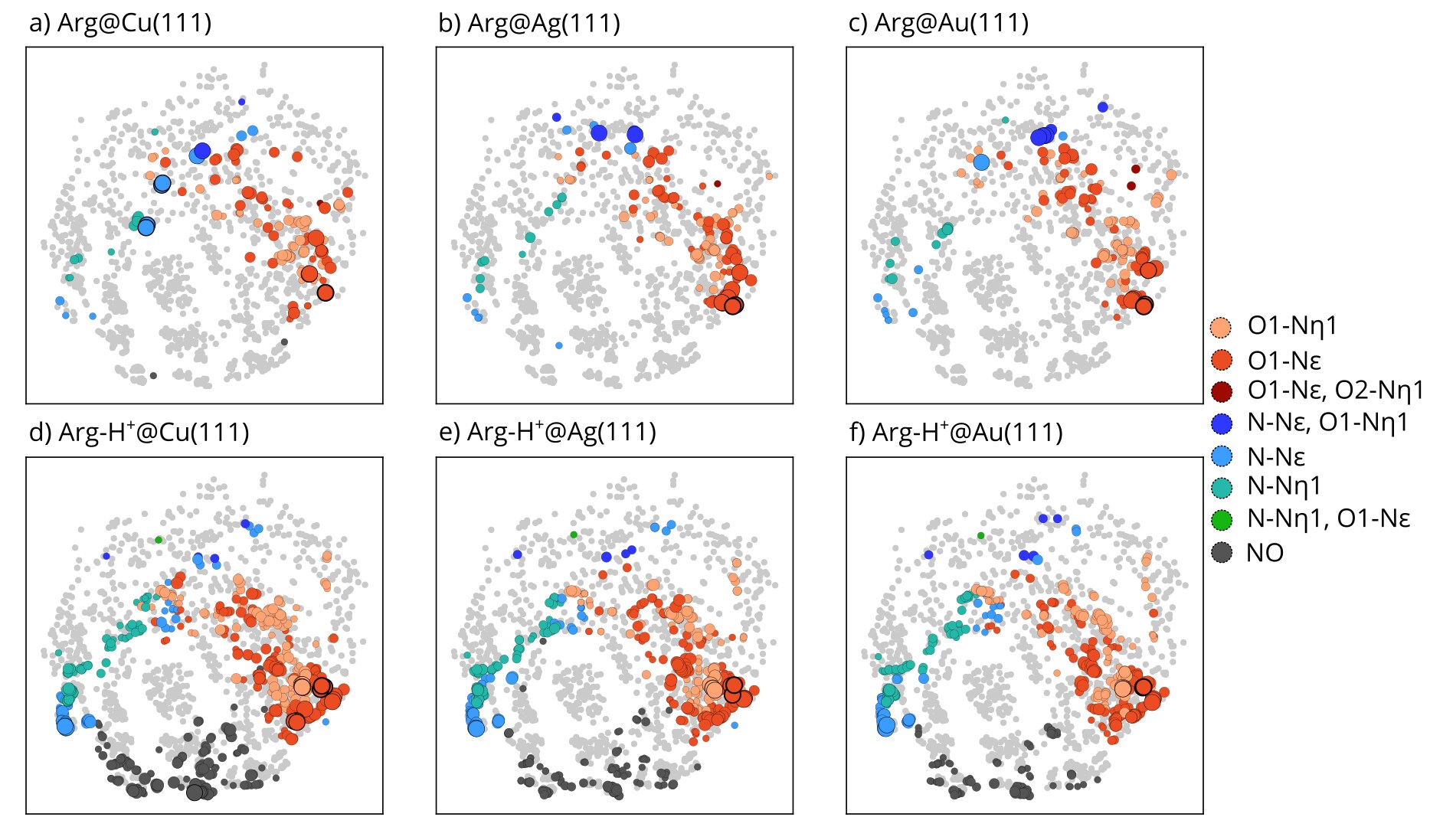}
\caption{Projection of Arg and Arg-H$^+$ conformers adsorbed on the different metalic surfaces on the low-dimensional map of gas-phase Arg, colored according to the H-bond pattern.}
\label{fig:hbonds}
\end{figure}

\begin{table}[]
\caption{Number of different families within 0.1/0.5 eV energy range for different systems.}
\begin{tabular}{|c|c|c|c|c|c|c|c|c|}
\hline
           & Arg    & Arg-H$^+$ & \multicolumn{3}{c|}{Arg} & \multicolumn{3}{c|}{Arg-H$^+$} \\ \hline
Atomnames    &        &      & Cu     & Ag     & Au     & Cu      & Ag     & Au     \\ \hline
NO         & 0 /599 & 0/2  & 0/2    & 0/0    & 0/0    & 1 /162  & 0/124  & 0/98   \\ \hline
N-N$\varepsilon$        & 0 /70  & 0/11 & 5/10   & 0/9    & 1/9    & 4 /80   & 3/79   & 3/77   \\ \hline
N-N$\eta1$        & 7 /87  & 0/20 & 0/11   & 0/12   & 0/13   & 0 /78   & 0/78   & 0/73   \\ \hline
N-N$\varepsilon$, O1-N$\eta1$   & 2 /11  & 2/4  & 1/2    & 2/6    & 4/6    & 0 /4    & 0/4    & 0/5    \\ \hline
O1-N$\varepsilon$        & 2 /115 & 0/31 & 6/56   & 8/62   & 9/56   & 11/152  & 6/146  & 6/140  \\ \hline
O1-N$\eta1$        & 16/237 & 0/37 & 0/66   & 0/70   & 0/71   & 5 /135  & 4/115  & 5/109  \\ \hline
O1-N$\varepsilon$, O1-N$\eta1$   & 5 /27  & 0/2  & 0/1    & 0/1    & 0/2    & 0 /0    & 0/0    & 0/0    \\ \hline
N-N$\eta1$, O1-N$\varepsilon$   & 0 /5   & 0/1  & 0/0    & 0/0    & 0/0    & 0 /1    & 0/1    & 0/1    \\ \hline
N-N$\eta1$, O1-N$\eta1$   & 0 /8   & 0/0  & 0/0    & 0/0    & 0/0    & 0 /0    & 0/0    & 0/0    \\ \hline
N-N$\varepsilon$, N-N$\eta1$   & 0 /8   & 0/0  & 0/0    & 0/0    & 0/0    & 0 /0    & 0/0    & 0/0    \\ \hline
N-N$\varepsilon$, O1-N$\varepsilon$   & 0 /7   & 0/0  & 0/0    & 0/0    & 0/0    & 0 /0    & 0/0    & 0/0    \\ \hline
N-N$\eta1$, O1-N$\eta1$ & 0 /2   & 0/0  & 0/0    & 0/0    & 0/0    & 0 /0    & 0/0    & 0/0    \\ \hline
O1-N$\eta1$, O2-N$\eta2$   & 0 /3   & 0/0  & 0/0    & 0/0    & 0/0    & 0 /0    & 0/0    & 0/0    \\ \hline
\end{tabular}
\end{table}

\clearpage
\section{Harmonic free energies}
Free energies were calculated in the harmonic approximation \cite{McQuarrie, Fultz2010} for selected molecules adsorbed on surfaces within 0.1 eV range.

$$
F(\textrm{T}) = E_{\textrm{PES}} + F_\textrm{vib}(\textrm{T}) + F_\textrm{rot}(\textrm{T}),
$$

where $E_{\textrm{PES}}$ is the total energy obtained with DFT (PBE+vdW$^{\textrm{surf}}$ functional), and

$$
F_{\mathrm{vib}}(\mathrm{T})=\sum_{i}^{3 N-6}\left[\frac{\hbar \omega_{i}}{2}+k_{\mathrm{B}} T \ln \left(1-e^{-\beta \hbar \omega_{i}}\right)\right],
$$

where N is the total number of atoms in the molecule (metal atoms were not displaced and were taken into account in external field), $k_\textrm{B}$ is Boltzmann constant, $T$ is the temperature, $\omega_i$ are vibrational frequencies obtained by diagonalization of Hessian matrix with use of developing version of phonopy-FHI-aims \cite{phonopy, phonopy-fidaynyan} and
 
$$
F_\textrm{rot}(\textrm{T}) =-k_\textrm{B} T \textrm{ln} \left[ \frac{\sqrt{\pi}}{\sigma} \left(  \frac{8\pi^2 I k_\textrm{B} T}{h^2}  \right) \right],
$$
where $I$ is the moment of inertia of the molecule obtained after diagonalization of the inertia tensor of the molecule. For the adsorbed conformers, rotational contributions are completely neglected since rotation around all principal axes of the molecule become internal vibrational modes of the system.
\begin{figure}[h]
\center
\includegraphics[width=0.9\textwidth]{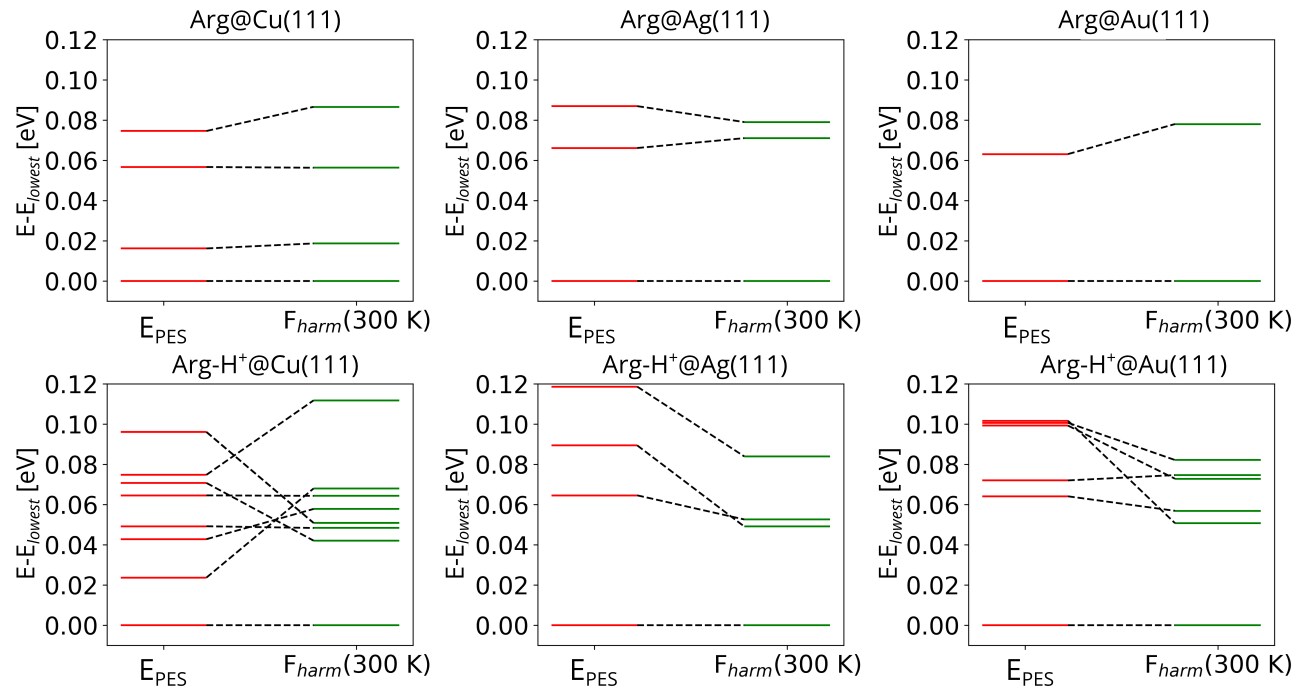}
\caption{Harmonic free energies calculated for adsorbed structures within the lowest 0.1 eV total-energy range. E$_{\textbf{PES}}$ corresponds to the total energy of the system obtained at DFT level and F$_{\textbf{harm}}$ corresponds to the free energy of the system at 300 K calculated as described above.}
\label{sup:Harmonic_free_energies_300}
\end{figure}

\begin{figure}[ht!]
\centering
\includegraphics[width=0.95\textwidth]{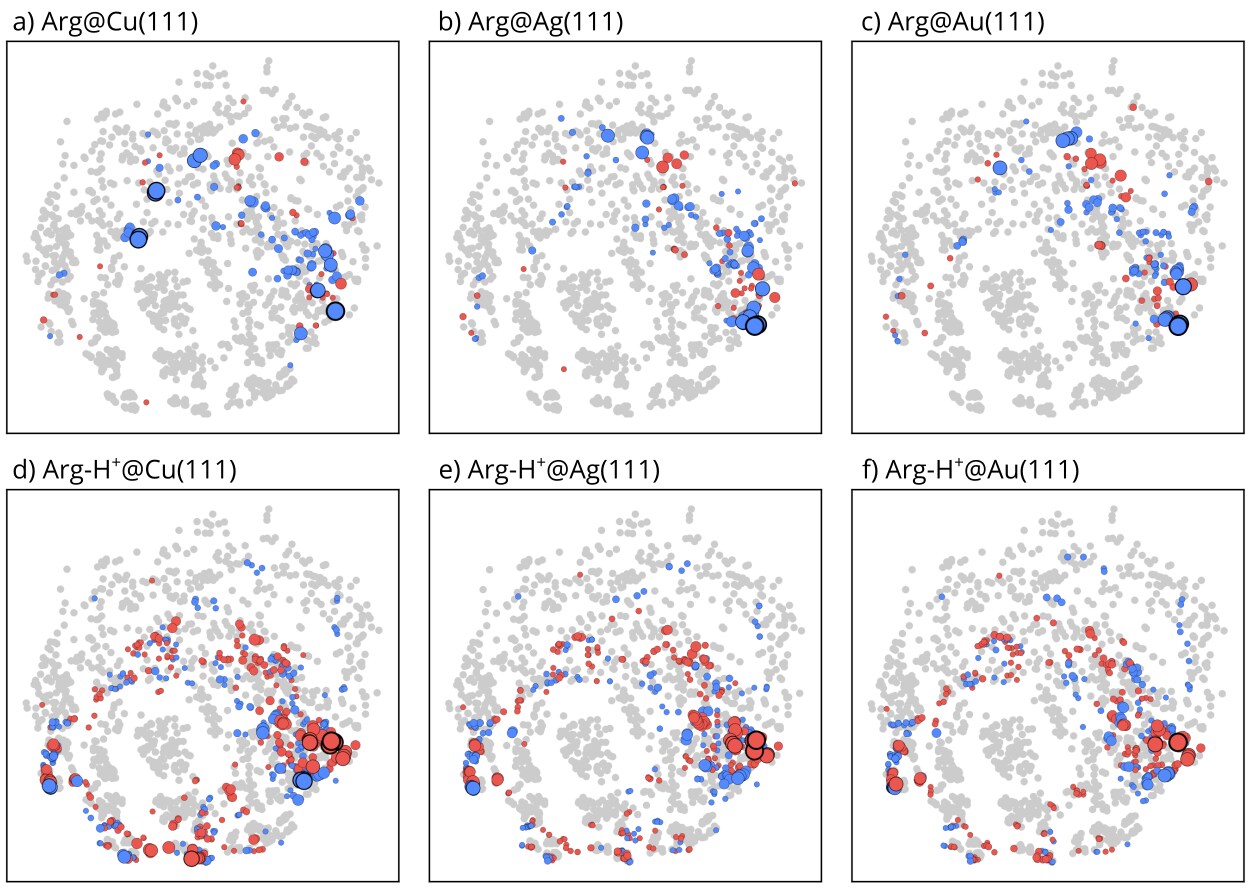}
\caption{Low dimensional maps of Arg and Arg-H$^+$ adsorbed on Cu(111), Ag(111) and Au(111) color-coded with respect to the orientation of the C$_\alpha$H group. Blue correspond to \textit{up} orientation and red correspond to \textit{down} orientation of the C$_\alpha$H group.}
\label{fig:UpDown_map}
\end{figure}

\clearpage
\section{Charge rearrangement while on the surface}

\begin{table}[ht]
\caption{Calculated charge on the molecule with use of Hirshfeld partial charge analysis and by integration of the electron density difference in the molecular region. Values are in electrons. Conformations are depicted in the following Figures S8-13.}
\center
\begin{tabular}{|l|l|l|l|l|l|}
\hline
Conformer & Hirshfeld & Integral & Conformer & Hirshfeld & Integral \\ \hline
\multicolumn{3}{|c|}{Arg@Cu}     & \multicolumn{3}{c|}{Arg-H$^+$@Cu}     \\ \hline
a     & 0.11      & 0.19     & a     & 0.29      & 0.85     \\ \hline
b     & 0.03      & 0.30     & b     & 0.30      & 0.85     \\ \hline
c     & 0.04      & 0.31     & c     & 0.31      & 0.84     \\ \hline
d     & 0.08      & 0.26     & d     & 0.43      & 0.88     \\ \hline
e     & 0.01      & 0.24     & e     & 0.46      & 0.85     \\ \hline
f     & 0.11      & 0.30     & f     & 0.38      & 0.82     \\ \hline
\multicolumn{3}{|c|}{Arg@Ag}     & \multicolumn{3}{c|}{Arg-H$^+$@Ag}     \\ \hline
a     & 0.04      & 0.15     & a     & 0.28      & 0.83     \\ \hline
b     & -0.08     & 0.23     & b     & 0.30      & 0.83     \\ \hline
c     & -0.03     & 0.24     & c     & 0.31      & 0.82     \\ \hline
d     & -0.06     & 0.21     & d     & 0.43      & 0.86     \\ \hline
e     & -0.13     & 0.16     & e     & 0.46      & 0.85     \\ \hline
f     & 0.05      & 0.14     & f     & 0.36      & 0.86     \\ \hline
\multicolumn{3}{|c|}{Arg@Au}     & \multicolumn{3}{c|}{Arg-H$^+$@Au}     \\ \hline
a     & 0.06      & 0.05     & a     & 0.32      & 0.86     \\ \hline
b     & -0.01     & 0.29     & b     & 0.29      & 0.86     \\ \hline
c     & 0.00      & 0.30     & c     & 0.34      & 0.85     \\ \hline
d     & -0.10     & 0.25     & d     & 0.48      & 0.91     \\ \hline
e     & 0.01      & 0.23     & e     & 0.49      & 0.90     \\ \hline
f     & 0.06      & 0.31     & f     & 0.43      & 0.92     \\ \hline
\end{tabular}
\end{table}

\begin{figure}[h]
\center
\includegraphics[width=0.8\textwidth]{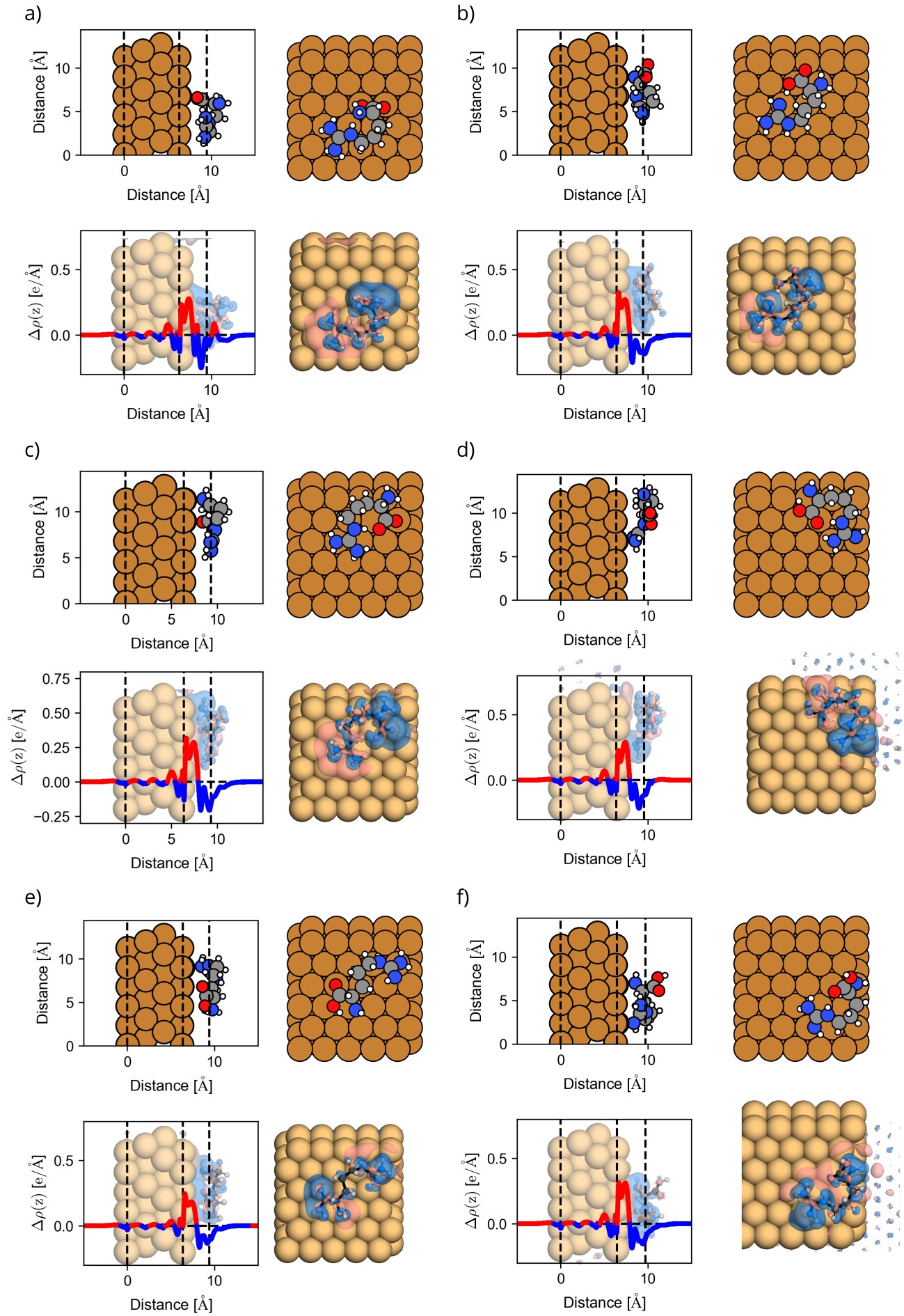}
\caption{Side and top views of the adsorbed structures of Arg on Cu(111). Dashed black lines correspond to: average z position of the atoms in the lowest layer of the surface (left), average z position of  atoms in the highest layer of the surface (middle), centre of the mass of the molecule (right). Red/blue solid lines (and also red/blue regions) correspond to the electron density accumulation/depletion, calculated as discussed in the manuscript in the section 2.3.}
\label{sup:Charge_rearrangement_ArgCu}
\end{figure}

\begin{figure}[h]
\center
\includegraphics[width=0.8\textwidth]{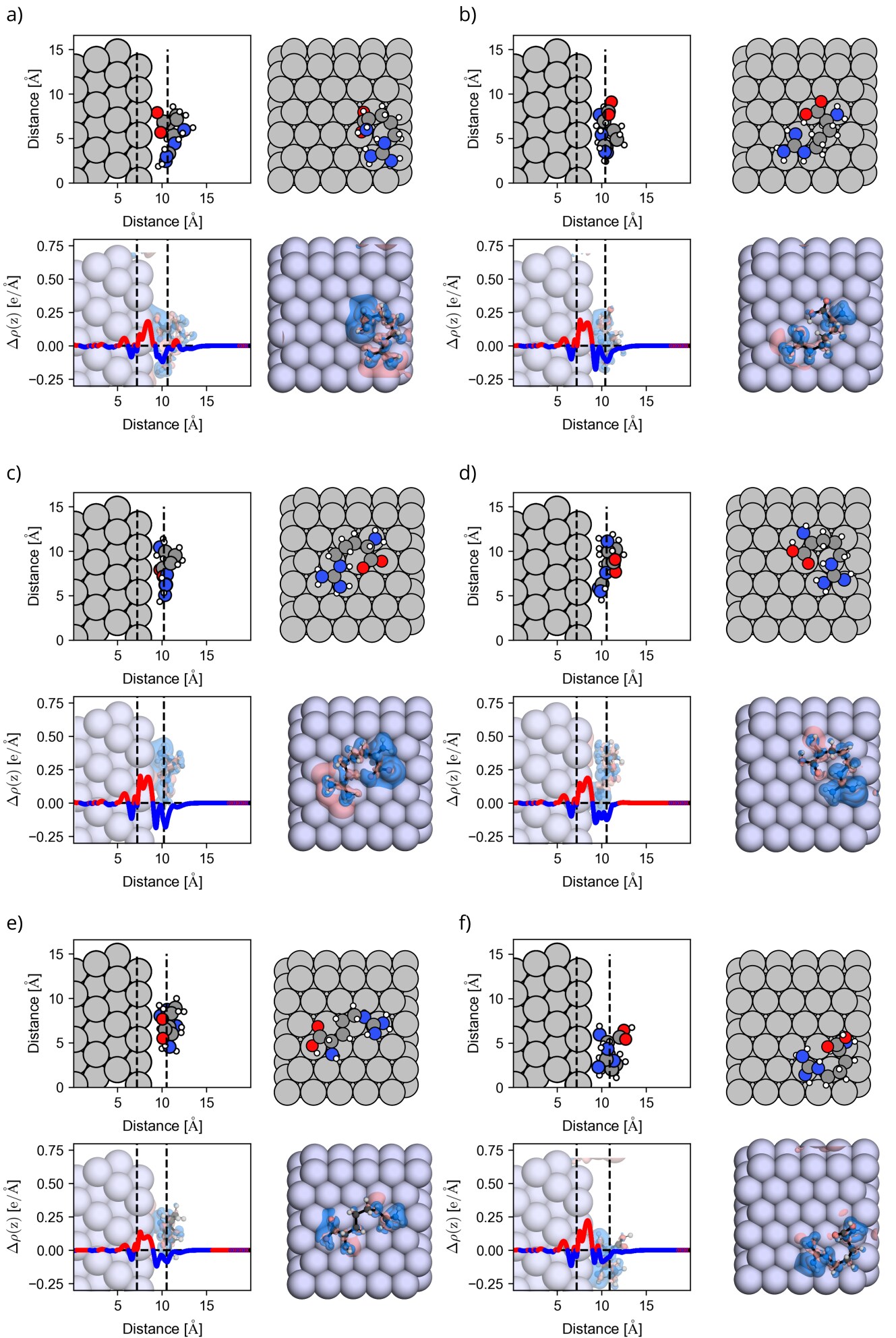}
\caption{Side and top views of the adsorbed structures of Arg on Ag(111). Dashed black lines correspond to: average z position of the atoms in the lowest layer of the surface (left), average z position of  atoms in the highest layer of the surface (middle), centre of the mass of the molecule (right). Red/blue solid lines (and also red/blue regions) correspond to the electron density accumulation/depletion, calculated as discussed in the manuscript in the section 2.3.}
\label{sup:Charge_rearrangement_ArgAg}
\end{figure}

\begin{figure}[h]
\center
\includegraphics[width=0.8\textwidth]{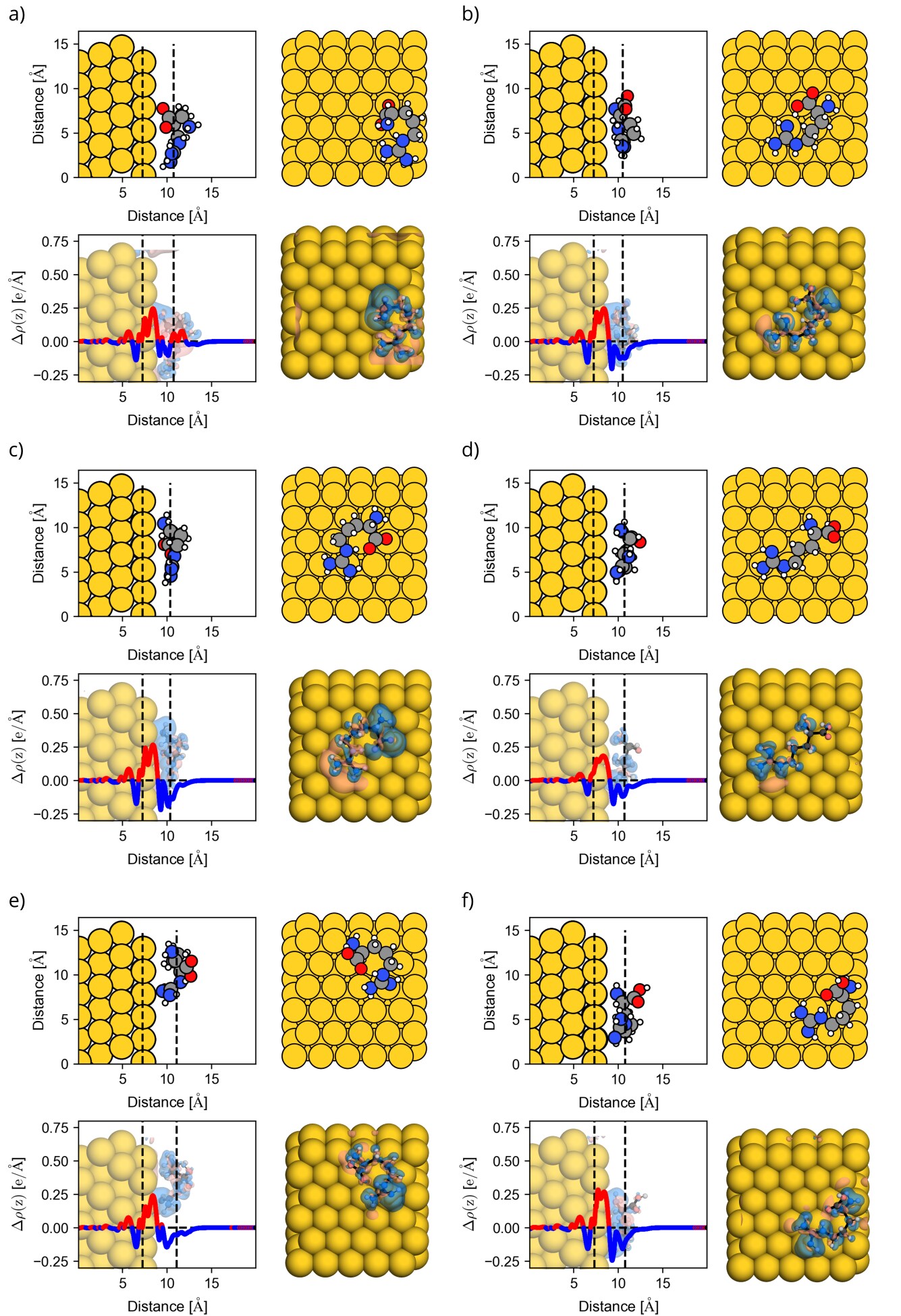}
\caption{Side and top views of the adsorbed structures of Arg on Au(111). Dashed black lines correspond to: average z position of the atoms in the lowest layer of the surface (left), average z position of  atoms in the highest layer of the surface (middle), centre of the mass of the molecule (right). Red/blue solid lines (and also red/blue regions) correspond to the electron density accumulation/depletion, calculated as discussed in the manuscript in the section 2.3.}
\label{sup:Charge_rearrangement_ArgAu}
\end{figure}

\begin{figure}[h]
\center
\includegraphics[width=0.8\textwidth]{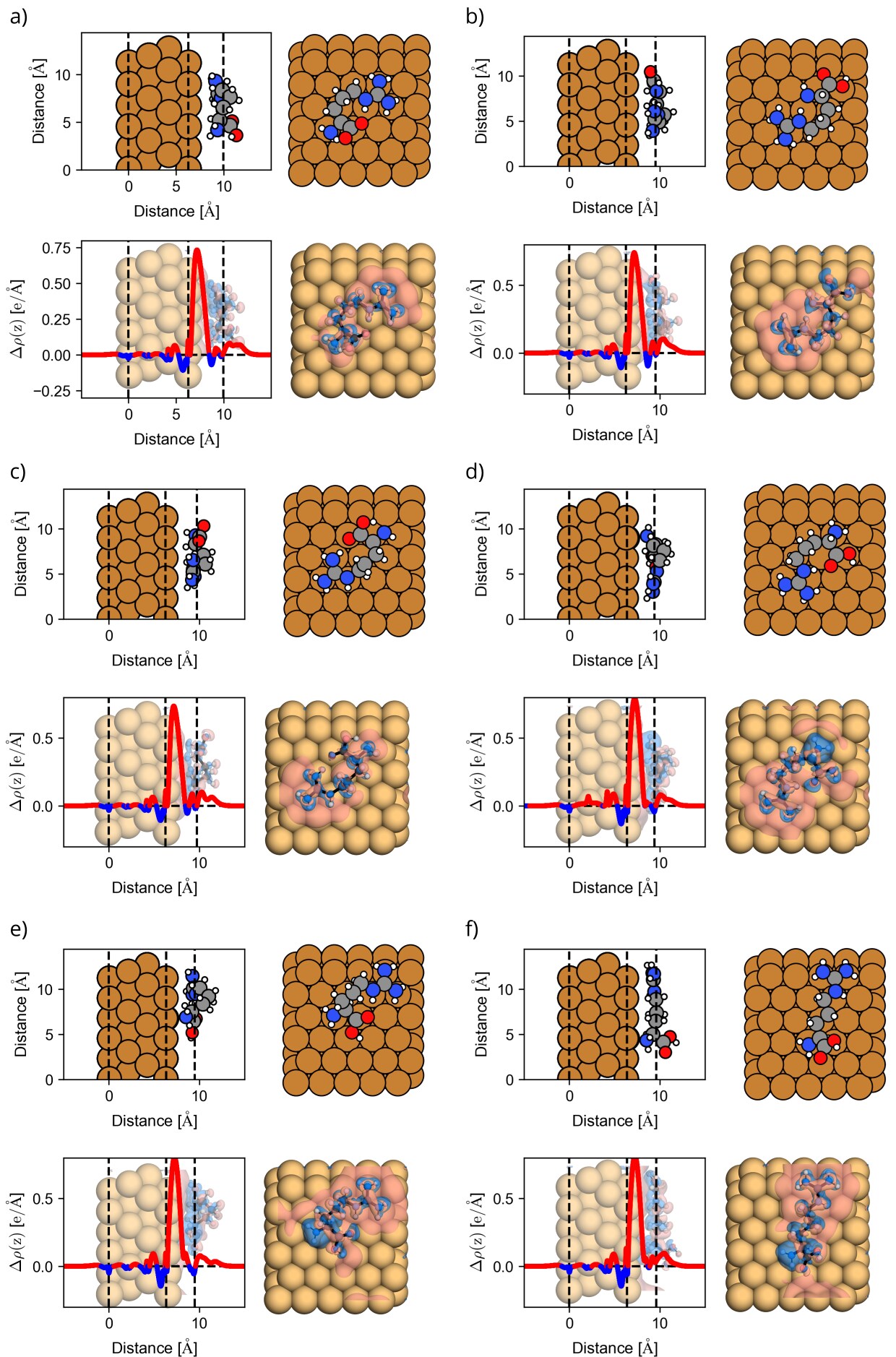}
\caption{Side and top views of the adsorbed structures of Arg-H$^+$ on Cu(111). Dashed black lines correspond to: average z position of the atoms in the lowest layer of the surface (left), average z position of  atoms in the highest layer of the surface (middle), centre of the mass of the molecule (right). Red/blue solid lines (and also red/blue regions) correspond to the electron density accumulation/depletion, calculated as discussed in the manuscript in the section 2.3.}
\label{sup:Charge_rearrangement_ArgHCu}
\end{figure}

\begin{figure}[h]
\center
\includegraphics[width=0.8\textwidth]{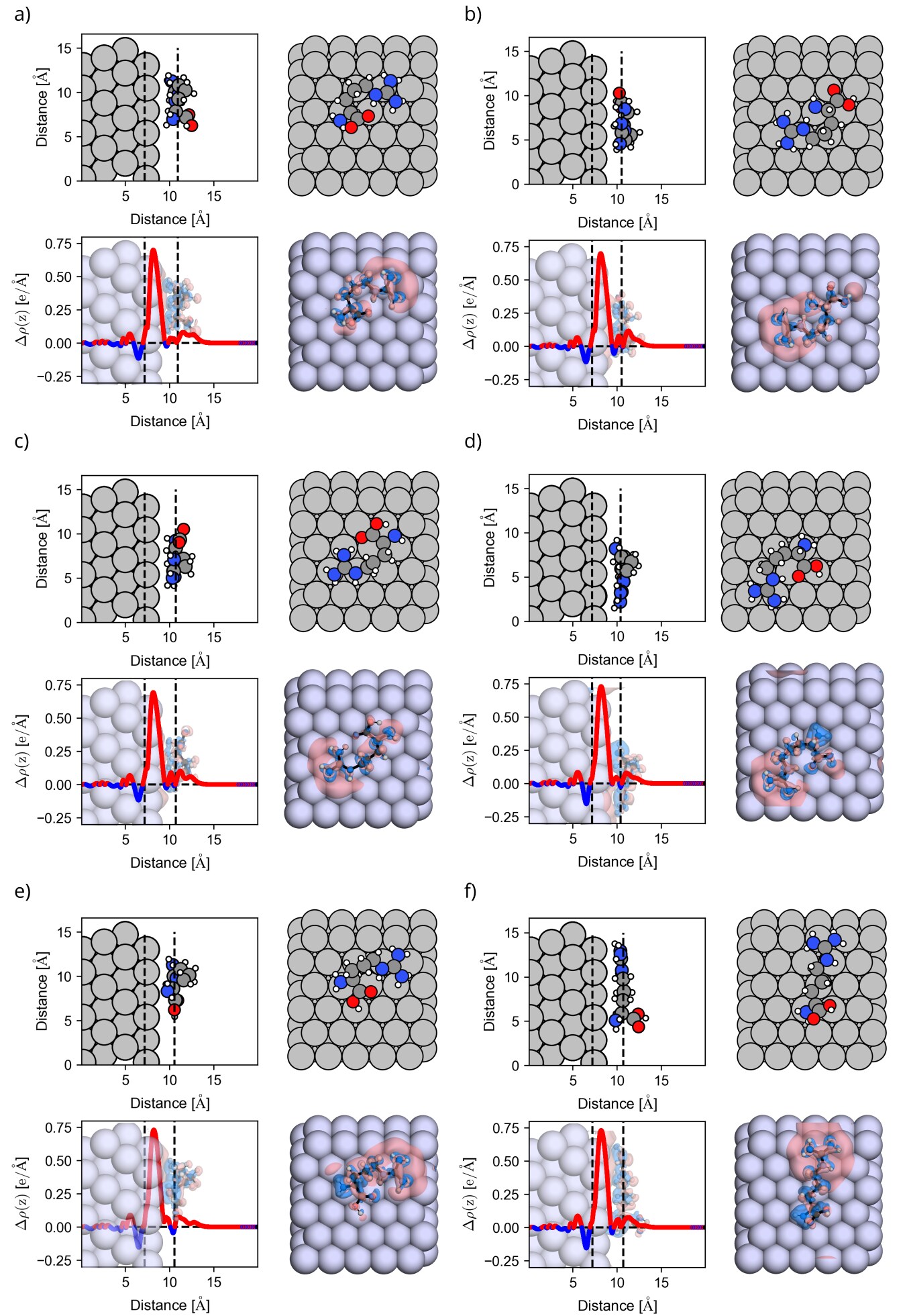}
\caption{Side and top views of the adsorbed structures of Arg-H$^+$ on Ag(111). Dashed black lines correspond to: average z position of the atoms in the lowest layer of the surface (left), average z position of  atoms in the highest layer of the surface (middle), centre of the mass of the molecule (right). Red/blue solid lines (and also red/blue regions) correspond to the electron density accumulation/depletion, calculated as discussed in the manuscript in the section 2.3.}
\label{sup:Charge_rearrangement_ArgHAg}
\end{figure}

\begin{figure}[h]
\center
\includegraphics[width=0.8\textwidth]{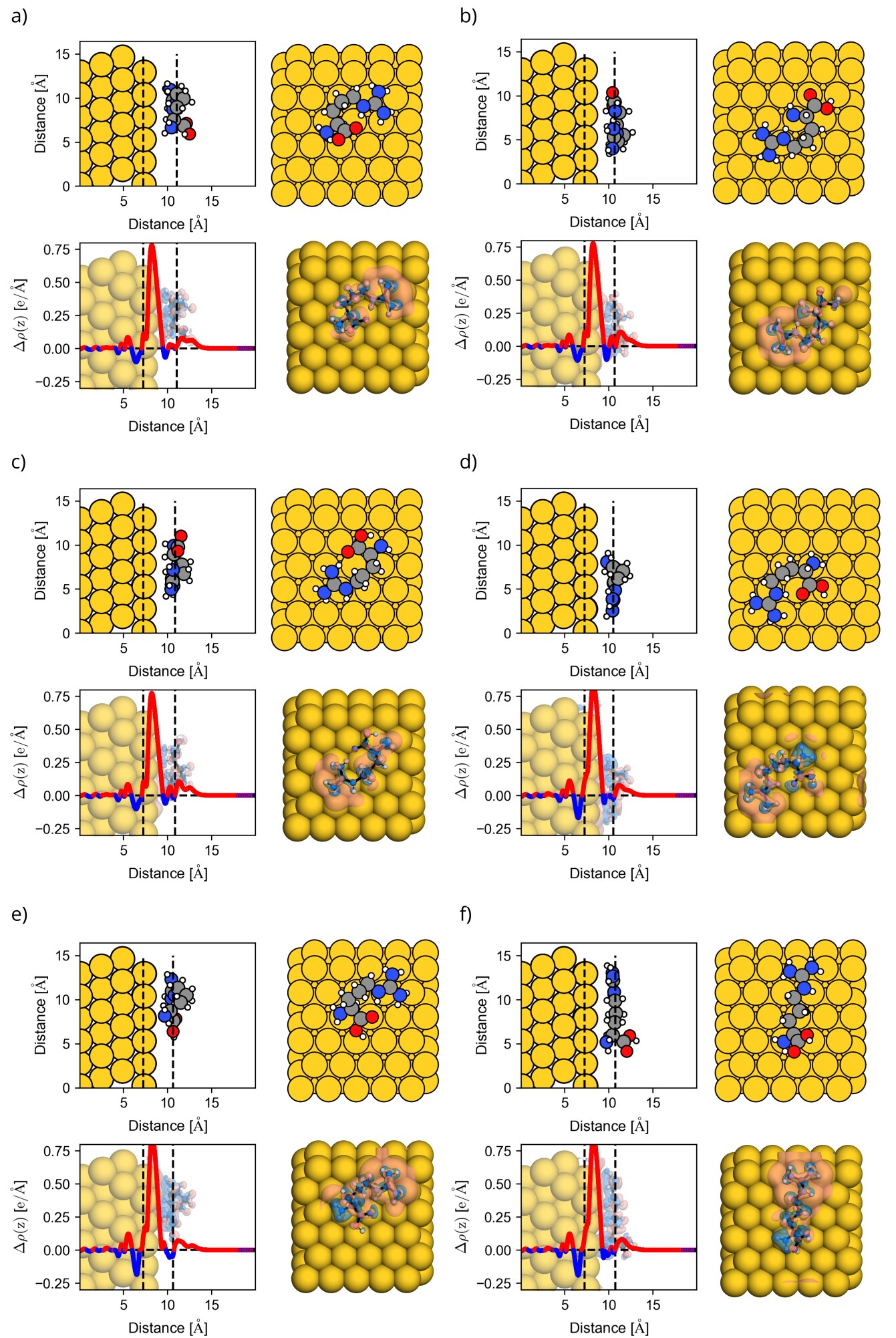}
\caption{Side and top views of the adsorbed structures of Arg-H$^+$ on Au(111). Dashed black lines correspond to: average z position of the atoms in the lowest layer of the surface (left), average z position of  atoms in the highest layer of the surface (middle), centre of the mass of the molecule (right). Red/blue solid lines (and also red/blue regions) correspond to the electron density accumulation/depletion, calculated as discussed in the manuscript in the section 2.3.}
\label{sup:Charge_rearrangement_ArgHAu}
\end{figure}

In order to take a look in electronic level alignments of interface system after adsorption the projected, angular-momentum resolved partial density of states (pDOS) averaged over all atoms of each species  were calculated and normalized per molecule and per surface respectively. For corresponding isolated molecular geometry HOMO and LUMO levels were calculated and plotted together with interface pDOS. These calculations were performed with higher number of k-grid points: 6x6x1. Gaussian broadening was chosen to be 0.05. 

\begin{figure}[h!]
\center
\includegraphics[width=0.8\textwidth]{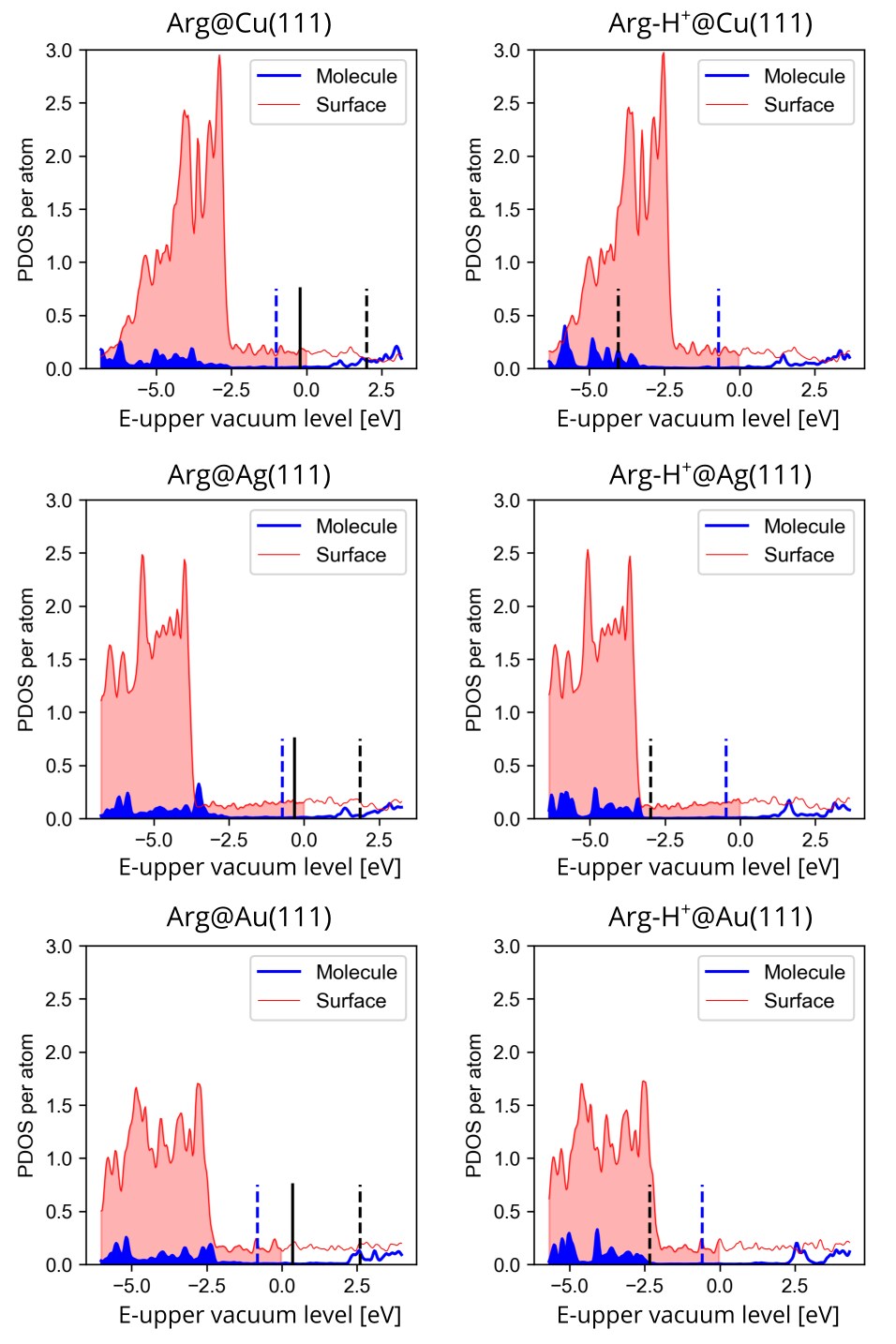}
\caption{Projected densities of states of the lowest energy structures on each surface. Filled area corresponds to the occupied states below highest occupied state (VBM) of the whole system. HOMO (black solid line) and LUMO (black dashed line) are the states of the corresponding gas-phase molecular conformer calculated with the same geometry as it adopts when adsorbed. The Fermi energy of the pristine slab is depicted with blue dashed line.} 
\label{sup:DOS_PDOS}
\end{figure}

\begin{figure}[h!]
\center
\includegraphics[width=1.0\textwidth]{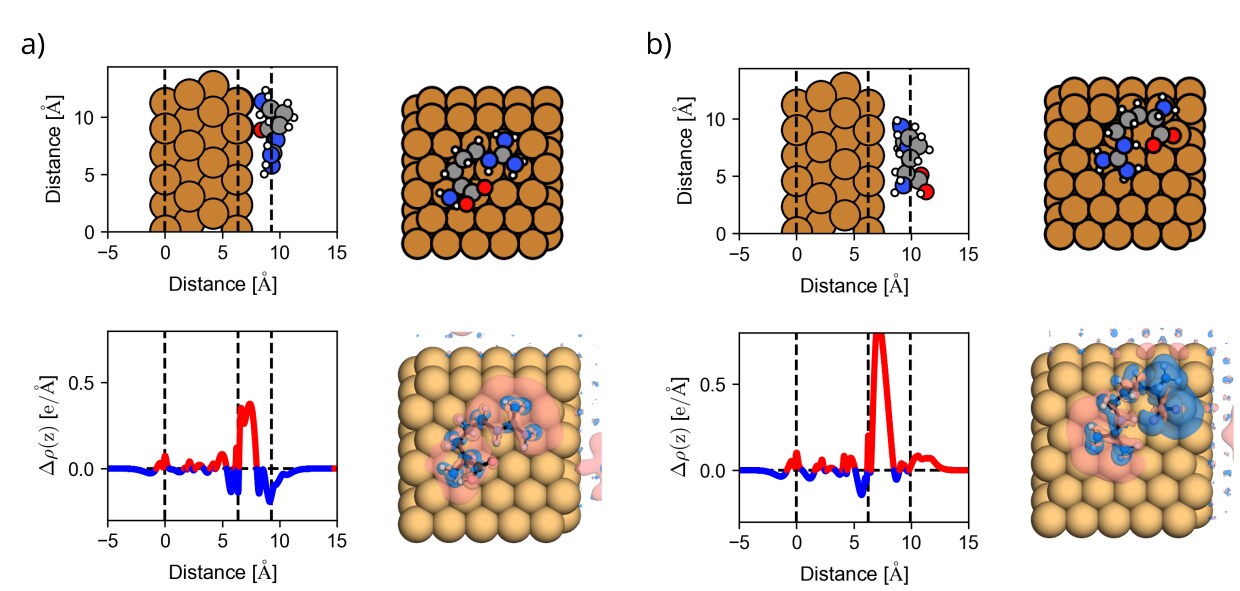}
\caption{Side and top views of the adsorbed structures of a) Arg on Cu(111) (conformer c in Fig. S8) and b) Arg-H$^+$ on Cu(111) (conformer a in Fig. S11). Dashed black lines correspond to: average z position of the atoms in the lowest layer of the surface (left), average z position of  atoms in the highest layer of the surface (middle), centre of the mass of the molecule (right). Red/blue solid lines (and also red/blue regions) correspond to the electron density accumulation/depletion, calculated as discussed in the manuscript in the section 2.3 with PBE0 functional.} 
\label{sup:DOS_PDOS}
\end{figure}

\clearpage
\section{Deprotonation on the surfaces}

In Arg, we found it most favorable to detach the proton from the guanidino group, while for Arg-H$^+$, it was most favorable to detach the proton from the carboxyl group. In both cases we note that the final adsorbed species is a hydrogen, i.e., it does not carry a positive charge. We chose three representative conformers  at each surface: the lowest energy structure and other two with different H-bonds within the molecule. We placed the detached proton at a distance of at least 2.5 \AA~ from the molecule and fully optimized the dissociated structures. Comparing the energy difference between the final and initial states gives a lower limit for the dissociation barrier.

\begin{figure}[h]
\center
\includegraphics[width=0.9\textwidth]{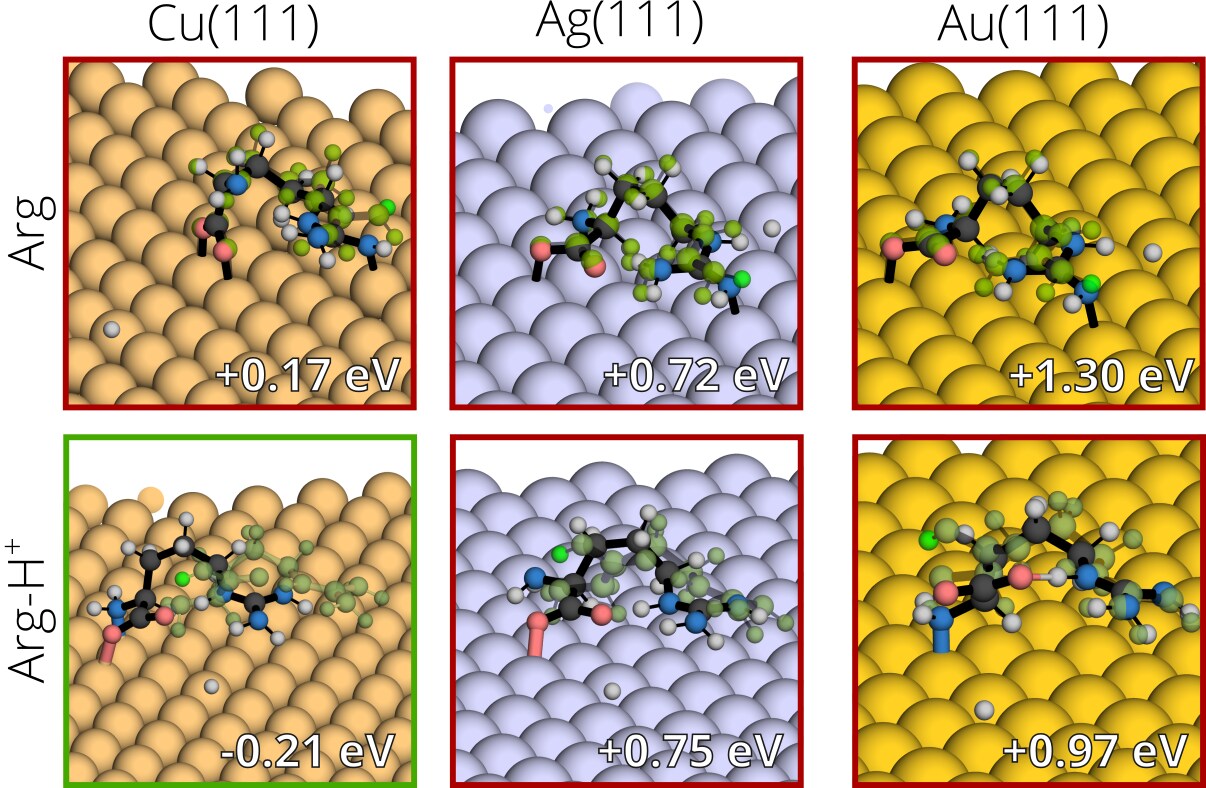}
\caption{Representative structures that were analyzed for the calculation of the deprotonation energies. Colored structures represent the deprotonated relaxed structure. The green translucent structures represent the initial structure from which hydrogen was removed and placed on surface. The hydrogen that was removed is highlighted in bright green. $\Delta E$ (see main text) is also reported in each panel.}
\label{sup:Deprotonation}
\end{figure}

\begin{figure}[h]
\includegraphics[width=0.9\textwidth]{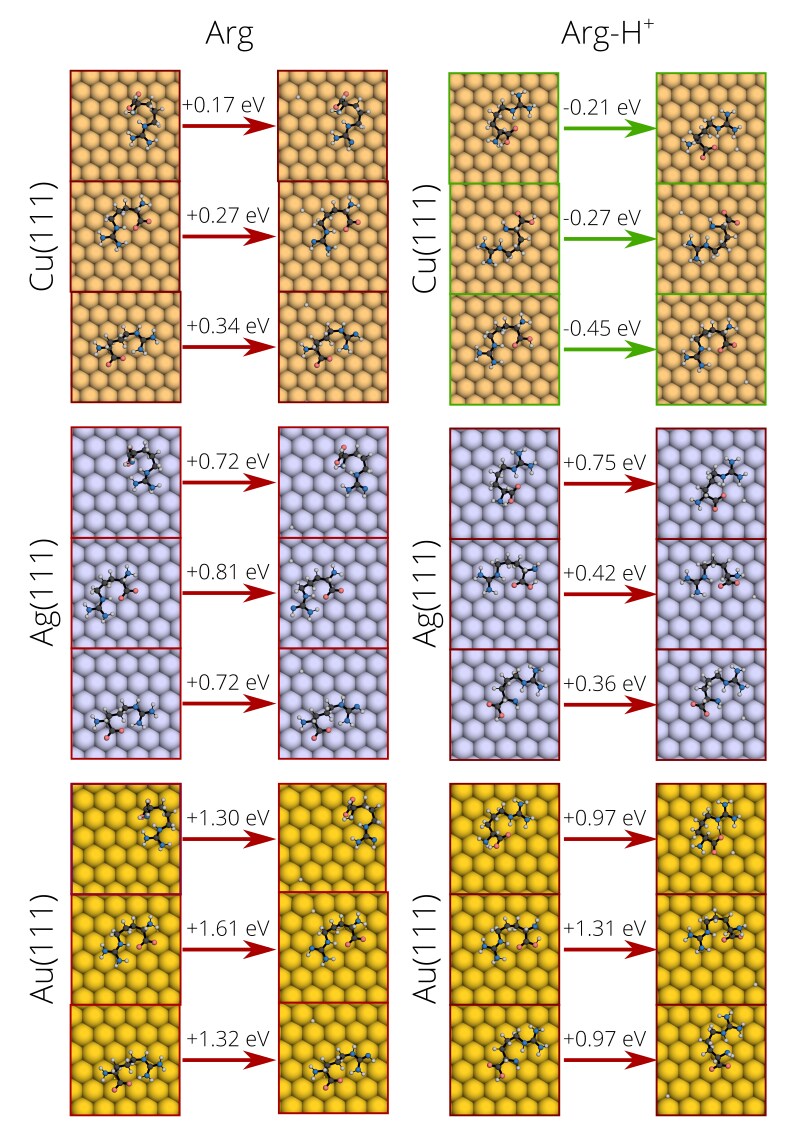}
\caption{All structures that were analyzed for the calculation of the deprotonation energies. $\Delta E$ (see main text) is also reported in each panel.}
\label{supfig:Deprotonation_v0}
\end{figure}

\clearpage

\section{Comparison of DFT with INTERFACE-FF}

\begin{figure}[htbp]
  \centering
    \includegraphics[width=1.05\textwidth]{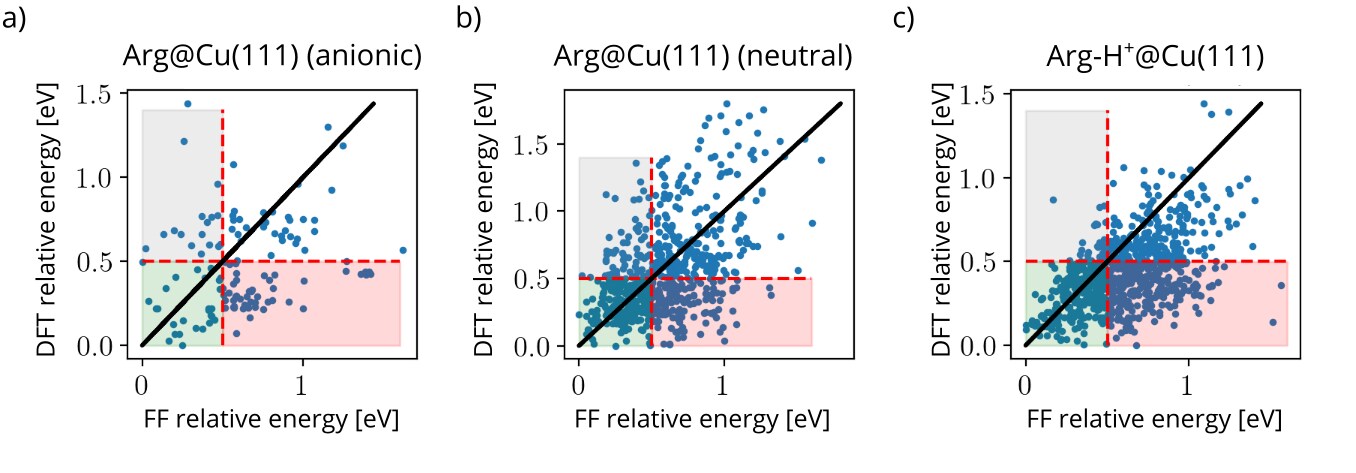}
  \caption{Comparison of the relative energies obtained from DFT optimized structures and the same structures after post-relaxation in with the INTERFACE force field. Dots on the diagonal line represent an optimal correlation. The red area marks structures that lie in the lower 0.5 eV energy range in DFT but above the 0.5 eV energy range in INTERFACE-FF. The green area marks the structures that are in the lower 0.5 eV energy range regardless of the level of theory. The grey area marks the structures that are above the 0.5 eV energy range in DFT but below the 0.5 eV energy range in INTERFACE-FF.}
  \label{sup:RelEn_ArgHCu_DFTvsFF}
\end{figure}

Selected local minima obtained at DFT level of theory were optimized with the  INTERFACE-FF\cite{Heinz:2016:Review} 
using the NAMD package \cite{Phillips2005}. Calculations were performed with periodic boundary conditions with the same cell size and number of Cu atoms as in the DFT calculations. 
We obtained parameters for certain protonation states as described in the following.
For the calculation of Arg, two protomers \textbf{P1} and \textbf{P3} (as denoted in the main text) had to be prepared.
They are called ``ARN'' (P1)  and  ``ARZ'' (P3) in the topology file that is provided. 

The parametrization of
``ARZ'' proceeded by taking the C-terminus in the deprotonated form (PRES CTER) and the N-terminus in the neutral form (PRES NNEU) from \texttt{top-all36-prot.rtf}, such that the protonation is COO-NH2 with total charge 0. The partial charge of the guanidino group is +1. Other parameters were taken from ARG (\texttt{top-all22-prot-metals.inp}) which is in the protonated form, by default, in CHARMM force field.

\begin{figure}[htbp]
  \centering
    \includegraphics[width=0.95\textwidth]{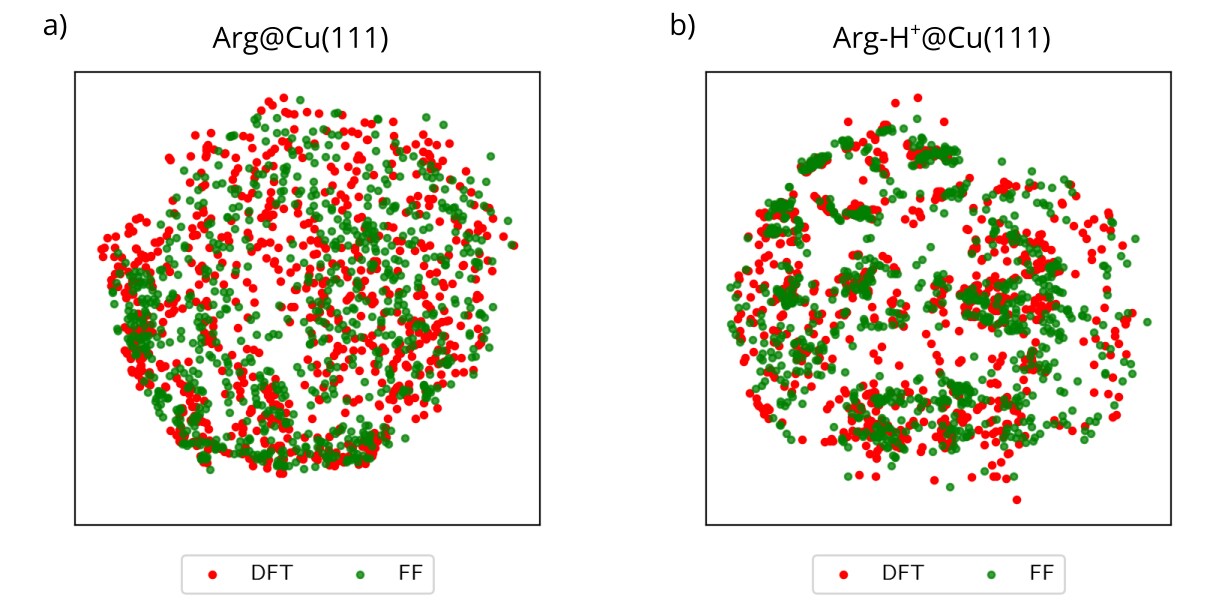}
  \caption{Low-dimensional map of the conformational space of the Arg and Arg-H$^+$ molecules adsorbed on the Cu(111) surface. The map was optimized considering all DFT and INTERFACE-FF structures. Green dots represent conformations obtained at DFT level of theory and red dots represent conformations obtained after geometry optimization with INTERFACE-FF. Close proximity of the dots reflects their structural similarity.}
  \label{sup:ArgCu_moleculeDFTvsFF_Sketchmap}
\end{figure}

The parametrization of ``ARN'' proceeded by taking the parameters for neutral C-terminus and N-terminus (PRES CNEU and PRES NNEU from \texttt{top-all36-prot.rtf}) and the deprotonated methyl-guanidinium group (RESI MGU2) from \texttt{par-all36-cgenff.rtf}. Atom types and parameters for MGU2 were copied from \texttt{par-all36-cgenff.rtf}. The missing parameters related to joining MGU2 with the rest of Arg were manually added to the topology file. They were obtained from the corresponding values appearing in the protonated Arg, assuming that CG331 == CT2, NG311 == NC2, HGPAM1 == HC, NG2D1 == NC2, CG2N1 == C, where needed. The partial charge of CD atom was manually adjusted (decreased by 0.1) in order to have a neutral molecule with neutral guanidino group. Parameters for the neutral C-terminus and N-terminus were taken from the  \texttt{top-all36-prot.rtf} file (PRES CNEU and PRES NNEU) and manually added to the customized file of topology.

ArgH named as ``ARX'' has total charge +1 (partial charge of guanidino group is also +1) and COOH-NH2 termini which is neutral. All the other parameters were directly taken from the INTERFACE-FF \texttt{all22-prot-metals} topology and parameter files. 

We conclude from Fig. \ref{sup:RelEn_ArgHCu_DFTvsFF} that DFT (PBE+vdW$^{\text{surf}}$) and the INTERFACE-FF yield very different energy hierarchies. However, from Fig. \ref{sup:ArgCu_moleculeDFTvsFF_Sketchmap}, we conclude that both levels of theory represent a similar conformational space. However, Table \ref{sup:table-ff-dft-adsorption-sites} shows that DFT and the FF yield different adsorption site preferences for the amino and carboxyl groups. In particular, DFT predicts that O will adsorb almost exclusively on top sites, consistent with the accepted adsorption site preference of CO groups on the pristine Cu(111) surface. The FF predicts a larger population of other adsorption sites, in particular hollow sites, compared to DFT. 

\begin{table}[htbp]
\center
\caption{Surface site adsorption preferences of chosen chemical groups in Arg and Arg-H$^+$. All numbers are reported as a percentage of the total number of conformers optimized with DFT (PBE+vdW$^{\text{surf}}$) and the INTERFACE force field. \label{sup:table-ff-dft-adsorption-sites}}
\begin{tabular}{|l|l|l|l|l|l|l|l|l|}
\hline
                & \multicolumn{4}{c|}{Arg@Cu(111)}               & \multicolumn{4}{c|}{Arg-H$^+$@Cu(111)}                     \\ \hline
                & \multicolumn{2}{c|}{Amino} &  \multicolumn{2}{c|}{Carboxyl}    & \multicolumn{2}{c|}{Amino} & \multicolumn{2}{c|}{Carboxyl} \\ \hline
Adsorption site & DFT          & FF          & DFT        & FF   & DFT          & FF          & DFT            & FF           \\ \hline
Top             & 80           & 53          & 76         & 48   & 59           & 50          & 70             & 45           \\ \hline
Bridge          & 9            & 18          & 14         & 18   & 18           & 20          & 15             & 22           \\ \hline
Hollow-FCC      & 5            & 13          & 4          & 17   & 13           & 15          & 7              & 16           \\ \hline
Hollow-HCP      & 6            & 16          & 5          & 17   & 10           & 15          & 9              & 18           \\ \hline
\end{tabular}
\end{table}

\clearpage
\bibliography{bibliography}